\documentclass[aps,prd,twocolumn,
               notitlepage,mathrsfs,
               amssymb,amsmath,amsfonts,
               superscriptaddress,
               longbibliography,
               nofootinbib,floatfix]{revtex4-2}
              
\usepackage[normalem]{ulem}
\usepackage{graphicx}
\usepackage[linktocpage,breaklinks]{hyperref}
\usepackage[capitalize]{cleveref}
\usepackage[usenames,dvipsnames]{xcolor}
\hypersetup{colorlinks=true,citecolor=NavyBlue,
            linkcolor=magenta,
            urlcolor=magenta}
\usepackage{multirow,array}
\DeclareMathAlphabet{\pazocal}{OMS}{zplm}{m}{n}

\newcommand{\dd}{\rm d}
\newcommand{\dedp}{\frac{\partial e}{\partial p}}

\begin{document}

\title{Accessing universal relations of binary neutron star waveforms in massive scalar-tensor theory}

\date{\today}

\author{Alan Tsz-Lok Lam}
\email{tszlok.lam@aei.mpg.de}
\affiliation{Max Planck Institute for Gravitational Physics (Albert Einstein Institute), 14476 Potsdam, Germany}

\author{Yong Gao}
\affiliation{Max Planck Institute for Gravitational Physics (Albert Einstein Institute), 14476 Potsdam, Germany}

\author{Hao-Jui Kuan}
\affiliation{Max Planck Institute for Gravitational Physics (Albert Einstein Institute), 14476 Potsdam, Germany}

\author{Masaru Shibata}
\affiliation{Max Planck Institute for Gravitational Physics (Albert Einstein Institute), 14476 Potsdam, Germany}
\affiliation{Center of Gravitational Physics and Quantum Information, Yukawa Institute for Theoretical Physics, Kyoto University, Kyoto, 606-8502, Japan} 

\author{Karim Van Aelst}
\affiliation{Max Planck Institute for Gravitational Physics (Albert Einstein Institute), 14476 Potsdam, Germany}

\author{Kenta Kiuchi}
\affiliation{Max Planck Institute for Gravitational Physics (Albert Einstein Institute), 14476 Potsdam, Germany}
\affiliation{Center of Gravitational Physics and Quantum Information, Yukawa Institute for Theoretical Physics, Kyoto University, Kyoto, 606-8502, Japan} 

\begin{abstract}
We investigate how the quasi-universal relations connecting tidal deformability with gravitational waveform characteristics and/or properties of individual neutron stars that were proposed in the literature within general relativity would be influenced in the massive Damour-Esposito-Farese-type scalar-tensor gravity.
For this purpose, we systematically perform numerical relativity simulations of $\sim120$ binary neutron star mergers with varying scalar coupling constants. 
Although only three neutron-star equations of state are adopted, a clear breach of universality can be observed in the data sets. 
In addition to presenting difficulties in constructing quasi-universal relations in alternative gravity theories, we also briefly compare the impacts of non-general-relativity physics on the waveform features and those due to the first order or cross-over quantum chromodynamical phase transition.
\end{abstract}
\maketitle

\section{Introduction}
Coalescence of binary neutron stars (BNSs) offers a unique avenue to test gravity in its strong regime and to probe thermodynamic states of matter at subatomic densities. The gravitational wave (GW) signal originating from such a process was detected for the first time in 2017 by LIGO and VIRGO observatories~\cite{LIGOScientific:2014pky,VIRGO:2014yos}, though only in the late-inspiral epoch. 
This event, GW170817 \cite{LIGOScientific:2017vwq,LIGOScientific:2018mvr,LIGOScientific:2018hze}, has led to certain constraints on gravitation \cite{LIGOScientific:2018dkp,Berti:2018cxi,Perkins:2020tra} and the equation of state (EOS) of nuclear matter \cite{LIGOScientific:2018cki,Raithel:2019uzi,Chatziioannou:2020pqz,Dietrich:2020eud}.
The analysis was conducted assuming general relativity (GR) as the underlying theory of gravity to agnostically bound the observation's deviation from the prediction of GR.
However, tests of a specific alternative theory of gravity require the development of waveform templates within the theory and may entail certain modifications in the data analysis formalism. 
Although analytic efforts in waveform modeling have been devoted to some theories, e.g., the scalar-tensor theory and the scalar-Gauss-Bonnet theory, a lot remains to be done to establish machinery at the same level of sophistication as that in GR to analyze GWs.

GWs emitted during and in the aftermath of the merger would lie in the frequency band of $2$--$4$~kHz if the system produces a hypermassive neutron star (HMNS) as a transient remnant~\cite{Shibata:2005mz, Bauswein:2011tp, Hotokezaka:2011dh}. 
The current ground-based GW detectors are less sensitive in these bands \cite{LIGOScientific:2017fdd,LIGOScientific:2018urg,LIGOScientific:2020aai}; in fact, even with the design sensitivity of Advanced LIGO, the postmerger waveform of a GW170817-like event might only have a SNR of $\sim2$--$3$, which can hardly be detected.
However, waveforms at a few kHz may be reachable with the next-generation detectors such as the Einstein Telescope~\cite{Hild:2010id, Punturo:2010zz, Branchesi:2023mws} and the Cosmic Explorer~\cite{LIGOScientific:2016wof, Reitze:2019iox, Evans:2021gyd}, for which the sensitivity is by a factor of $\agt 10$ higher than those of current detectors. 

Postmerger waveforms are informative of the dynamics of remnant systems. Of particular interest are the mergers that lead to an HMNS temporarily supported by differential rotations \cite{Baumgarte:1999cq, Shibata:2005ss} and high thermal pressure \cite{Bauswein:2010dn, Hotokezaka:2013iia,  Kaplan:2013wra,Muhammed:2024zmk}.
The fluid motions within these remnants will emit a loud GW transient over $\sim 10$--20\,ms with characteristic frequencies corresponding to the oscillation modes excited in the remnant massive NS \cite{Shibata:2003ga,Shibata:2005ss,Shibata:2005xz,Oechslin:2007gn,Bauswein:2011tp,Hotokezaka:2013iia,Takami:2014zpa,Takami:2014tva}. 
The dominant peak in the spectrum can be related to the fundamental mode of the remnant, whose frequency depends sensitively on both the EOS and the underlying gravitational theory \cite{Sotani:2004rq,Kruger:2021yay,Blazquez-Salcedo:2021exm}.
Therefore, the measurement of this frequency provides combined information about the nature of gravity and supranuclear matters.

However, to what extent we can learn about the gravitation and the EOS is subject to at least these technical and theoretical challenges:
(i) the morphology of the postmerger waveforms are qualitatively different from that of inspiral, requiring different modeling and analysis strategies \cite{Clark:2014wua,Chatziioannou:2017ixj,Pesios:2024bve}, and 
(ii) the influences of the microphysics and the gravitational aspects of the problem on waveforms are strongly degenerate \cite{Shao:2019gjj}, which hinders a clear determination of matter effects and deviations from GR.
One of the cogent proposals to address the latter issue appeals to quasi-universal relations that connect the spectral properties of postmerger waveforms with properties of cold stars in isolation or participating in a coalescing binary.

Within GR, these quasi-universal relations
are leveraged to infer quantities that are not directly observable \cite{Yagi:2013awa,Yagi:2013bca,Maselli:2013mva,Yagi:2015pkc,Chatziioannou:2018vzf,Torres-Rivas:2018svp}, facilitate efficient Bayesian analysis \cite{Breschi:2019srl,Breschi:2021xrx,Wijngaarden:2022sah,Puecher:2022oiz,Suleiman:2024ztn}, and develop phenomenological waveform models by reducing the matter’s degrees of freedom \cite{Takami:2014tva,Bernuzzi:2014kca,Yagi:2016bkt,Bauswein:2017aur}.
In alternative theories, the EOS-insensitive feature of these relations will be useful in disentangling the EOS effects from gravity, and thus can help to distinguish non-GR imprints from the uncertainties of EOS.
However, this method requires a cautious evaluation of the reliability of these relations in the gravity theory under study to prevent any contamination in the inference.
Taking the massive Damour-Esposito-Farese-type (DEF; \cite{Damour:1992we,Damour:1993hw,Damour:1995kt,Damour:1998jk}) scalar-tensor theory of gravity as an example, whose action is given as \cite{Kuroda:2023zbz,Kuan:2023hrh,Lam:2024wpq},
\begin{align}
    S =& \frac{1}{16\pi}\int d^4x \sqrt{-g}
    \left[ \phi {\cal R} - \frac{\omega(\phi)}{\phi}
    \nabla_a \phi \nabla^a \phi - \frac{2m_\phi^2\varphi^2\phi^2}{B} \right] \nonumber\\
    &- S_{\rm matter},
\end{align}
we illustrate in the remainder of this article that many (if not all) of the quasi-universal relations on the market are actually breached, hinting at a strong caveat of using them for Bayesian analysis.
Here, ${\cal R}$ is the Ricci scalar associated with metric $g_{ab}$, $g$ is the determinant of the metric, $\phi$ is the scalar field, $S_{\rm matter}$ is the action for matter, $\omega(\phi)$ is defined via $[\omega(\phi)+3/2]^{-1}= B\, \ln \phi$ for the scalar coupling constant $B$, and $\varphi=\sqrt{2\ln\phi}$ is an auxiliary scalar field.
The scalar mass has been constrained by pulsar observations \cite{Shao:2017gwu,Anderson:2019eay,Zhao:2022vig} as $m_\phi>10^{-15}$~eV \cite{Ramazanoglu:2016kul,Yazadjiev:2016pcb}.
In addition, GW170817 can tentatively suggest a lower bound on scalar mass as $m_\phi>10^{-12}$~eV \cite{Zhang:2021mks,Kuan:2023hrh}.
In this article, we will consider $m_\phi=1.33\times10^{-11}$~eV (Compton wavelength of $\approx 15$\,km), which suffices to demonstrate the main conclusion: we will emphasize the violation of the quasi-universal relations, which can only be more profound for smaller $m_\phi$.

For the simulations considered in this article, the coupling constant of the scalar field in the DEF theory has been chosen such that the non-GR effects can only marginally appear during inspiral to respect the observation of GW170817 in~\cite{Lam:2024wpq}.
In particular, the radius-mass and tidal deformability-mass relations for the considered scalar coupling constants are shown in \cref{fig:TOV} (see the supplemental material for the equations for computing tidal deformability in the massive DEF theory).
We can see a qualitative difference between the sequence of the H4 EOS and those of the other two EOS: the scalarized sequence of equilibria of static, spherical stars does not merge into the GR branch in the high-density regime. 
The steep softening behavior of the H4 EOS at the high density prevents the revealing of a core that features a negative trace of the energy-stress tensor, staving off the conditions for descalarization (see, e.g., the discussion in Sec.~III of~\cite{Shibata:2013pra}).
The complete catalog of the simulated system is listed in the supplemental material, while the details of numerical schemes and setups can be found in \cite{Lam:2024wpq} as well as in \cite{Shibata:2013pra,Taniguchi:2014fqa}.
We also note that the simulations included in this article focus on the post-merger evolution and thus the quasi-equilibrium states of binaries were prepared at $<5$ orbits before the merger as initial data.

Throughout this article, we adopt the geometrical units $c=G=1$, and denote the ratio between the masses of binary as $q=m_2/m_1\le 1$, 
the instantaneous frequency of GWs at the merger as $f_{\rm peak}$, the GW amplitude at the merger as $h_{\rm peak}$ (here the merger time is defined as the moment when the GW amplitude reaches the maximal), 
the threshold mass for prompt collapse to a black hole as $M_{\rm thr}$, and the frequency of the dominant peak in the post-merger waveform as $f_2$.
The numerical results presented here are limited to simulations of equal-mass binaries, including those performed in the recent work~\cite{Lam:2024wpq} using theory-consistent quasi-equilibrium states as initial data~\cite{Kuan:2023hrh} and some simulations within GR newly performed here.

\begin{figure}
    \centering
    \includegraphics[width=\columnwidth]{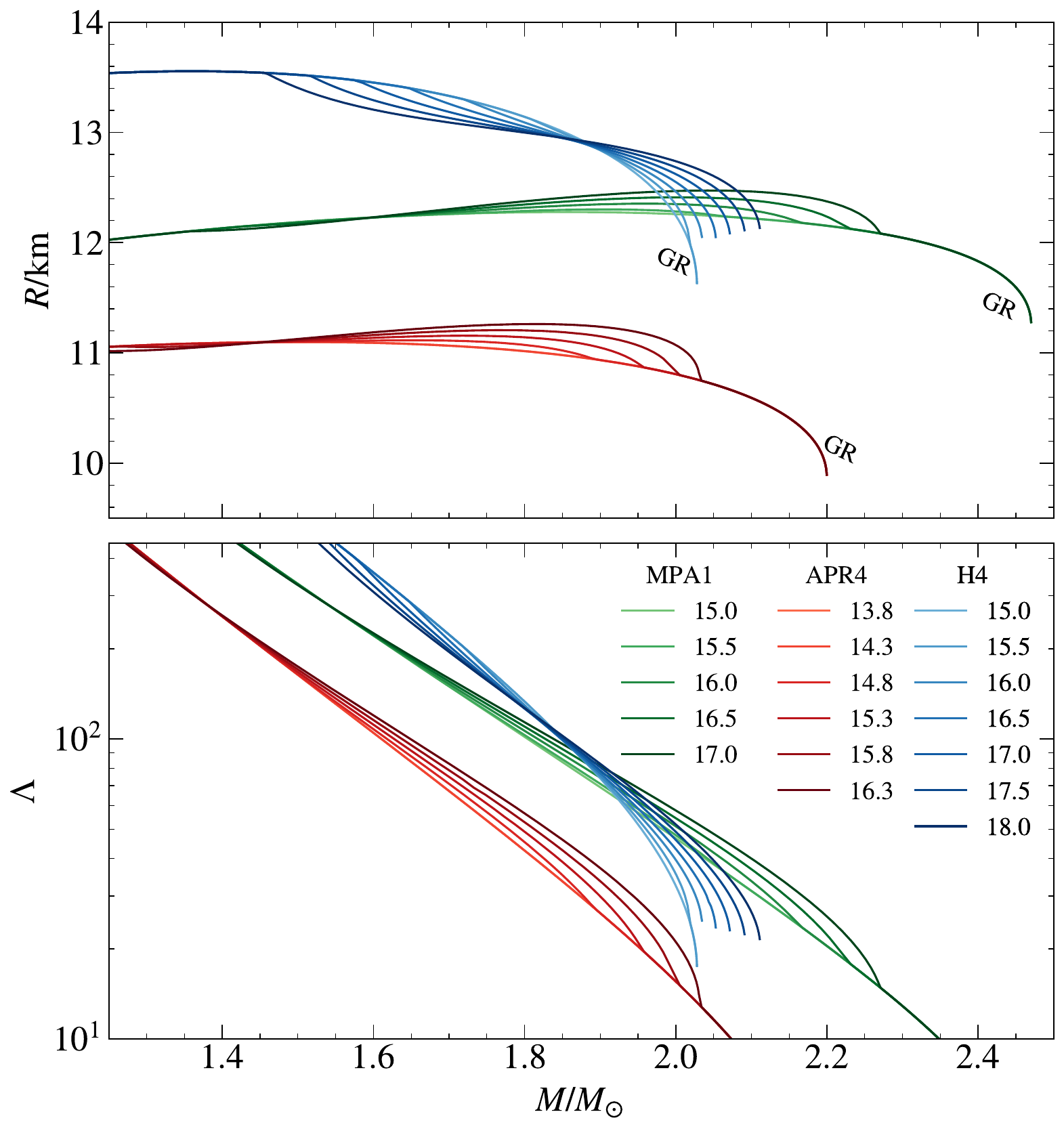}
    \caption{The upper panel displays the radius-mass relation, while the lower panel illustrates the tidal deformability-mass relation. The lines labeled "GR" represent the cases identical to those in General Relativity. Other lines deviating from the GR curves indicate the occurrence of spontaneous scalarization. Three EOSs are considered in the piecewise-polytropic approximation \cite{Read:2008iy}. For each EOS, a variety of scalar coupling constants (labels on the plot) are adopted while fixing the scalar mass as $m_\phi=1.33\times10^{-11}$~eV.
    }
    \label{fig:TOV}
\end{figure}

\begin{figure}
    \centering
    \includegraphics[width=\columnwidth]{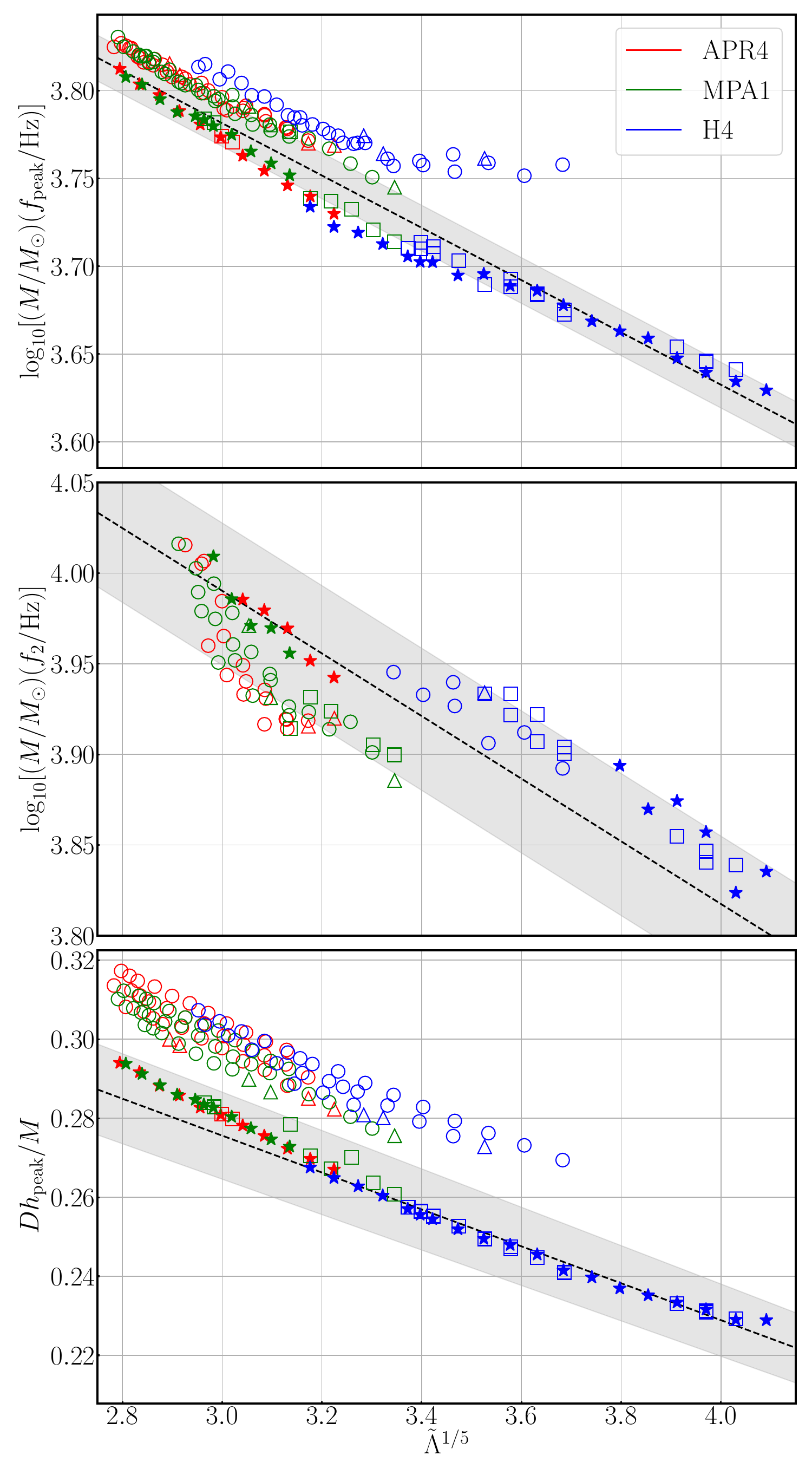}
    \caption{Relations between binary tidal deformability $\tilde{\Lambda}$ and the GW's frequency at the moment of merger (top), the frequency of the dominant peak of the postmerger waveform (middle), and the maximal strain of emitted GWs (bottom). The filled stars are the results of our simulations in GR, while open circles, triangles, and squares are for models with spontaneous scalarization, dynamical scalarization and no scalarization in the inspiral phase, respectively. The dashed lines are the fitting formula proposed in \cite{Kiuchi:2019kzt}. }
    \label{fig:UR}
\end{figure}

\section{Correlations between $\tilde{\Lambda}$ and GW characteristics}
The main tidal signature in inspiral waveforms depends predominantly on the binary tidal deformability $\tilde{\Lambda}=\frac{13}{16}(m_1+12m_2)m_1^4\Lambda_1/13M^5+(1\leftrightarrow 2)$, where $M=m_1+m_2$ and the tidal deformability of the individual stars are $\Lambda_1$ and $\Lambda_2$, respectively~\cite{Flanagan:2007ix,Hinderer:2007mb,Favata:2013rwa,Wade:2014vqa,Hu:2021tyw}. 
The estimate on $\tilde{\Lambda}$ for GW170817 yielded, though loosely, the first constraints on the yet unknown EOS of NS while assuming GR as the gravitational theory.
On the observation front, measurability of $\tilde{\Lambda}$ is within the uncertainty of $\sigma_{\tilde{\Lambda}}\sim400$ at the $2\sigma$ level with current detectors and is expected to be improved to $\sigma_{\tilde{\Lambda}}\lesssim50$ at the $1\sigma$ level in the fifth observation mission.
It is owing to this dominant role of $\tilde{\Lambda}$ in affecting the phasing of waveforms that several quasi-universal relations have been proposed to relate it with GW properties as introduced as follows. 

Using numerical simulations, a quasi-universal relation between $\tilde{\Lambda}$ and $f_{\rm peak}$ is found for $1.35+1.35\,M_\odot$ irrotational binaries \cite{Read:2013zra,Bernuzzi:2014kca}. 
The validity of this relation is extended in \cite{Takami:2014tva,Rezzolla:2016nxn,Bernuzzi:2014owa} to binaries with individual NSs having a mass of $1.2$--$1.65\,M_\odot$ while keeping binaries as symmetric and irrotational.
Aside from reading off the numerical results, \citet{Bernuzzi:2014kca} also discovers this universality by inspecting effective-one-body waveform models, where the mass range is further extended to include the mass close to the Tolman–Oppenheimer–Volkoff limit for the respective EOS and includes a small spin up to $|\chi|=0.1$.
The influence of mass-ratio on this relation is pointed out later on, which is evidenced by the simulations of asymmetric, irrotational binaries with varying mass ratios between 0.734--1 \cite{Kiuchi:2017pte,Kiuchi:2019kzt}. 
This motivates \citet{Kiuchi:2019kzt} to generalize the relation to capturing the effect of asymmetry, and subsequently, the coefficients of the fitting formula acquire a $q$-dependence.

The top panel of \cref{fig:UR} shows our numerical results (the filled markers denote the GR data) together with the relation established in \cite{Kiuchi:2019kzt} when setting the coefficients for equal-mass binaries (dashed line).
First we see that the GR data deviates slightly from the fitting formula, but this is within the uncertainty of the fitting formula itself ($4\%$; shaded area).
The largest deviation is found as $\lesssim5.8\%$ in the middle range of $\tilde{\Lambda}$, which is near the low (high) end of our H4 (APR4 and MPA1) samples.
We can also notice that the binaries that do not exhibit scalarization before the merger (squares) obey well the quasi-universal relation, which can be expected since the inspiral dynamics leading up to the merger are equivalent to in GR for these cases.

On the other hand, the relation tends to underestimate $f_{\rm peak}$ for a given $\tilde{\Lambda}$ for either spontaneously (circles) or dynamically (triangles) scalarized mergers, indicating that the orbital frequency right before the merger is systematically enhanced compared to the case where the scalar phenomenon is silent.
Although the deviation is still within the formula's uncertainty and does not show a decisive violation, the mergers with large $\tilde{\Lambda}$ (i.e., the stiff EOS H4) display a clear disagreement with the formula.
In particular, $f_{\rm peak}$ for the EOS H4 is roughly constant for $\tilde{\Lambda}^{1/5}\gtrsim3.4$, and thus differ further from the relation to the right of the plot.

On top of the GW frequency at the merger, Refs.~\cite{Bernuzzi:2015rla,Kiuchi:2019kzt,Breschi:2019srl,Breschi:2022xnc} demonstrated that $\tilde{\Lambda}$ can also be quasi-universally related to $f_{2}$ for a quite wide range of mass ratios ($0.67\le q\le1$) while commenting on a possible violation of the universality when including spinning and/or magnetized binaries. 
The relation is also proposed in \cite{Takami:2014tva,Rezzolla:2016nxn}, while their simulations were limited to nearly equal-mass binaries.
Our data together with the formula in \cite{Kiuchi:2019kzt} are shown in the middle panel of \cref{fig:UR}, where the shaded area presents the fitting uncertain of $9\%$. 
We note that mergers promptly collapsing into a black hole are not shown here since no information of $f_2$ can be extracted.
For the GR cases, data points with the APR4 and MPA1 EOSs lie on the line within a minor deviation of $<1\%$, while those with the H4 EOS are on the boundary of the fitting uncertainty. 
In contrast to the $\tilde{\Lambda}$-$Mf_{\rm peak}$ relation, the scalar field is always activated in the aftermath of the merger for the adopted coupling constants.
Therefore, $f_2$ is naturally expected to be different from what would be predicted in GR.
Indeed, we observe a systematic reduction in $f_2$ when $\tilde{\Lambda}^{1/5}\lesssim3.4$, for the chosen samples with the soft EOSs APR4 and MPA1.
However, the cases with the H4 EOS are quite consistent between GR and the considered DEF theories, though there is still a difference: a prompt collapse realizes for $\tilde{\Lambda}^{1/5}\lesssim3.8$ in GR, while an HMNS can still be formed until $\tilde{\Lambda}^{1/5}\lesssim3.4$ depending on $B$. 

\citet{Kiuchi:2019kzt} further provides relations of $\tilde{\Lambda}$ to $h_{\rm peak}$.
We again compare our numerical data of $h_{\rm peak}$ with their formula, shown in the bottom panel of \cref{fig:UR}. 
Our GR results progressively exceed the fitting formula for lower $\tilde{\Lambda}$, and the deviation reaches $\lesssim3.3\%$ to the left side of the plot.
In general, cases that are not scalarized in the inspiral epoch, including those in the DEF theory with weak scalar coupling and those in GR, align well with quasi-universal relations. 
However, binaries endowed with a scalar cloud during inspiral exhibit a systematic upward shift from this trend.

\section{Correlations between $f_2$ and properties of individual NS}
On top of the above relations, the frequency of the dominant mode in postmerger waveforms can also be universally connected to the certain properties of a cold spherical neutron star in isolation, e.g., the Love number ($\Lambda_{1.6}$) and radius ($R_{1.6}$) of the $1.6\,M_\odot$ NS, assuming no strong phase transitions.
In particular, Bauswein \textit{et al.} proposed a $f_2$-$R_{1.6}$ relation \cite{Bauswein:2012ya,Bauswein:2014qla} (see also \cite{Bauswein:2011tp}) from their
simulations of $1.35+1.35\,M_\odot$ binaries while adopting the conformal flatness condition (CFC).
The data set for seeking such a relation has been significantly extended by including different $M$ while keeping $q=1$ in \cite{Lioutas:2021jbl}. 
In the above work, the authors found different relations for each $M$ and this dependence on $M$ is also found later in \cite{Chatziioannou:2017ixj}.
On the other hand, focusing on binaries with similar total binary mass (viz.~2.7 and 2.6~$M_\odot$) for mass ratios $0.8\le q\le1$, Refs.~\cite{Hotokezaka:2013iia,Bauswein:2015vxa} showed a consistent fitting, while the data spread broader away from the fitting formula as quantified in \cite{Kiuchi:2019kzt}.
This relation is substantially revised by including also the chirp mass as additional fitting parameter in \cite{Vretinaris:2019spn}. 
In that work, the authors adopted the combined numerical results of equal-mass binary mergers under CFC with individual NS' mass ranging from 1.2--1.9~$M_\odot$, and the simulations withdrawing CFC of unequal-mass binaries with $q\ge0.49$ for a mass range of 0.94--1.94~$M_\odot$ released in the \textsc{CoRe} database \cite{Dietrich:2018phi}.

\begin{figure}
    \centering
    \includegraphics[width=\columnwidth]{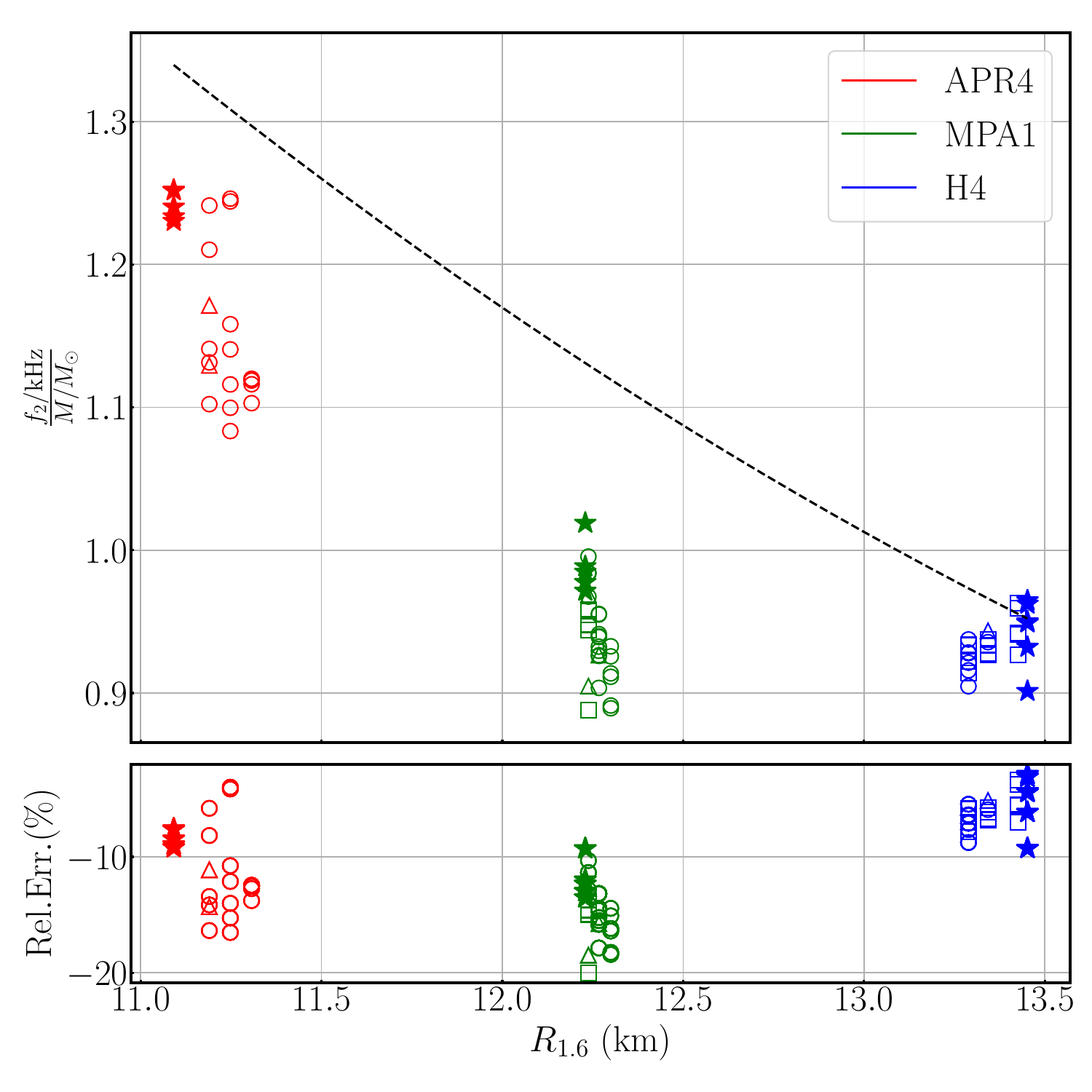}
    \caption{Correlation between $f_2/M$ and $R_{1.6}$ for the considered EOS (see plot legends) and various of $B$. 
    The numerical results are denoted in the same manner as in \cref{fig:UR}. The quasi-universal relation proposed in \cite{Bauswein:2015vxa} is shown as the dashed curve, while the relative deviation of our numerical results to the formula is given in the bottom panel.}
    \label{fig:R16}
\end{figure}

In \cref{fig:R16}, we show the comparison with the quasi-universal relation obtained in \cite{Bauswein:2015vxa}. 
Even in GR, the formula can only approximately describe the cases with the EOS H4, while the systems with the other softer EOSs are significantly below.
The relative deviation is depicted in the bottom panel, where we see that the formula tends to overestimate $f_2$ frequency by $\gtrsim10\%$ for the APR4 and MPA1 EOSs.
Focusing on the numerical data, it can be noticed that $R_{1.6}$ is larger in the DEF theories for the APR4 and MPA1 EOSs, while the trend is reversed for the H4 EOS.
The overall reduced value of $f_2$ in the DEF theories can seemingly be explained by the effective stiffening for the APR4 and MPA1 EOSs.
However, such a rationale does not apply to the H4 EOS, indicating that the interplay between gravity and matter is non-trivial and more investigation is needed to understand their competition in determining the stellar structure.

We have also compared the numerical results with the $f_2$-$\Lambda_{1.6}$ relation in \citet{Lioutas:2021jbl}.
The situation is more or less the same as the comparison with $f_2$-$R_{1.6}$ relation, and therefore we do not present it here. 

\section{Degeneracy with QCD phase transition}
Certain caveats have already been raised that the tightness of quasi-universal relations can be broadened by including a wider set of EOS \cite{Raithel:2022orm} or violated by either a strong, first-order \cite{Most:2018eaw,Bauswein:2018bma,Weih:2019xvw,Blacker:2020nlq,Liebling:2020dhf,Prakash:2021wpz,Raithel:2022efm,Fujimoto:2022xhv,Haque:2022dsc,Prakash:2023afe} or cross-over phase transition \cite{Huang:2022mqp}.
Consequently, an inconsistency between the inference on the EOS from the inspiral and postmerger waveforms is speculated as an indicator of phase transitions occurring during the merger process. 
In particular, the $f_2$ peak will have a higher frequency than what would be predicted by the quasi-universal relations for EOS with first-order phase transition since matters will be softened when the new degree of freedom emerges.
On the other hand, matters will experience a stiffening at 3--4~$n_0$ followed by a softening at 4--5~$n_0$ for the cross-over phase transition scenario \cite{Baym:2017whm,Baym:2019iky}, leading to a reduced $f_2$. Here $n_0=0.16$~fm$^{-3}$ is the nuclear saturation number density.

However, the connection between the violation in the quasi-universal relations and matter phase transition should be carefully revisited as it can also arise from a modification in the underlying gravitational theory, as shown in this article.
The similarity between modified gravity and the quantum chromodynamic (QCD) phase transition in terms of postmerger waveforms does not end here. 
After baryons crush to form exotic particles, the EOS can be stiffened or softened depending on the nature of the QCD phase transition (see above). In turn, the core can become less or more compact thereby adjusting the frequency of fluid oscillation and the associated GWs \cite{Weih:2019xvw,Blacker:2024tet}.
This process can also manifest in scalarized HMNSs (cf.~\cref{fig:UR}). 
In particular, the scalar activity in HMNSs pertaining to H4 can lead to a higher $f_2$ than the prediction by the quasi-universal relation (see the deviation in $f_2$ for $3.3 \lesssim \tilde{\Lambda}^{1/5} \lesssim 3.7$ in \cref{fig:UR}), reminiscent of the influence of a first-order nuclear phase transition. 
On the other hand, the coupling between the scalar field and matter tends to reduce $f_2$ for the EOSs APR4 and MPA1, mimicking the cross-over phase transition (see the deviation in $f_2$ for $\tilde{\Lambda}^{1/5} \lesssim 3.4$ in \cref{fig:UR}).

There is a distinction between the QCD phase transition and the gravitational transition of states: an interface (e.g., quark-hadron) will reveal in the former process, supporting a class of oscillation modes ($i$-mode) that may leave certain imprints in GW signals  \cite{Tonetto:2020bie,Lau:2020bfq,Zhu:2022pja}.
On the other hand, there is a class of mode linked to the scalar field, i.e., $\phi$-mode \cite{Mendes:2018qwo,Blazquez-Salcedo:2021exm}.
In principle, the quadrupole member of $\phi$-mode can emit GWs as a result of the entrained fluid motions.
Both the $i$- and $\phi$-modes have typically the frequency of several hundred Hz, and the largely overlapped frequency band makes it non-trivial to tell them apart even if this weak emission could be detected.

\section{Conclusion}
We systematically performed numerical simulations of BNS mergers in GR and DEF theories to solve for the waveforms throughout inspiral up to the merger, where the considered scalar coupling constants are summarized in \cref{fig:TOV}.
Based on the numerical data, we examine several quasi-universal relations connecting the binary tidal deformability to waveform characteristics.
For the mergers that scalarization does not realize before merger, the GW's frequency and amplitude at the merger in the DEF theories aligned well with the fitting formula valid in GR (cf.~the top and bottom panels of \cref{fig:UR}).
These two relations can, however, be significantly violated if scalarization occurs in the inspiral phase.

Although we only take three EOSs into account, our results already suggest a serious caveat when applying quasi-universal relations established in GR to probe the EOS and gravity in modified gravity.
In particular, we demonstrated that a gravitational effect like scalarization could also lead to a violation in quasi-universal, mimicking the similar violation that could be caused by a strong phase transition.
This indicates that one cannot take the future disagreement between the detected GW signal and the predictions of quasi-universal relations as a smoking-gun of either effect.
Much more investigation remains to be done to further discriminate one effect from the other. 
Also, thermal effects in postmerger signals still remain to be explored \cite{Shibata:2005ss,Bauswein:2010dn,Figura:2020fkj,Raithel:2023gct,Fields:2023bhs,Miravet-Tenes:2024vba,Bernuzzi:2015rla}, and thus the quasi-universal relations are to be inspected even for EOS without phase transition and within GR.

\section*{Acknowledgements}
Numerical computations were performed on the clusters Sakura and Raven at the Max Planck Computing and Data Facility. 
This work was in part supported by Grant-in-Aid for Scientific Research (grant Nos.~20H00158, 23H04900, and 23H01172) of Japanese MEXT/JSPS. 


\appendix

\section{Tidal deformability of neutron stars in massive scalar tensor gravity}

We assume a perfect fluid, for which the energy-momentum tensor in the Jordan frame is
\begin{equation}
{T}^{a b}=({e}+{p}) {u}^{a} {u}^{b}+{g}^{a b}{p}\,,
\end{equation}
where $u^{a}$, $e$, and $p$ are the four-velocity, the energy density, and the pressure of the fluid element respectively. 
The field equations are conveniently formulated using the metric $\tilde{g}_{a b}$ in the Einstein frame. 
This metric is related to the Jordan frame metric $g_{a b}$ (the one defined in the main text) through a conformal transformation given by $g_{a b} = A^{2}(\bar{\varphi})\tilde{g}_{a b}$, where $A(\bar{\varphi})$ is the coupling function, and $\bar{\varphi}$ represents the scalar field in the Einstein frame~\cite{Kuan:2023hrh}.

The line element of the spherical background in the Einstein frame can be written as 
\begin{align}
    \label{eqn:static}
       {\dd} s^{2}=\tilde{g}_{\mu\nu}{\dd} x^{\mu}{\dd} x^{\nu}&= -e^{\nu(r)} {\dd} t^{2}+\frac{1}{1-2m(r)/r}
       {\dd}r^{2} \nonumber\\
       &+r^{2}\left({\dd} \theta^{2}+\sin ^{2} \theta {\dd}\phi^{2}\right) \,,
\end{align}
In general relativity, the metric component $\tilde{g}_{tt}$ of a tidally-deformed neutron star can be expanded as  
\begin{align}
\label{eq:metric_expansion}
-\frac{\left(1+\tilde{g}_{t t}\right)}{2}&=  -\frac{M}{r}-\frac{3 Q_{i j}}{2 r^3}\left(n^i n^j-\frac{1}{3} \delta^{i j}\right)+O\left(\frac{1}{r^4}\right) \nonumber\\
 &+\frac{1}{2} \mathcal{E}_{i j} x^i x^j+O\left(r^3\right)\,,
\end{align}
where $\mathcal{E}_{ij}$ is the tidal field generated by the companion star, $Q_{ij}$ is the quadrupolar moment of the neutron star induced by the tidal field, and $n^i \equiv x^i / r$.
To the linear order of $\mathcal{E}_{ij}$, the induced quadrupole $Q_{ij}$ can be written as 
\begin{equation}
    Q_{ij}=-\lambda \mathcal{E}_{ij}\,.
\end{equation}
Here the parameter $\lambda$ measures the tidal deformability of the star and can be related to the $l=2$ tidal Love number via $k_2=3 \lambda R^{-5}/2$ for $R$ the circumferential radius of the star \cite{Hinderer:2007mb,Flanagan:2007ix}.

In the massive Damour-Esposito-Farese-type scalar-tensor theory considered in the main text, both the scalar and tensor fields respond to the external tidal field induced by either tensor or scalar fields. 
Therefore, we should also expand the metric in terms of perturbations in the scalar field besides $\mathcal{E}_{ij}$ to determine the tidal deformability. 
In addition to the tensor and scalar tidal deformabilities, one also needs to consider a third kind of tidal deformability describing the tensor/scalar multipole moments induced by a scalar/tensor field as noted by \citet{Creci:2023cfx}.
However, the asymptotic behavior of a massive scalar field $\bar{\varphi}$ and its perturbation $\delta\bar{\varphi}$ read 
\begin{equation}
    \bar{\varphi},\ \delta \bar{\varphi} \rightarrow \frac{1}{r} e^{-r / \lambda_{\bar{\varphi}}},
\end{equation}
where $\lambda_{\bar{\varphi}} = 2\pi\hbar c/m_{\bar{\varphi}}$ is the Compton length scale of the Yukawa suppression associated with the scalar mass $m_{\bar{\varphi}}$.
Such exponential decay does not affect the expansion in \cref{eq:metric_expansion} on any order of $1/r$, which is different from the massless case~\cite{Pani:2014jra,Brown:2022kbw,Creci:2023cfx}. 
We thus only need to consider the tensor tidal deformability of scalarized NSs for the massive theories considered in the present work.

Restricting to the even-parity, static, and $l=2$ perturbations in the Reege-Wheeler gauge, the perturbation in the metric and the scalar field can be, respectively, written as 
\begin{align}
\label{eq:metric_per}
\delta \tilde{g}_{\mu \nu}^{(2 m)}=&Y_{2 m}(\theta, \phi)\left[\begin{array}{cccc}
e^\nu H_0 & H_1 & 0 & 0 \\
H_1 & H_2 /\left(1-\frac{2 m(r)}{r}\right) & 0 & 0 \\
0 & 0 & r^2 K & 0 \\
0 & 0 & 0 & r^2 \sin ^2 \theta K
\end{array}\right]\,,
\end{align}
and
\begin{align}
\label{eq:scalar_per}
\delta \bar{\varphi}^{(2m)} =& \delta \bar{\varphi} Y_{2m}(\theta,\phi)\,,
\end{align}
where $H_{0}$, $H_{2}$, $K$, and $\delta \bar{\varphi}$ are perturbed quantities depending $r$, and $Y_{2m}$ are the spherical harmonics.

Following the procedure in~\citet{Hinderer:2007mb}, we substitute Eqs.~(\ref{eq:metric_per}-\ref{eq:scalar_per}) into the linearized Einstein equations $\delta G_\nu^\mu=8 \pi \delta T_\nu^\mu$ and denote $H_0=-H_2=H(r)$ to obtain 
\begin{align}
    & H_1 = 0\,,\\
    & K^{\prime}=-H^{\prime}-\nu^{\prime}H-4\delta\bar{\varphi} \bar{\varphi}^{\prime}\,, \\
    \label{eq:H}
    & H^{\prime \prime}+c_1 H^{\prime}+c_0 H=c_s \delta \bar{\varphi}, \\
    \label{eq:phi}
    & \delta \bar{\varphi}^{\prime \prime}+d_1 \delta \bar{\varphi}^{\prime}+d_0 \delta \bar{\varphi}=d_s H.
\end{align}
Here $(^{\prime})$ denotes the derivative respect to $r$, and 
\begin{widetext}
\begin{eqnarray}
  c_1&=&d_1=-\frac{r^3 \left(8 \pi  A^4 (e-p)+V\right)+4 m-4 r}{2 r (r-2 m)}\,,\\
  c_s&=&4d_s=\frac{r^2 \left(-8 \pi  \alpha  A^4 \left[(\dedp-1) e+(\dedp-9) p\right]+2 r \bar{\varphi}^{\prime}
   \left(16 \pi  A^4 p-V\right)+4 (r-2 m) \bar{\varphi}^{\prime 3}-\frac{\dd V}{\dd \bar{\varphi}}\right)+8 m \bar{\varphi}^{\prime}}{r (r-2 m)}\,,\\
    c_0&=& \frac{4 \pi  A^4 r (e+p)}{r-2 m}\dedp  +\frac{\left[r^3 \left(V-16 \pi  A^4 p\right)-4 m\right]\bar{\varphi}^{\prime 2}}{r-2 m}-r^2 \bar{\varphi}^{\prime 4}+\frac{r \left(8 \pi r^3 A^4 p +4 m-r\right)V}{(r-2 m)^2}-\frac{r^4 V^2}{4 (r-2 m)^2}\nonumber\\
    && + \frac{-64 \pi ^2 A^8 p^2 r^6+4 \pi  A^4 r^3 (5 e (r-2 m)-26 m p+9 p r)-4 m^2+12 m r-6
   r^2}{r^2 (r-2 m)^2}\,,\\
  d_0&=&\frac{4 \pi  \alpha ^2 A^4 r (e+p)}{r-2 m}\dedp  -\frac{96 \pi  \alpha ^2 A^4 r^2 (e-p)+16 \pi  A^3 r^2 \frac{{\dd^2} A}{{\dd} \bar{\varphi}^2} (e-3 p)+16 r (r-2 m)
   \bar{\varphi} '^2+r^2 \frac{{\dd^2} V}{{\dd} \bar{\varphi}^2}+24}{4 r (r-2 m)}\,.
\end{eqnarray}
\end{widetext}
In general relativity, we have $c_s=d_s=0$, and the perturbations for the tensor field and the scalar field decouple.

Eqs.~(\ref{eq:H}-\ref{eq:phi}) are linear equations for $H$ and $\delta \bar{\varphi}$. Thus, to solve them, one can integrate the coupled equations twice with the following initial values~\cite{Hu:2021tyw,Pani:2014jra}
\begin{align}
\left.H\right|_{r_0}=&r_0^2,\left.\quad H^{\prime}\right|_{r_0}=2 r_0,\left.\quad \delta \bar{\varphi}\right|_{r_0}=0,\left.\quad \delta \bar{\varphi}^{\prime}\right|_{r_0}=0 \\
\left.H\right|_{r_0}=&0,\left.\quad H^{\prime}\right|_{r_0}=0,\left.\quad \delta \bar{\varphi}\right|_{r_0}=r_0^2,\left.\quad \delta \bar{\varphi}^{\prime}\right|_{r_0}=2 r_0
\end{align}
for $r_0$ a tiny radius near the center of the star. Then one makes a linear combination of the two integrated results to construct a solution whose asymptotic value of $\delta\bar{\varphi}$ vanishes at $r\gg \lambda_{\bar{\varphi}}$.

In practice, we integrate the system to a sufficiently large radius $r=r_{\rm b}$ and ignore the scalar field for $r>r_{\rm b}$. 
Defining
\begin{equation}
    C=\left.\frac{m}{r}\right|_{r \rightarrow r_{\rm b}}, \,\,\text{and} \quad y=\left.\frac{r H^{\prime}}{H}\right|_{r \rightarrow r_{\rm b}}\,,
\end{equation}
the tidal deformability, which is in the same form as the GR case, can be calculated as (e.g., \cite{Hinderer:2007mb})
\begin{widetext}
\begin{align}
\lambda= & \frac{2}{3} r_{\rm b}^5 \frac{8 C^5}{5}(1-2 C)^2[2+2 C(y-1)-y] 
 \times\left\{4 C^3\left[13-11 y+C(3 y-2)+2 C^2(1+y)\right]\right. \\
& +3(1-2 C)^2[2-y+2 C(y-1)] \ln (1-2 C)  +2 C[6-3 y+3 C(5 y-8)]\}^{-1} .
\end{align}
\end{widetext}
The tidal deformability used in the main text is then obtained as $\Lambda=\lambda/M^5$ for $M$ the mass of the star.

\section{Catalog of simulated binaries and theories}

We list in \cref{tab:outcomes_APR4,tab:outcomes_MPA1,tab:outcomes_H4} the information for every simulations, including the ADM mass of coalescing neutron stars, binary tidal deformability, parameters of the DEF theories, the status of scalar field before merger, spectral properties of GWs.
\cref{tab:outcomes_GR} shows the same information aside from the status of $\varphi$ for all simulations in GR.

\begin{table*}
    \centering
    \caption{Summary of outcomes for the BNS mergers in the massive DEF theory with APR4 EOS. The first column lists the model name which combines EOS, coupling strength $B$, and baryon mass of each NS in units of $M_\odot$. The second column shows the ADM mass $M_{\rm ADM}$ of each isolated NS. The third column shows the state of pre-merger scalarization with symbols $\times$, $\bigtriangleup$ and $\bigcirc$ corresponding to no scalarization, dynamical scalarization, and spontaneous scalarization in the pre-merger phase, respectively. The fourth column lists the binary tidal deformability $\tilde \Lambda$.
    The last three columns summarize the properties of GWs.
}
    \begin{tabular}{
        >{\raggedright}p{0.3\columnwidth} 
	>{\centering}p{0.26\columnwidth} 
        >{\centering}p{0.2\columnwidth} 
	>{\centering}p{0.2\columnwidth} 
        >{\centering}p{0.2\columnwidth} 
        >{\centering}p{0.2\columnwidth} 
        >{\centering\arraybackslash}p{0.3\columnwidth} 
    }
        \hline\hline \\[-.8em]
        Model name
        &  $M_{\rm ADM}$ ($M_\odot$) 
        & Inspiral $\bar{\varphi}$
        & $\tilde{\Lambda}$ 
        & $f_{\rm peak} $ (kHz)
        & $f_{2} $ (kHz)
        & $D h_{\rm peak} / M$ \\
        \hline
\texttt{APR4\_B13.8\_M1.57} & 1.4041 & $\times$ & 251.5 & 2.100 & - & 0.280 \\
\texttt{APR4\_B13.8\_M1.58} & 1.4119 & $\times$ & 242.8 & 2.105 & - & 0.281 \\
\hline
\texttt{APR4\_B14.3\_M1.57} & 1.4041 & $\times$ & 251.5 & 2.101 & - & 0.280 \\
\texttt{APR4\_B14.3\_M1.58} & 1.4119 & $\times$ & 242.8 & 2.105 & - & 0.281 \\
\hline
\texttt{APR4\_B14.8\_M1.62} & 1.4436 & $\bigtriangleup$ & 210.5 & 2.231 & - & 0.298 \\
\texttt{APR4\_B14.8\_M1.63} & 1.4514 & $\bigtriangleup$ & 203.2 & 2.252 & - & 0.300 \\
\hline
\texttt{APR4\_B15.3\_M1.48} & 1.3323 & $\bigtriangleup$ & 348.8 & 2.204 & 3.121 & 0.282 \\
\texttt{APR4\_B15.3\_M1.50} & 1.3500 & $\bigtriangleup$ & 321.6 & 2.181 & 3.049 & 0.285 \\
\texttt{APR4\_B15.3\_M1.52} & 1.3643 & $\bigcirc$ & 300.7 & 2.199 & 3.008 & 0.288 \\
\texttt{APR4\_B15.3\_M1.54} & 1.3802 & $\bigcirc$ & 279.6 & 2.213 & 3.123 & 0.292 \\
\texttt{APR4\_B15.3\_M1.56} & 1.3961 & $\bigcirc$ & 260.4 & 2.201 & 3.186 & 0.294 \\
\texttt{APR4\_B15.3\_M1.58} & 1.4118 & $\bigcirc$ & 242.9 & 2.215 & 3.417 & 0.298 \\
\texttt{APR4\_B15.3\_M1.60} & 1.4277 & $\bigcirc$ & 226.7 & 2.232 & 3.544 & 0.300 \\
\texttt{APR4\_B15.3\_M1.62} & 1.4433 & $\bigcirc$ & 212.0 & 2.226 & - & 0.303 \\
\texttt{APR4\_B15.3\_M1.64} & 1.4590 & $\bigcirc$ & 198.4 & 2.237 & - & 0.304 \\
\texttt{APR4\_B15.3\_M1.65} & 1.4668 & $\bigcirc$ & 192.0 & 2.224 & - & 0.305 \\
\texttt{APR4\_B15.3\_M1.66} & 1.4747 & $\bigcirc$ & 185.7 & 2.220 & - & 0.307 \\
\texttt{APR4\_B15.3\_M1.68} & 1.4900 & $\bigcirc$ & 174.3 & 2.246 & - & 0.308 \\
\hline
\texttt{APR4\_B15.8\_M1.50} & 1.3481 & $\bigcirc$ & 321.4 & 2.192 & 3.075 & 0.290 \\
\texttt{APR4\_B15.8\_M1.52} & 1.3642 & $\bigcirc$ & 299.3 & 2.204 & 3.045 & 0.294 \\
\texttt{APR4\_B15.8\_M1.54} & 1.3800 & $\bigcirc$ & 279.3 & 2.217 & 2.990 & 0.296 \\
\texttt{APR4\_B15.8\_M1.56} & 1.3960 & $\bigcirc$ & 260.9 & 2.209 & 3.070 & 0.299 \\
\texttt{APR4\_B15.8\_M1.58} & 1.4116 & $\bigcirc$ & 244.3 & 2.183 & 3.270 & 0.301 \\
\texttt{APR4\_B15.8\_M1.60} & 1.4274 & $\bigcirc$ & 228.8 & 2.203 & 3.557 & 0.304 \\
\texttt{APR4\_B15.8\_M1.62} & 1.4430 & $\bigcirc$ & 214.5 & 2.221 & 3.591 & 0.305 \\
\texttt{APR4\_B15.8\_M1.64} & 1.4586 & $\bigcirc$ & 201.3 & 2.218 & - & 0.308 \\
\texttt{APR4\_B15.8\_M1.66} & 1.4742 & $\bigcirc$ & 189.0 & 2.222 & - & 0.310 \\
\texttt{APR4\_B15.8\_M1.67} & 1.4819 & $\bigcirc$ & 183.3 & 2.222 & - & 0.311 \\
\texttt{APR4\_B15.8\_M1.68} & 1.4897 & $\bigcirc$ & 177.7 & 2.239 & - & 0.312 \\
\texttt{APR4\_B15.8\_M1.70} & 1.5052 & $\bigcirc$ & 167.0 & 2.220 & - & 0.314 \\
\hline
\texttt{APR4\_B16.3\_M1.52} & 1.3637 & $\bigcirc$ & 300.1 & 2.204 & 3.044 & 0.297 \\
\texttt{APR4\_B16.3\_M1.54} & 1.3797 & $\bigcirc$ & 280.6 & 2.196 & 3.090 & 0.299 \\
\texttt{APR4\_B16.3\_M1.56} & 1.3955 & $\bigcirc$ & 263.0 & 2.216 & 3.123 & 0.302 \\
\texttt{APR4\_B16.3\_M1.58} & 1.4111 & $\bigcirc$ & 246.8 & 2.179 & 3.113 & 0.304 \\
\texttt{APR4\_B16.3\_M1.60} & 1.4268 & $\bigcirc$ & 231.9 & 2.212 & 3.196 & 0.307 \\
\texttt{APR4\_B16.3\_M1.62} & 1.4425 & $\bigcirc$ & 217.9 & 2.205 & - & 0.309 \\
\texttt{APR4\_B16.3\_M1.64} & 1.4580 & $\bigcirc$ & 205.0 & 2.203 & - & 0.311 \\
\texttt{APR4\_B16.3\_M1.66} & 1.4737 & $\bigcirc$ & 193.0 & 2.231 & - & 0.313 \\
\texttt{APR4\_B16.3\_M1.68} & 1.4891 & $\bigcirc$ & 181.8 & 2.215 & - & 0.315 \\
\texttt{APR4\_B16.3\_M1.69} & 1.4967 & $\bigcirc$ & 176.6 & 2.229 & - & 0.316 \\
\texttt{APR4\_B16.3\_M1.70} & 1.5046 & $\bigcirc$ & 171.4 & 2.231 & - & 0.317 \\
        \hline\hline
    \end{tabular}
    \label{tab:outcomes_APR4}
\end{table*}
\begin{table*}
    \centering
    \caption{Same as \cref{tab:outcomes_APR4} but for the MPA1 EOS.}
    \begin{tabular}{
        >{\raggedright}p{0.3\columnwidth} 
	>{\centering}p{0.26\columnwidth} 
        >{\centering}p{0.2\columnwidth} 
	>{\centering}p{0.2\columnwidth} 
        >{\centering}p{0.2\columnwidth} 
        >{\centering}p{0.2\columnwidth} 
        >{\centering\arraybackslash}p{0.3\columnwidth} 
    }
        \hline\hline \\[-.8em]
        Model name
        &  $M_{\rm ADM}$ ($M_\odot$) 
        & Inspiral $\bar{\varphi}$
        & $\tilde{\Lambda}$ 
        & $f_{\rm peak} $ (kHz)
        & $f_{2} $ (kHz)
        & $D h_{\rm peak}/M $ \\
        \hline 
\texttt{MPA1\_B15.0\_M1.78} & 1.5830 & $\times$ & 236.4 & 1.911 & - & 0.283 \\
\texttt{MPA1\_B15.0\_M1.79} & 1.5910 & $\times$ & 229.1 & 1.911 & - & 0.284 \\
\hline 
\texttt{MPA1\_B15.5\_M1.78} & 1.5830 & $\times$ & 236.4 & 1.911 & - & 0.283 \\
\texttt{MPA1\_B15.5\_M1.79} & 1.5910 & $\times$ & 229.1 & 1.910 & - & 0.284 \\
\hline 
\texttt{MPA1\_B16.0\_M1.60} & 1.4399 & $\times$ & 419.3 & 1.798 & 2.757 & 0.261 \\
\texttt{MPA1\_B16.0\_M1.62} & 1.4559 & $\times$ & 392.9 & 1.806 & 2.761 & 0.264 \\
\texttt{MPA1\_B16.0\_M1.64} & 1.4721 & $\times$ & 368.0 & 1.834 & - & 0.270 \\
\texttt{MPA1\_B16.0\_M1.66} & 1.4880 & $\times$ & 345.1 & 1.834 & 2.820 & 0.267 \\
\texttt{MPA1\_B16.0\_M1.68} & 1.5040 & $\times$ & 323.6 & 1.821 & 2.839 & 0.270 \\
\texttt{MPA1\_B16.0\_M1.70} & 1.5200 & $\times$ & 303.5 & 1.966 & 2.699 & 0.278 \\
\texttt{MPA1\_B16.0\_M1.72} & 1.5358 & $\bigtriangleup$ & 285.0 & 1.964 & 2.779 & 0.287 \\
\texttt{MPA1\_B16.0\_M1.74} & 1.5537 & $\bigtriangleup$ & 265.5 & 1.989 & 3.011 & 0.290 \\
\texttt{MPA1\_B16.0\_M1.76} & 1.5674 & $\bigcirc$ & 251.4 & 2.001 & 3.033 & 0.292 \\
\texttt{MPA1\_B16.0\_M1.78} & 1.5832 & $\bigcirc$ & 236.4 & 1.978 & 3.116 & 0.294 \\
\texttt{MPA1\_B16.0\_M1.80} & 1.5989 & $\bigcirc$ & 222.5 & 1.987 & 3.146 & 0.296 \\
\texttt{MPA1\_B16.0\_M1.82} & 1.6145 & $\bigcirc$ & 209.6 & 1.977 & 3.214 & 0.299 \\
\texttt{MPA1\_B16.0\_M1.84} & 1.6299 & $\bigcirc$ & 197.7 & 1.984 & - & 0.302 \\
\texttt{MPA1\_B16.0\_M1.85} & 1.6377 & $\bigcirc$ & 192.0 & 2.000 & - & 0.303 \\
\texttt{MPA1\_B16.0\_M1.86} & 1.6456 & $\bigcirc$ & 186.4 & 2.006 & - & 0.304 \\
\hline 
\texttt{MPA1\_B16.5\_M1.60} & 1.4399 & $\bigtriangleup$ & 419.3 & 1.930 & 2.668 & 0.276 \\
\texttt{MPA1\_B16.5\_M1.62} & 1.4559 & $\bigcirc$ & 391.9 & 1.934 & 2.735 & 0.277 \\
\texttt{MPA1\_B16.5\_M1.64} & 1.4719 & $\bigcirc$ & 366.7 & 1.948 & 2.812 & 0.280 \\
\texttt{MPA1\_B16.5\_M1.66} & 1.4880 & $\bigcirc$ & 343.3 & 1.965 & 2.756 & 0.284 \\
\texttt{MPA1\_B16.5\_M1.68} & 1.5039 & $\bigcirc$ & 322.0 & 1.970 & 2.787 & 0.286 \\
\texttt{MPA1\_B16.5\_M1.70} & 1.5197 & $\bigcirc$ & 302.5 & 1.955 & 2.746 & 0.288 \\
\texttt{MPA1\_B16.5\_M1.72} & 1.5357 & $\bigcirc$ & 284.2 & 1.965 & 2.864 & 0.291 \\
\texttt{MPA1\_B16.5\_M1.74} & 1.5514 & $\bigcirc$ & 267.6 & 1.970 & 2.916 & 0.294 \\
\texttt{MPA1\_B16.5\_M1.76} & 1.5673 & $\bigcirc$ & 251.9 & 1.971 & 2.914 & 0.296 \\
\texttt{MPA1\_B16.5\_M1.78} & 1.5830 & $\bigcirc$ & 237.5 & 1.966 & 2.980 & 0.298 \\
\texttt{MPA1\_B16.5\_M1.80} & 1.5986 & $\bigcirc$ & 224.2 & 1.975 & 3.053 & 0.301 \\
\texttt{MPA1\_B16.5\_M1.82} & 1.6142 & $\bigcirc$ & 211.6 & 1.973 & - & 0.303 \\
\texttt{MPA1\_B16.5\_M1.84} & 1.6296 & $\bigcirc$ & 200.1 & 1.979 & - & 0.304 \\
\texttt{MPA1\_B16.5\_M1.86} & 1.6453 & $\bigcirc$ & 189.1 & 1.988 & - & 0.306 \\
\texttt{MPA1\_B16.5\_M1.87} & 1.6531 & $\bigcirc$ & 183.9 & 1.996 & - & 0.307 \\
\texttt{MPA1\_B16.5\_M1.88} & 1.6607 & $\bigcirc$ & 178.9 & 2.000 & - & 0.308 \\
\texttt{MPA1\_B16.5\_M1.90} & 1.6761 & $\bigcirc$ & 169.4 & 2.019 & - & 0.310 \\
\hline 
\texttt{MPA1\_B17.0\_M1.70} & 1.5196 & $\bigcirc$ & 302.2 & 1.967 & 2.777 & 0.292 \\
\texttt{MPA1\_B17.0\_M1.72} & 1.5353 & $\bigcirc$ & 285.0 & 1.951 & 2.842 & 0.294 \\
\texttt{MPA1\_B17.0\_M1.74} & 1.5511 & $\bigcirc$ & 268.8 & 1.946 & 2.759 & 0.297 \\
\texttt{MPA1\_B17.0\_M1.76} & 1.5670 & $\bigcirc$ & 253.7 & 1.953 & 2.856 & 0.300 \\
\texttt{MPA1\_B17.0\_M1.78} & 1.5824 & $\bigcirc$ & 239.9 & 1.972 & 2.821 & 0.302 \\
\texttt{MPA1\_B17.0\_M1.80} & 1.5981 & $\bigcirc$ & 226.8 & 1.968 & 2.981 & 0.303 \\
\texttt{MPA1\_B17.0\_M1.82} & 1.6139 & $\bigcirc$ & 214.4 & 1.969 & - & 0.305 \\
\texttt{MPA1\_B17.0\_M1.84} & 1.6293 & $\bigcirc$ & 203.1 & 1.990 & - & 0.307 \\
\texttt{MPA1\_B17.0\_M1.86} & 1.6447 & $\bigcirc$ & 192.4 & 1.999 & - & 0.309 \\
\texttt{MPA1\_B17.0\_M1.87} & 1.6525 & $\bigcirc$ & 187.3 & 1.998 & - & 0.310 \\
\texttt{MPA1\_B17.0\_M1.88} & 1.6601 & $\bigcirc$ & 182.5 & 1.992 & - & 0.311 \\
\texttt{MPA1\_B17.0\_M1.90} & 1.6757 & $\bigcirc$ & 172.9 & 1.994 & - & 0.312 \\
        \hline\hline
    \end{tabular}
    \label{tab:outcomes_MPA1}
\end{table*}
\begin{table*}
    \centering
    \caption{Same as \cref{tab:outcomes_APR4} but for the H4 EOS.}
    \begin{tabular}{
        >{\raggedright}p{0.3\columnwidth} 
	>{\centering}p{0.26\columnwidth} 
        >{\centering}p{0.2\columnwidth} 
	>{\centering}p{0.2\columnwidth} 
        >{\centering}p{0.2\columnwidth} 
        >{\centering}p{0.2\columnwidth} 
        >{\centering\arraybackslash}p{0.3\columnwidth} 
    }
        \hline\hline \\[-.8em]
        Model name
        &  $M_{\rm ADM}$ ($M_\odot$) 
        & Inspiral $\bar{\varphi}$
        & $\tilde{\Lambda}$ 
        & $f_{\rm peak} $ (kHz)
        & $f_{2} $ (kHz)
        & $D h_{\rm peak} / M$ \\
        \hline 
\texttt{H4\_B15.0\_M1.70} & 1.5414 & $\times$ & 470.0 & 1.653 & - & 0.255 \\
\texttt{H4\_B15.0\_M1.71} & 1.5494 & $\times$ & 453.2 & 1.655 & - & 0.256 \\
\hline 
\texttt{H4\_B15.5\_M1.70} & 1.5414 & $\times$ & 470.0 & 1.654 & - & 0.255 \\
\texttt{H4\_B15.5\_M1.71} & 1.5494 & $\times$ & 453.2 & 1.655 & - & 0.256 \\
\hline 
\texttt{H4\_B16.0\_M1.70} & 1.5414 & $\times$ & 470.0 & 1.654 & - & 0.255 \\
\texttt{H4\_B16.0\_M1.71} & 1.5494 & $\times$ & 453.2 & 1.655 & - & 0.256 \\
\hline 
\texttt{H4\_B16.5\_M1.70} & 1.5414 & $\times$ & 470.0 & 1.668 & - & 0.255 \\
\texttt{H4\_B16.5\_M1.71} & 1.5494 & $\times$ & 453.2 & 1.669 & - & 0.257 \\
\hline 
\texttt{H4\_B17.0\_M1.50} & 1.3762 & $\times$ & 986.3 & 1.607 & 2.551 & 0.231 \\
\texttt{H4\_B17.0\_M1.60} & 1.4594 & $\times$ & 680.3 & 1.612 & 2.747 & 0.241 \\
\texttt{H4\_B17.0\_M1.62} & 1.4758 & $\times$ & 632.1 & 1.637 & 2.831 & 0.245 \\
\texttt{H4\_B17.0\_M1.64} & 1.4923 & $\times$ & 587.0 & 1.634 & 2.874 & 0.247 \\
\texttt{H4\_B17.0\_M1.66} & 1.5087 & $\times$ & 545.2 & 1.621 & 2.842 & 0.249 \\
\texttt{H4\_B17.0\_M1.68} & 1.5251 & $\times$ & 506.1 & 1.655 & - & 0.253 \\
\texttt{H4\_B17.0\_M1.70} & 1.5414 & $\times$ & 470.0 & 1.653 & - & 0.255 \\
\texttt{H4\_B17.0\_M1.72} & 1.5575 & $\times$ & 436.6 & 1.648 & - & 0.257 \\
\texttt{H4\_B17.0\_M1.74} & 1.5738 & $\bigtriangleup$ & 405.1 & 1.846 & - & 0.280 \\
\texttt{H4\_B17.0\_M1.75} & 1.5867 & $\bigtriangleup$ & 381.9 & 1.874 & - & 0.281 \\
\texttt{H4\_B17.0\_M1.76} & 1.5899 & $\bigcirc$ & 370.3 & 1.851 & - & 0.283 \\
\texttt{H4\_B17.0\_M1.78} & 1.6060 & $\bigcirc$ & 336.8 & 1.868 & - & 0.286 \\
\texttt{H4\_B17.0\_M1.80} & 1.6220 & $\bigcirc$ & 307.6 & 1.878 & - & 0.289 \\
\hline 
\texttt{H4\_B17.5\_M1.50} & 1.3762 & $\times$ & 986.3 & 1.606 & 2.553 & 0.231 \\
\texttt{H4\_B17.5\_M1.60} & 1.4594 & $\times$ & 680.3 & 1.622 & 2.725 & 0.241 \\
\texttt{H4\_B17.5\_M1.62} & 1.4758 & $\times$ & 632.1 & 1.636 & 2.736 & 0.245 \\
\texttt{H4\_B17.5\_M1.64} & 1.4923 & $\times$ & 587.0 & 1.651 & 2.798 & 0.247 \\
\texttt{H4\_B17.5\_M1.66} & 1.5087 & $\bigtriangleup$ & 545.2 & 1.913 & 2.847 & 0.273 \\
\texttt{H4\_B17.5\_M1.68} & 1.5251 & $\bigcirc$ & 498.2 & 1.903 & 2.854 & 0.275 \\
\texttt{H4\_B17.5\_M1.70} & 1.5413 & $\bigcirc$ & 451.2 & 1.867 & - & 0.279 \\
\texttt{H4\_B17.5\_M1.72} & 1.5575 & $\bigcirc$ & 410.4 & 1.852 & - & 0.283 \\
\texttt{H4\_B17.5\_M1.74} & 1.5736 & $\bigcirc$ & 374.8 & 1.872 & - & 0.287 \\
\texttt{H4\_B17.5\_M1.75} & 1.5818 & $\bigcirc$ & 358.3 & 1.863 & - & 0.288 \\
\texttt{H4\_B17.5\_M1.76} & 1.5898 & $\bigcirc$ & 343.1 & 1.877 & - & 0.289 \\
\texttt{H4\_B17.5\_M1.78} & 1.6058 & $\bigcirc$ & 315.3 & 1.877 & - & 0.291 \\
\texttt{H4\_B17.5\_M1.80} & 1.6216 & $\bigcirc$ & 290.6 & 1.910 & - & 0.294 \\
\texttt{H4\_B17.5\_M1.82} & 1.6377 & $\bigcirc$ & 268.0 & 1.915 & - & 0.297 \\
\texttt{H4\_B17.5\_M1.84} & 1.6535 & $\bigcirc$ & 247.9 & 1.956 & - & 0.301 \\
\texttt{H4\_B17.5\_M1.86} & 1.6694 & $\bigcirc$ & 229.5 & 1.957 & - & 0.304 \\
\hline 
\texttt{H4\_B18.0\_M1.48} & 1.3594 & $\times$ & 1062.8 & 1.610 & 2.539 & 0.229 \\
\texttt{H4\_B18.0\_M1.50} & 1.3762 & $\times$ & 986.3 & 1.609 & 2.516 & 0.231 \\
\texttt{H4\_B18.0\_M1.52} & 1.3929 & $\times$ & 916.2 & 1.618 & 2.569 & 0.233 \\
\texttt{H4\_B18.0\_M1.60} & 1.4594 & $\bigcirc$ & 677.4 & 1.962 & 2.673 & 0.269 \\
\texttt{H4\_B18.0\_M1.62} & 1.4758 & $\bigcirc$ & 609.6 & 1.912 & 2.767 & 0.273 \\
\texttt{H4\_B18.0\_M1.64} & 1.4922 & $\bigcirc$ & 551.2 & 1.923 & 2.700 & 0.276 \\
\texttt{H4\_B18.0\_M1.66} & 1.5086 & $\bigcirc$ & 500.5 & 1.880 & 2.800 & 0.279 \\
\texttt{H4\_B18.0\_M1.68} & 1.5249 & $\bigcirc$ & 456.4 & 1.877 & 2.809 & 0.283 \\
\texttt{H4\_B18.0\_M1.70} & 1.5410 & $\bigcirc$ & 417.9 & 1.855 & 2.861 & 0.286 \\
\texttt{H4\_B18.0\_M1.72} & 1.5572 & $\bigcirc$ & 383.6 & 1.892 & - & 0.289 \\
\texttt{H4\_B18.0\_M1.74} & 1.5733 & $\bigcirc$ & 353.0 & 1.890 & - & 0.292 \\
\texttt{H4\_B18.0\_M1.76} & 1.5894 & $\bigcirc$ & 325.7 & 1.899 & - & 0.294 \\
\texttt{H4\_B18.0\_M1.77} & 1.5973 & $\bigcirc$ & 313.3 & 1.907 & - & 0.295 \\
\texttt{H4\_B18.0\_M1.78} & 1.6053 & $\bigcirc$ & 301.3 & 1.903 & - & 0.297 \\
\texttt{H4\_B18.0\_M1.80} & 1.6212 & $\bigcirc$ & 279.3 & 1.931 & - & 0.300 \\
\texttt{H4\_B18.0\_M1.82} & 1.6371 & $\bigcirc$ & 259.2 & 1.947 & - & 0.302 \\
\texttt{H4\_B18.0\_M1.84} & 1.6529 & $\bigcirc$ & 241.0 & 1.937 & - & 0.305 \\
\texttt{H4\_B18.0\_M1.86} & 1.6687 & $\bigcirc$ & 224.3 & 1.950 & - & 0.307 \\
        \hline\hline
    \end{tabular}
    \label{tab:outcomes_H4}
\end{table*}

\begin{table*}
    \centering
    \caption{Same as \cref{tab:outcomes_APR4} but for the APR4, MPA1 and H4 EOSs in GR.
    The first column lists the model name which combines EOS and baryon mass of each NS in units of $M_\odot$.}
    \begin{tabular}{
        >{\raggedright}p{0.3\columnwidth} 
	>{\centering}p{0.26\columnwidth} 
	>{\centering}p{0.2\columnwidth} 
        >{\centering}p{0.2\columnwidth} 
        >{\centering}p{0.2\columnwidth} 
        >{\centering\arraybackslash}p{0.3\columnwidth} 
    }
        \hline\hline \\[-.8em]
        Model name
        &  $M_{\rm ADM}$ ($M_\odot$) 
        & $\tilde{\Lambda}$ 
        & $f_{\rm peak} $ (kHz)
        & $f_{2} $ (kHz)
        & $D h_{\rm peak} / M$ \\
        \hline 
\texttt{APR4\_M1.48} & 1.3323  & 348.5 & 2.015 & 3.287 & 0.267 \\
\texttt{APR4\_M1.50} & 1.3483  & 323.5 & 2.037 & 3.318 & 0.270 \\
\texttt{APR4\_M1.52} & 1.3643  & 300.9 & 2.042 & 3.417 & 0.272 \\
\texttt{APR4\_M1.54} & 1.3802  & 279.2 & 2.058 & 3.456 & 0.276 \\
\texttt{APR4\_M1.56} & 1.3961  & 260.2 & 2.075 & 3.463 & 0.278 \\
\texttt{APR4\_M1.58} & 1.4119  & 241.8 & 2.102 & - & 0.281 \\
\texttt{APR4\_M1.60} & 1.4277  & 225.7 & 2.114 & - & 0.283 \\
\texttt{APR4\_M1.62} & 1.4434  & 210.2 & 2.128 & - & 0.286 \\
\texttt{APR4\_M1.64} & 1.4591  & 196.1 & 2.150 & - & 0.288 \\
\texttt{APR4\_M1.66} & 1.4747  & 182.9 & 2.158 & - & 0.292 \\
\texttt{APR4\_M1.68} & 1.4903  & 170.5 & 2.178 & - & 0.294 \\
\hline 
\texttt{MPA1\_M1.70} & 1.5198  & 303.0 & 1.858 & 2.971 & 0.273 \\
\texttt{MPA1\_M1.72} & 1.5357  & 285.4 & 1.868 & 3.036 & 0.275 \\
\texttt{MPA1\_M1.74} & 1.5515  & 267.3 & 1.878 & 3.014 & 0.277 \\
\texttt{MPA1\_M1.76} & 1.5673  & 250.9 & 1.900 & 3.087 & 0.280 \\
\texttt{MPA1\_M1.78} & 1.5831  & 236.0 & 1.903 & 3.226 & 0.283 \\
\texttt{MPA1\_M1.79} & 1.5909  & 228.5 & 1.909 & - & 0.283 \\
\texttt{MPA1\_M1.80} & 1.5988  & 221.9 & 1.909 & - & 0.285 \\
\texttt{MPA1\_M1.82} & 1.6144  & 208.6 & 1.900 & - & 0.286 \\
\texttt{MPA1\_M1.84} & 1.6300  & 196.6 & 1.914 & - & 0.288 \\
\texttt{MPA1\_M1.86} & 1.6456  & 184.6 & 1.933 & - & 0.291 \\
\texttt{MPA1\_M1.88} & 1.6611  & 174.2 & 1.934 & - & 0.294 \\
\hline 
\texttt{H4\_M1.46} & 1.3426  & 1145.9 & 1.586 & 2.549 & 0.229 \\
\texttt{H4\_M1.48} & 1.3594  & 1062.8 & 1.585 & 2.451 & 0.229 \\
\texttt{H4\_M1.50} & 1.3762  & 986.3 & 1.584 & 2.615 & 0.232 \\
\texttt{H4\_M1.52} & 1.3929  & 916.2 & 1.595 & 2.688 & 0.233 \\
\texttt{H4\_M1.54} & 1.4096  & 850.3 & 1.617 & 2.628 & 0.235 \\
\texttt{H4\_M1.56} & 1.4262  & 789.4 & 1.614 & 2.745 & 0.237 \\
\texttt{H4\_M1.58} & 1.4428  & 732.9 & 1.616 & - & 0.240 \\
\texttt{H4\_M1.60} & 1.4593  & 678.9 & 1.632 & - & 0.241 \\
\texttt{H4\_M1.62} & 1.4758  & 631.7 & 1.644 & - & 0.246 \\
\texttt{H4\_M1.64} & 1.4923  & 586.0 & 1.637 & - & 0.248 \\
\texttt{H4\_M1.66} & 1.5086  & 543.9 & 1.644 & - & 0.250 \\
\texttt{H4\_M1.68} & 1.5250  & 505.1 & 1.624 & - & 0.252 \\
\texttt{H4\_M1.70} & 1.5413  & 469.2 & 1.635 & - & 0.254 \\
\texttt{H4\_M1.71} & 1.5494  & 452.4 & 1.627 & - & 0.256 \\
\texttt{H4\_M1.72} & 1.5576  & 436.0 & 1.630 & - & 0.257 \\
\texttt{H4\_M1.74} & 1.5738  & 404.5 & 1.640 & - & 0.260 \\
\texttt{H4\_M1.76} & 1.5899  & 375.4 & 1.647 & - & 0.263 \\
\texttt{H4\_M1.78} & 1.6061  & 348.4 & 1.643 & - & 0.265 \\
\texttt{H4\_M1.80} & 1.6221  & 323.4 & 1.670 & - & 0.268 \\
        \hline\hline
    \end{tabular}
    \label{tab:outcomes_GR}
\end{table*}

\bibliography{references}

\begin{thebibliography}{116}%
\makeatletter
\providecommand \@ifxundefined [1]{%
 \@ifx{#1\undefined}
}%
\providecommand \@ifnum [1]{%
 \ifnum #1\expandafter \@firstoftwo
 \else \expandafter \@secondoftwo
 \fi
}%
\providecommand \@ifx [1]{%
 \ifx #1\expandafter \@firstoftwo
 \else \expandafter \@secondoftwo
 \fi
}%
\providecommand \natexlab [1]{#1}%
\providecommand \enquote  [1]{``#1''}%
\providecommand \bibnamefont  [1]{#1}%
\providecommand \bibfnamefont [1]{#1}%
\providecommand \citenamefont [1]{#1}%
\providecommand \href@noop [0]{\@secondoftwo}%
\providecommand \href [0]{\begingroup \@sanitize@url \@href}%
\providecommand \@href[1]{\@@startlink{#1}\@@href}%
\providecommand \@@href[1]{\endgroup#1\@@endlink}%
\providecommand \@sanitize@url [0]{\catcode `\\12\catcode `\$12\catcode
  `\&12\catcode `\#12\catcode `\^12\catcode `\_12\catcode `\%12\relax}%
\providecommand \@@startlink[1]{}%
\providecommand \@@endlink[0]{}%
\providecommand \url  [0]{\begingroup\@sanitize@url \@url }%
\providecommand \@url [1]{\endgroup\@href {#1}{\urlprefix }}%
\providecommand \urlprefix  [0]{URL }%
\providecommand \Eprint [0]{\href }%
\providecommand \doibase [0]{https://doi.org/}%
\providecommand \selectlanguage [0]{\@gobble}%
\providecommand \bibinfo  [0]{\@secondoftwo}%
\providecommand \bibfield  [0]{\@secondoftwo}%
\providecommand \translation [1]{[#1]}%
\providecommand \BibitemOpen [0]{}%
\providecommand \bibitemStop [0]{}%
\providecommand \bibitemNoStop [0]{.\EOS\space}%
\providecommand \EOS [0]{\spacefactor3000\relax}%
\providecommand \BibitemShut  [1]{\csname bibitem#1\endcsname}%
\let\auto@bib@innerbib\@empty
\bibitem [{\citenamefont {Aasi}\ \emph {et~al.}(2015)\citenamefont {Aasi} \emph
  {et~al.}}]{LIGOScientific:2014pky}%
  \BibitemOpen
  \bibfield  {author} {\bibinfo {author} {\bibfnamefont {J.}~\bibnamefont
  {Aasi}} \emph {et~al.} (\bibinfo {collaboration} {LIGO Scientific}),\
  }\bibfield  {title} {\bibinfo {title} {{Advanced LIGO}},\ }\href
  {https://doi.org/10.1088/0264-9381/32/7/074001} {\bibfield  {journal}
  {\bibinfo  {journal} {Class. Quant. Grav.}\ }\textbf {\bibinfo {volume}
  {32}},\ \bibinfo {pages} {074001} (\bibinfo {year} {2015})},\ \Eprint
  {https://arxiv.org/abs/1411.4547} {arXiv:1411.4547 [gr-qc]} \BibitemShut
  {NoStop}%
\bibitem [{\citenamefont {Acernese}\ \emph {et~al.}(2015)\citenamefont
  {Acernese} \emph {et~al.}}]{VIRGO:2014yos}%
  \BibitemOpen
  \bibfield  {author} {\bibinfo {author} {\bibfnamefont {F.}~\bibnamefont
  {Acernese}} \emph {et~al.} (\bibinfo {collaboration} {VIRGO}),\ }\bibfield
  {title} {\bibinfo {title} {{Advanced Virgo: a second-generation
  interferometric gravitational wave detector}},\ }\href
  {https://doi.org/10.1088/0264-9381/32/2/024001} {\bibfield  {journal}
  {\bibinfo  {journal} {Class. Quant. Grav.}\ }\textbf {\bibinfo {volume}
  {32}},\ \bibinfo {pages} {024001} (\bibinfo {year} {2015})},\ \Eprint
  {https://arxiv.org/abs/1408.3978} {arXiv:1408.3978 [gr-qc]} \BibitemShut
  {NoStop}%
\bibitem [{\citenamefont {Abbott}\ \emph
  {et~al.}(2017{\natexlab{a}})\citenamefont {Abbott} \emph
  {et~al.}}]{LIGOScientific:2017vwq}%
  \BibitemOpen
  \bibfield  {author} {\bibinfo {author} {\bibfnamefont {B.~P.}\ \bibnamefont
  {Abbott}} \emph {et~al.} (\bibinfo {collaboration} {LIGO Scientific,
  Virgo}),\ }\bibfield  {title} {\bibinfo {title} {{GW170817: Observation of
  Gravitational Waves from a Binary Neutron Star Inspiral}},\ }\href
  {https://doi.org/10.1103/PhysRevLett.119.161101} {\bibfield  {journal}
  {\bibinfo  {journal} {Phys. Rev. Lett.}\ }\textbf {\bibinfo {volume} {119}},\
  \bibinfo {pages} {161101} (\bibinfo {year} {2017}{\natexlab{a}})},\ \Eprint
  {https://arxiv.org/abs/1710.05832} {arXiv:1710.05832 [gr-qc]} \BibitemShut
  {NoStop}%
\bibitem [{\citenamefont {Abbott}\ \emph
  {et~al.}(2019{\natexlab{a}})\citenamefont {Abbott} \emph
  {et~al.}}]{LIGOScientific:2018mvr}%
  \BibitemOpen
  \bibfield  {author} {\bibinfo {author} {\bibfnamefont {B.~P.}\ \bibnamefont
  {Abbott}} \emph {et~al.} (\bibinfo {collaboration} {LIGO Scientific,
  Virgo}),\ }\bibfield  {title} {\bibinfo {title} {{GWTC-1: A
  Gravitational-Wave Transient Catalog of Compact Binary Mergers Observed by
  LIGO and Virgo during the First and Second Observing Runs}},\ }\href
  {https://doi.org/10.1103/PhysRevX.9.031040} {\bibfield  {journal} {\bibinfo
  {journal} {Phys. Rev. X}\ }\textbf {\bibinfo {volume} {9}},\ \bibinfo {pages}
  {031040} (\bibinfo {year} {2019}{\natexlab{a}})},\ \Eprint
  {https://arxiv.org/abs/1811.12907} {arXiv:1811.12907 [astro-ph.HE]}
  \BibitemShut {NoStop}%
\bibitem [{\citenamefont {Abbott}\ \emph
  {et~al.}(2019{\natexlab{b}})\citenamefont {Abbott} \emph
  {et~al.}}]{LIGOScientific:2018hze}%
  \BibitemOpen
  \bibfield  {author} {\bibinfo {author} {\bibfnamefont {B.~P.}\ \bibnamefont
  {Abbott}} \emph {et~al.} (\bibinfo {collaboration} {LIGO Scientific,
  Virgo}),\ }\bibfield  {title} {\bibinfo {title} {{Properties of the binary
  neutron star merger GW170817}},\ }\href
  {https://doi.org/10.1103/PhysRevX.9.011001} {\bibfield  {journal} {\bibinfo
  {journal} {Phys. Rev. X}\ }\textbf {\bibinfo {volume} {9}},\ \bibinfo {pages}
  {011001} (\bibinfo {year} {2019}{\natexlab{b}})},\ \Eprint
  {https://arxiv.org/abs/1805.11579} {arXiv:1805.11579 [gr-qc]} \BibitemShut
  {NoStop}%
\bibitem [{\citenamefont {Abbott}\ \emph
  {et~al.}(2019{\natexlab{c}})\citenamefont {Abbott} \emph
  {et~al.}}]{LIGOScientific:2018dkp}%
  \BibitemOpen
  \bibfield  {author} {\bibinfo {author} {\bibfnamefont {B.~P.}\ \bibnamefont
  {Abbott}} \emph {et~al.} (\bibinfo {collaboration} {LIGO Scientific,
  Virgo}),\ }\bibfield  {title} {\bibinfo {title} {{Tests of General Relativity
  with GW170817}},\ }\href {https://doi.org/10.1103/PhysRevLett.123.011102}
  {\bibfield  {journal} {\bibinfo  {journal} {Phys. Rev. Lett.}\ }\textbf
  {\bibinfo {volume} {123}},\ \bibinfo {pages} {011102} (\bibinfo {year}
  {2019}{\natexlab{c}})},\ \Eprint {https://arxiv.org/abs/1811.00364}
  {arXiv:1811.00364 [gr-qc]} \BibitemShut {NoStop}%
\bibitem [{\citenamefont {Berti}\ \emph {et~al.}(2018)\citenamefont {Berti},
  \citenamefont {Yagi},\ and\ \citenamefont {Yunes}}]{Berti:2018cxi}%
  \BibitemOpen
  \bibfield  {author} {\bibinfo {author} {\bibfnamefont {E.}~\bibnamefont
  {Berti}}, \bibinfo {author} {\bibfnamefont {K.}~\bibnamefont {Yagi}},\ and\
  \bibinfo {author} {\bibfnamefont {N.}~\bibnamefont {Yunes}},\ }\bibfield
  {title} {\bibinfo {title} {{Extreme Gravity Tests with Gravitational Waves
  from Compact Binary Coalescences: (I) Inspiral-Merger}},\ }\href
  {https://doi.org/10.1007/s10714-018-2362-8} {\bibfield  {journal} {\bibinfo
  {journal} {Gen. Rel. Grav.}\ }\textbf {\bibinfo {volume} {50}},\ \bibinfo
  {pages} {46} (\bibinfo {year} {2018})},\ \Eprint
  {https://arxiv.org/abs/1801.03208} {arXiv:1801.03208 [gr-qc]} \BibitemShut
  {NoStop}%
\bibitem [{\citenamefont {Perkins}\ \emph {et~al.}(2021)\citenamefont
  {Perkins}, \citenamefont {Yunes},\ and\ \citenamefont
  {Berti}}]{Perkins:2020tra}%
  \BibitemOpen
  \bibfield  {author} {\bibinfo {author} {\bibfnamefont {S.~E.}\ \bibnamefont
  {Perkins}}, \bibinfo {author} {\bibfnamefont {N.}~\bibnamefont {Yunes}},\
  and\ \bibinfo {author} {\bibfnamefont {E.}~\bibnamefont {Berti}},\ }\bibfield
   {title} {\bibinfo {title} {{Probing Fundamental Physics with Gravitational
  Waves: The Next Generation}},\ }\href
  {https://doi.org/10.1103/PhysRevD.103.044024} {\bibfield  {journal} {\bibinfo
   {journal} {Phys. Rev. D}\ }\textbf {\bibinfo {volume} {103}},\ \bibinfo
  {pages} {044024} (\bibinfo {year} {2021})},\ \Eprint
  {https://arxiv.org/abs/2010.09010} {arXiv:2010.09010 [gr-qc]} \BibitemShut
  {NoStop}%
\bibitem [{\citenamefont {Abbott}\ \emph {et~al.}(2018)\citenamefont {Abbott}
  \emph {et~al.}}]{LIGOScientific:2018cki}%
  \BibitemOpen
  \bibfield  {author} {\bibinfo {author} {\bibfnamefont {B.~P.}\ \bibnamefont
  {Abbott}} \emph {et~al.} (\bibinfo {collaboration} {LIGO Scientific,
  Virgo}),\ }\bibfield  {title} {\bibinfo {title} {{GW170817: Measurements of
  neutron star radii and equation of state}},\ }\href
  {https://doi.org/10.1103/PhysRevLett.121.161101} {\bibfield  {journal}
  {\bibinfo  {journal} {Phys. Rev. Lett.}\ }\textbf {\bibinfo {volume} {121}},\
  \bibinfo {pages} {161101} (\bibinfo {year} {2018})},\ \Eprint
  {https://arxiv.org/abs/1805.11581} {arXiv:1805.11581 [gr-qc]} \BibitemShut
  {NoStop}%
\bibitem [{\citenamefont {Raithel}(2019)}]{Raithel:2019uzi}%
  \BibitemOpen
  \bibfield  {author} {\bibinfo {author} {\bibfnamefont {C.~A.}\ \bibnamefont
  {Raithel}},\ }\bibfield  {title} {\bibinfo {title} {{Constraints on the
  Neutron Star Equation of State from GW170817}},\ }\href
  {https://doi.org/10.1140/epja/i2019-12759-5} {\bibfield  {journal} {\bibinfo
  {journal} {Eur. Phys. J. A}\ }\textbf {\bibinfo {volume} {55}},\ \bibinfo
  {pages} {80} (\bibinfo {year} {2019})},\ \Eprint
  {https://arxiv.org/abs/1904.10002} {arXiv:1904.10002 [astro-ph.HE]}
  \BibitemShut {NoStop}%
\bibitem [{\citenamefont {Chatziioannou}(2020)}]{Chatziioannou:2020pqz}%
  \BibitemOpen
  \bibfield  {author} {\bibinfo {author} {\bibfnamefont {K.}~\bibnamefont
  {Chatziioannou}},\ }\bibfield  {title} {\bibinfo {title} {{Neutron star tidal
  deformability and equation of state constraints}},\ }\href
  {https://doi.org/10.1007/s10714-020-02754-3} {\bibfield  {journal} {\bibinfo
  {journal} {Gen. Rel. Grav.}\ }\textbf {\bibinfo {volume} {52}},\ \bibinfo
  {pages} {109} (\bibinfo {year} {2020})},\ \Eprint
  {https://arxiv.org/abs/2006.03168} {arXiv:2006.03168 [gr-qc]} \BibitemShut
  {NoStop}%
\bibitem [{\citenamefont {Dietrich}\ \emph {et~al.}(2021)\citenamefont
  {Dietrich}, \citenamefont {Hinderer},\ and\ \citenamefont
  {Samajdar}}]{Dietrich:2020eud}%
  \BibitemOpen
  \bibfield  {author} {\bibinfo {author} {\bibfnamefont {T.}~\bibnamefont
  {Dietrich}}, \bibinfo {author} {\bibfnamefont {T.}~\bibnamefont {Hinderer}},\
  and\ \bibinfo {author} {\bibfnamefont {A.}~\bibnamefont {Samajdar}},\
  }\bibfield  {title} {\bibinfo {title} {{Interpreting Binary Neutron Star
  Mergers: Describing the Binary Neutron Star Dynamics, Modelling Gravitational
  Waveforms, and Analyzing Detections}},\ }\href
  {https://doi.org/10.1007/s10714-020-02751-6} {\bibfield  {journal} {\bibinfo
  {journal} {Gen. Rel. Grav.}\ }\textbf {\bibinfo {volume} {53}},\ \bibinfo
  {pages} {27} (\bibinfo {year} {2021})},\ \Eprint
  {https://arxiv.org/abs/2004.02527} {arXiv:2004.02527 [gr-qc]} \BibitemShut
  {NoStop}%
\bibitem [{\citenamefont {Shibata}\ \emph {et~al.}(2006)\citenamefont
  {Shibata}, \citenamefont {Duez}, \citenamefont {Liu}, \citenamefont
  {Shapiro},\ and\ \citenamefont {Stephens}}]{Shibata:2005mz}%
  \BibitemOpen
  \bibfield  {author} {\bibinfo {author} {\bibfnamefont {M.}~\bibnamefont
  {Shibata}}, \bibinfo {author} {\bibfnamefont {M.~D.}\ \bibnamefont {Duez}},
  \bibinfo {author} {\bibfnamefont {Y.~T.}\ \bibnamefont {Liu}}, \bibinfo
  {author} {\bibfnamefont {S.~L.}\ \bibnamefont {Shapiro}},\ and\ \bibinfo
  {author} {\bibfnamefont {B.~C.}\ \bibnamefont {Stephens}},\ }\bibfield
  {title} {\bibinfo {title} {{Magnetized hypermassive neutron star collapse: A
  Central engine for short gamma-ray bursts}},\ }\href
  {https://doi.org/10.1103/PhysRevLett.96.031102} {\bibfield  {journal}
  {\bibinfo  {journal} {Phys. Rev. Lett.}\ }\textbf {\bibinfo {volume} {96}},\
  \bibinfo {pages} {031102} (\bibinfo {year} {2006})},\ \Eprint
  {https://arxiv.org/abs/astro-ph/0511142} {arXiv:astro-ph/0511142}
  \BibitemShut {NoStop}%
\bibitem [{\citenamefont {Bauswein}\ and\ \citenamefont
  {Janka}(2012)}]{Bauswein:2011tp}%
  \BibitemOpen
  \bibfield  {author} {\bibinfo {author} {\bibfnamefont {A.}~\bibnamefont
  {Bauswein}}\ and\ \bibinfo {author} {\bibfnamefont {H.~T.}\ \bibnamefont
  {Janka}},\ }\bibfield  {title} {\bibinfo {title} {{Measuring neutron-star
  properties via gravitational waves from binary mergers}},\ }\href
  {https://doi.org/10.1103/PhysRevLett.108.011101} {\bibfield  {journal}
  {\bibinfo  {journal} {Phys. Rev. Lett.}\ }\textbf {\bibinfo {volume} {108}},\
  \bibinfo {pages} {011101} (\bibinfo {year} {2012})},\ \Eprint
  {https://arxiv.org/abs/1106.1616} {arXiv:1106.1616 [astro-ph.SR]}
  \BibitemShut {NoStop}%
\bibitem [{\citenamefont {Hotokezaka}\ \emph {et~al.}(2011)\citenamefont
  {Hotokezaka}, \citenamefont {Kyutoku}, \citenamefont {Okawa}, \citenamefont
  {Shibata},\ and\ \citenamefont {Kiuchi}}]{Hotokezaka:2011dh}%
  \BibitemOpen
  \bibfield  {author} {\bibinfo {author} {\bibfnamefont {K.}~\bibnamefont
  {Hotokezaka}}, \bibinfo {author} {\bibfnamefont {K.}~\bibnamefont {Kyutoku}},
  \bibinfo {author} {\bibfnamefont {H.}~\bibnamefont {Okawa}}, \bibinfo
  {author} {\bibfnamefont {M.}~\bibnamefont {Shibata}},\ and\ \bibinfo {author}
  {\bibfnamefont {K.}~\bibnamefont {Kiuchi}},\ }\bibfield  {title} {\bibinfo
  {title} {{Binary Neutron Star Mergers: Dependence on the Nuclear Equation of
  State}},\ }\href {https://doi.org/10.1103/PhysRevD.83.124008} {\bibfield
  {journal} {\bibinfo  {journal} {Phys. Rev. D}\ }\textbf {\bibinfo {volume}
  {83}},\ \bibinfo {pages} {124008} (\bibinfo {year} {2011})},\ \Eprint
  {https://arxiv.org/abs/1105.4370} {arXiv:1105.4370 [astro-ph.HE]}
  \BibitemShut {NoStop}%
\bibitem [{\citenamefont {Abbott}\ \emph
  {et~al.}(2017{\natexlab{b}})\citenamefont {Abbott} \emph
  {et~al.}}]{LIGOScientific:2017fdd}%
  \BibitemOpen
  \bibfield  {author} {\bibinfo {author} {\bibfnamefont {B.~P.}\ \bibnamefont
  {Abbott}} \emph {et~al.} (\bibinfo {collaboration} {LIGO Scientific,
  Virgo}),\ }\bibfield  {title} {\bibinfo {title} {{Search for Post-merger
  Gravitational Waves from the Remnant of the Binary Neutron Star Merger
  GW170817}},\ }\href {https://doi.org/10.3847/2041-8213/aa9a35} {\bibfield
  {journal} {\bibinfo  {journal} {Astrophys. J. Lett.}\ }\textbf {\bibinfo
  {volume} {851}},\ \bibinfo {pages} {L16} (\bibinfo {year}
  {2017}{\natexlab{b}})},\ \Eprint {https://arxiv.org/abs/1710.09320}
  {arXiv:1710.09320 [astro-ph.HE]} \BibitemShut {NoStop}%
\bibitem [{\citenamefont {Abbott}\ \emph
  {et~al.}(2019{\natexlab{d}})\citenamefont {Abbott} \emph
  {et~al.}}]{LIGOScientific:2018urg}%
  \BibitemOpen
  \bibfield  {author} {\bibinfo {author} {\bibfnamefont {B.~P.}\ \bibnamefont
  {Abbott}} \emph {et~al.} (\bibinfo {collaboration} {LIGO Scientific,
  Virgo}),\ }\bibfield  {title} {\bibinfo {title} {{Search for gravitational
  waves from a long-lived remnant of the binary neutron star merger
  GW170817}},\ }\href {https://doi.org/10.3847/1538-4357/ab0f3d} {\bibfield
  {journal} {\bibinfo  {journal} {Astrophys. J.}\ }\textbf {\bibinfo {volume}
  {875}},\ \bibinfo {pages} {160} (\bibinfo {year} {2019}{\natexlab{d}})},\
  \Eprint {https://arxiv.org/abs/1810.02581} {arXiv:1810.02581 [gr-qc]}
  \BibitemShut {NoStop}%
\bibitem [{\citenamefont {Abbott}\ \emph {et~al.}(2020)\citenamefont {Abbott}
  \emph {et~al.}}]{LIGOScientific:2020aai}%
  \BibitemOpen
  \bibfield  {author} {\bibinfo {author} {\bibfnamefont {B.~P.}\ \bibnamefont
  {Abbott}} \emph {et~al.} (\bibinfo {collaboration} {LIGO Scientific,
  Virgo}),\ }\bibfield  {title} {\bibinfo {title} {{GW190425: Observation of a
  Compact Binary Coalescence with Total Mass $\sim 3.4 M_{\odot}$}},\ }\href
  {https://doi.org/10.3847/2041-8213/ab75f5} {\bibfield  {journal} {\bibinfo
  {journal} {Astrophys. J. Lett.}\ }\textbf {\bibinfo {volume} {892}},\
  \bibinfo {pages} {L3} (\bibinfo {year} {2020})},\ \Eprint
  {https://arxiv.org/abs/2001.01761} {arXiv:2001.01761 [astro-ph.HE]}
  \BibitemShut {NoStop}%
\bibitem [{\citenamefont {Hild}\ \emph {et~al.}(2011)\citenamefont {Hild} \emph
  {et~al.}}]{Hild:2010id}%
  \BibitemOpen
  \bibfield  {author} {\bibinfo {author} {\bibfnamefont {S.}~\bibnamefont
  {Hild}} \emph {et~al.},\ }\bibfield  {title} {\bibinfo {title} {{Sensitivity
  Studies for Third-Generation Gravitational Wave Observatories}},\ }\href
  {https://doi.org/10.1088/0264-9381/28/9/094013} {\bibfield  {journal}
  {\bibinfo  {journal} {Class. Quant. Grav.}\ }\textbf {\bibinfo {volume}
  {28}},\ \bibinfo {pages} {094013} (\bibinfo {year} {2011})}\BibitemShut
  {NoStop}%
\bibitem [{\citenamefont {Punturo}\ \emph {et~al.}(2010)\citenamefont {Punturo}
  \emph {et~al.}}]{Punturo:2010zz}%
  \BibitemOpen
  \bibfield  {author} {\bibinfo {author} {\bibfnamefont {M.}~\bibnamefont
  {Punturo}} \emph {et~al.},\ }\bibfield  {title} {\bibinfo {title} {{The
  Einstein Telescope: A third-generation gravitational wave observatory}},\
  }\href {https://doi.org/10.1088/0264-9381/27/19/194002} {\bibfield  {journal}
  {\bibinfo  {journal} {Class. Quant. Grav.}\ }\textbf {\bibinfo {volume}
  {27}},\ \bibinfo {pages} {194002} (\bibinfo {year} {2010})}\BibitemShut
  {NoStop}%
\bibitem [{\citenamefont {Branchesi}\ \emph {et~al.}(2023)\citenamefont
  {Branchesi} \emph {et~al.}}]{Branchesi:2023mws}%
  \BibitemOpen
  \bibfield  {author} {\bibinfo {author} {\bibfnamefont {M.}~\bibnamefont
  {Branchesi}} \emph {et~al.},\ }\bibfield  {title} {\bibinfo {title} {{Science
  with the Einstein Telescope: a comparison of different designs}},\ }\href
  {https://doi.org/10.1088/1475-7516/2023/07/068} {\bibfield  {journal}
  {\bibinfo  {journal} {JCAP}\ }\textbf {\bibinfo {volume} {07}},\ \bibinfo
  {pages} {068}},\ \Eprint {https://arxiv.org/abs/2303.15923} {arXiv:2303.15923
  [gr-qc]} \BibitemShut {NoStop}%
\bibitem [{\citenamefont {Abbott}\ \emph
  {et~al.}(2017{\natexlab{c}})\citenamefont {Abbott} \emph
  {et~al.}}]{LIGOScientific:2016wof}%
  \BibitemOpen
  \bibfield  {author} {\bibinfo {author} {\bibfnamefont {B.~P.}\ \bibnamefont
  {Abbott}} \emph {et~al.} (\bibinfo {collaboration} {LIGO Scientific}),\
  }\bibfield  {title} {\bibinfo {title} {{Exploring the Sensitivity of Next
  Generation Gravitational Wave Detectors}},\ }\href
  {https://doi.org/10.1088/1361-6382/aa51f4} {\bibfield  {journal} {\bibinfo
  {journal} {Class. Quant. Grav.}\ }\textbf {\bibinfo {volume} {34}},\ \bibinfo
  {pages} {044001} (\bibinfo {year} {2017}{\natexlab{c}})},\ \Eprint
  {https://arxiv.org/abs/1607.08697} {arXiv:1607.08697 [astro-ph.IM]}
  \BibitemShut {NoStop}%
\bibitem [{\citenamefont {Reitze}\ \emph {et~al.}(2019)\citenamefont {Reitze}
  \emph {et~al.}}]{Reitze:2019iox}%
  \BibitemOpen
  \bibfield  {author} {\bibinfo {author} {\bibfnamefont {D.}~\bibnamefont
  {Reitze}} \emph {et~al.},\ }\bibfield  {title} {\bibinfo {title} {{Cosmic
  Explorer: The U.S. Contribution to Gravitational-Wave Astronomy beyond
  LIGO}},\ }\href@noop {} {\bibfield  {journal} {\bibinfo  {journal} {Bull. Am.
  Astron. Soc.}\ }\textbf {\bibinfo {volume} {51}},\ \bibinfo {pages} {035}
  (\bibinfo {year} {2019})},\ \Eprint {https://arxiv.org/abs/1907.04833}
  {arXiv:1907.04833 [astro-ph.IM]} \BibitemShut {NoStop}%
\bibitem [{\citenamefont {Evans}\ \emph {et~al.}(2021)\citenamefont {Evans}
  \emph {et~al.}}]{Evans:2021gyd}%
  \BibitemOpen
  \bibfield  {author} {\bibinfo {author} {\bibfnamefont {M.}~\bibnamefont
  {Evans}} \emph {et~al.},\ }\bibfield  {title} {\bibinfo {title} {{A Horizon
  Study for Cosmic Explorer: Science, Observatories, and Community}},\
  }\href@noop {} {\bibfield  {journal} {\bibinfo  {journal}
  {{arXiv:2109.09882}}\ } (\bibinfo {year} {2021})},\ \Eprint
  {https://arxiv.org/abs/2109.09882} {arXiv:2109.09882 [astro-ph.IM]}
  \BibitemShut {NoStop}%
\bibitem [{\citenamefont {Baumgarte}\ \emph {et~al.}(2000)\citenamefont
  {Baumgarte}, \citenamefont {Shapiro},\ and\ \citenamefont
  {Shibata}}]{Baumgarte:1999cq}%
  \BibitemOpen
  \bibfield  {author} {\bibinfo {author} {\bibfnamefont {T.~W.}\ \bibnamefont
  {Baumgarte}}, \bibinfo {author} {\bibfnamefont {S.~L.}\ \bibnamefont
  {Shapiro}},\ and\ \bibinfo {author} {\bibfnamefont {M.}~\bibnamefont
  {Shibata}},\ }\bibfield  {title} {\bibinfo {title} {{On the maximum mass of
  differentially rotating neutron stars}},\ }\href
  {https://doi.org/10.1086/312425} {\bibfield  {journal} {\bibinfo  {journal}
  {Astrophys. J. Lett.}\ }\textbf {\bibinfo {volume} {528}},\ \bibinfo {pages}
  {L29} (\bibinfo {year} {2000})},\ \Eprint
  {https://arxiv.org/abs/astro-ph/9910565} {arXiv:astro-ph/9910565}
  \BibitemShut {NoStop}%
\bibitem [{\citenamefont {Shibata}\ \emph {et~al.}(2005)\citenamefont
  {Shibata}, \citenamefont {Taniguchi},\ and\ \citenamefont
  {Uryu}}]{Shibata:2005ss}%
  \BibitemOpen
  \bibfield  {author} {\bibinfo {author} {\bibfnamefont {M.}~\bibnamefont
  {Shibata}}, \bibinfo {author} {\bibfnamefont {K.}~\bibnamefont {Taniguchi}},\
  and\ \bibinfo {author} {\bibfnamefont {K.}~\bibnamefont {Uryu}},\ }\bibfield
  {title} {\bibinfo {title} {{Merger of binary neutron stars with realistic
  equations of state in full general relativity}},\ }\href
  {https://doi.org/10.1103/PhysRevD.71.084021} {\bibfield  {journal} {\bibinfo
  {journal} {Phys. Rev. D}\ }\textbf {\bibinfo {volume} {71}},\ \bibinfo
  {pages} {084021} (\bibinfo {year} {2005})},\ \Eprint
  {https://arxiv.org/abs/gr-qc/0503119} {arXiv:gr-qc/0503119} \BibitemShut
  {NoStop}%
\bibitem [{\citenamefont {Bauswein}\ \emph {et~al.}(2010)\citenamefont
  {Bauswein}, \citenamefont {Janka},\ and\ \citenamefont
  {Oechslin}}]{Bauswein:2010dn}%
  \BibitemOpen
  \bibfield  {author} {\bibinfo {author} {\bibfnamefont {A.}~\bibnamefont
  {Bauswein}}, \bibinfo {author} {\bibfnamefont {H.~T.}\ \bibnamefont
  {Janka}},\ and\ \bibinfo {author} {\bibfnamefont {R.}~\bibnamefont
  {Oechslin}},\ }\bibfield  {title} {\bibinfo {title} {{Testing Approximations
  of Thermal Effects in Neutron Star Merger Simulations}},\ }\href
  {https://doi.org/10.1103/PhysRevD.82.084043} {\bibfield  {journal} {\bibinfo
  {journal} {Phys. Rev. D}\ }\textbf {\bibinfo {volume} {82}},\ \bibinfo
  {pages} {084043} (\bibinfo {year} {2010})},\ \Eprint
  {https://arxiv.org/abs/1006.3315} {arXiv:1006.3315 [astro-ph.SR]}
  \BibitemShut {NoStop}%
\bibitem [{\citenamefont {Hotokezaka}\ \emph {et~al.}(2013)\citenamefont
  {Hotokezaka}, \citenamefont {Kiuchi}, \citenamefont {Kyutoku}, \citenamefont
  {Muranushi}, \citenamefont {Sekiguchi}, \citenamefont {Shibata},\ and\
  \citenamefont {Taniguchi}}]{Hotokezaka:2013iia}%
  \BibitemOpen
  \bibfield  {author} {\bibinfo {author} {\bibfnamefont {K.}~\bibnamefont
  {Hotokezaka}}, \bibinfo {author} {\bibfnamefont {K.}~\bibnamefont {Kiuchi}},
  \bibinfo {author} {\bibfnamefont {K.}~\bibnamefont {Kyutoku}}, \bibinfo
  {author} {\bibfnamefont {T.}~\bibnamefont {Muranushi}}, \bibinfo {author}
  {\bibfnamefont {Y.-i.}\ \bibnamefont {Sekiguchi}}, \bibinfo {author}
  {\bibfnamefont {M.}~\bibnamefont {Shibata}},\ and\ \bibinfo {author}
  {\bibfnamefont {K.}~\bibnamefont {Taniguchi}},\ }\bibfield  {title} {\bibinfo
  {title} {{Remnant massive neutron stars of binary neutron star mergers:
  Evolution process and gravitational waveform}},\ }\href
  {https://doi.org/10.1103/PhysRevD.88.044026} {\bibfield  {journal} {\bibinfo
  {journal} {Phys. Rev. D}\ }\textbf {\bibinfo {volume} {88}},\ \bibinfo
  {pages} {044026} (\bibinfo {year} {2013})},\ \Eprint
  {https://arxiv.org/abs/1307.5888} {arXiv:1307.5888 [astro-ph.HE]}
  \BibitemShut {NoStop}%
\bibitem [{\citenamefont {Kaplan}\ \emph {et~al.}(2014)\citenamefont {Kaplan},
  \citenamefont {Ott}, \citenamefont {O'Connor}, \citenamefont {Kiuchi},
  \citenamefont {Roberts},\ and\ \citenamefont {Duez}}]{Kaplan:2013wra}%
  \BibitemOpen
  \bibfield  {author} {\bibinfo {author} {\bibfnamefont {J.~D.}\ \bibnamefont
  {Kaplan}}, \bibinfo {author} {\bibfnamefont {C.~D.}\ \bibnamefont {Ott}},
  \bibinfo {author} {\bibfnamefont {E.~P.}\ \bibnamefont {O'Connor}}, \bibinfo
  {author} {\bibfnamefont {K.}~\bibnamefont {Kiuchi}}, \bibinfo {author}
  {\bibfnamefont {L.}~\bibnamefont {Roberts}},\ and\ \bibinfo {author}
  {\bibfnamefont {M.}~\bibnamefont {Duez}},\ }\bibfield  {title} {\bibinfo
  {title} {{The Influence of Thermal Pressure on Equilibrium Models of
  Hypermassive Neutron Star Merger Remnants}},\ }\href
  {https://doi.org/10.1088/0004-637X/790/1/19} {\bibfield  {journal} {\bibinfo
  {journal} {Astrophys. J.}\ }\textbf {\bibinfo {volume} {790}},\ \bibinfo
  {pages} {19} (\bibinfo {year} {2014})},\ \Eprint
  {https://arxiv.org/abs/1306.4034} {arXiv:1306.4034 [astro-ph.HE]}
  \BibitemShut {NoStop}%
\bibitem [{\citenamefont {Muhammed}\ \emph {et~al.}(2024)\citenamefont
  {Muhammed}, \citenamefont {Duez}, \citenamefont {Chawhan}, \citenamefont
  {Ghadiri}, \citenamefont {Buchman}, \citenamefont {Foucart}, \citenamefont
  {Cheong}, \citenamefont {Kidder}, \citenamefont {Pfeiffer},\ and\
  \citenamefont {Scheel}}]{Muhammed:2024zmk}%
  \BibitemOpen
  \bibfield  {author} {\bibinfo {author} {\bibfnamefont {N.}~\bibnamefont
  {Muhammed}}, \bibinfo {author} {\bibfnamefont {M.~D.}\ \bibnamefont {Duez}},
  \bibinfo {author} {\bibfnamefont {P.}~\bibnamefont {Chawhan}}, \bibinfo
  {author} {\bibfnamefont {N.}~\bibnamefont {Ghadiri}}, \bibinfo {author}
  {\bibfnamefont {L.~T.}\ \bibnamefont {Buchman}}, \bibinfo {author}
  {\bibfnamefont {F.}~\bibnamefont {Foucart}}, \bibinfo {author} {\bibfnamefont
  {P.~C.-K.}\ \bibnamefont {Cheong}}, \bibinfo {author} {\bibfnamefont {L.~E.}\
  \bibnamefont {Kidder}}, \bibinfo {author} {\bibfnamefont {H.~P.}\
  \bibnamefont {Pfeiffer}},\ and\ \bibinfo {author} {\bibfnamefont {M.~A.}\
  \bibnamefont {Scheel}},\ }\bibfield  {title} {\bibinfo {title} {{Stability of
  hypermassive neutron stars with realistic rotation and entropy profiles}},\
  }\href@noop {} {\bibfield  {journal} {\bibinfo  {journal} {arXiv:
  2403.05642}\ } (\bibinfo {year} {2024})},\ \Eprint
  {https://arxiv.org/abs/2403.05642} {arXiv:2403.05642 [gr-qc]} \BibitemShut
  {NoStop}%
\bibitem [{\citenamefont {Shibata}\ \emph {et~al.}(2003)\citenamefont
  {Shibata}, \citenamefont {Taniguchi},\ and\ \citenamefont
  {Uryu}}]{Shibata:2003ga}%
  \BibitemOpen
  \bibfield  {author} {\bibinfo {author} {\bibfnamefont {M.}~\bibnamefont
  {Shibata}}, \bibinfo {author} {\bibfnamefont {K.}~\bibnamefont {Taniguchi}},\
  and\ \bibinfo {author} {\bibfnamefont {K.}~\bibnamefont {Uryu}},\ }\bibfield
  {title} {\bibinfo {title} {{Merger of binary neutron stars of unequal mass in
  full general relativity}},\ }\href
  {https://doi.org/10.1103/PhysRevD.68.084020} {\bibfield  {journal} {\bibinfo
  {journal} {Phys. Rev. D}\ }\textbf {\bibinfo {volume} {68}},\ \bibinfo
  {pages} {084020} (\bibinfo {year} {2003})},\ \Eprint
  {https://arxiv.org/abs/gr-qc/0310030} {arXiv:gr-qc/0310030} \BibitemShut
  {NoStop}%
\bibitem [{\citenamefont {Shibata}(2005)}]{Shibata:2005xz}%
  \BibitemOpen
  \bibfield  {author} {\bibinfo {author} {\bibfnamefont {M.}~\bibnamefont
  {Shibata}},\ }\bibfield  {title} {\bibinfo {title} {{Constraining nuclear
  equations of state using gravitational waves from hypermassive neutron
  stars}},\ }\href {https://doi.org/10.1103/PhysRevLett.94.201101} {\bibfield
  {journal} {\bibinfo  {journal} {Phys. Rev. Lett.}\ }\textbf {\bibinfo
  {volume} {94}},\ \bibinfo {pages} {201101} (\bibinfo {year} {2005})},\
  \Eprint {https://arxiv.org/abs/gr-qc/0504082} {arXiv:gr-qc/0504082}
  \BibitemShut {NoStop}%
\bibitem [{\citenamefont {Oechslin}\ and\ \citenamefont
  {Janka}(2007)}]{Oechslin:2007gn}%
  \BibitemOpen
  \bibfield  {author} {\bibinfo {author} {\bibfnamefont {R.}~\bibnamefont
  {Oechslin}}\ and\ \bibinfo {author} {\bibfnamefont {H.~T.}\ \bibnamefont
  {Janka}},\ }\bibfield  {title} {\bibinfo {title} {{Gravitational waves from
  relativistic neutron star mergers with nonzero-temperature equations of
  state}},\ }\href {https://doi.org/10.1103/PhysRevLett.99.121102} {\bibfield
  {journal} {\bibinfo  {journal} {Phys. Rev. Lett.}\ }\textbf {\bibinfo
  {volume} {99}},\ \bibinfo {pages} {121102} (\bibinfo {year} {2007})},\
  \Eprint {https://arxiv.org/abs/astro-ph/0702228} {arXiv:astro-ph/0702228}
  \BibitemShut {NoStop}%
\bibitem [{\citenamefont {Takami}\ \emph {et~al.}(2014)\citenamefont {Takami},
  \citenamefont {Rezzolla},\ and\ \citenamefont {Baiotti}}]{Takami:2014zpa}%
  \BibitemOpen
  \bibfield  {author} {\bibinfo {author} {\bibfnamefont {K.}~\bibnamefont
  {Takami}}, \bibinfo {author} {\bibfnamefont {L.}~\bibnamefont {Rezzolla}},\
  and\ \bibinfo {author} {\bibfnamefont {L.}~\bibnamefont {Baiotti}},\
  }\bibfield  {title} {\bibinfo {title} {{Constraining the Equation of State of
  Neutron Stars from Binary Mergers}},\ }\href
  {https://doi.org/10.1103/PhysRevLett.113.091104} {\bibfield  {journal}
  {\bibinfo  {journal} {Phys. Rev. Lett.}\ }\textbf {\bibinfo {volume} {113}},\
  \bibinfo {pages} {091104} (\bibinfo {year} {2014})},\ \Eprint
  {https://arxiv.org/abs/1403.5672} {arXiv:1403.5672 [gr-qc]} \BibitemShut
  {NoStop}%
\bibitem [{\citenamefont {Takami}\ \emph {et~al.}(2015)\citenamefont {Takami},
  \citenamefont {Rezzolla},\ and\ \citenamefont {Baiotti}}]{Takami:2014tva}%
  \BibitemOpen
  \bibfield  {author} {\bibinfo {author} {\bibfnamefont {K.}~\bibnamefont
  {Takami}}, \bibinfo {author} {\bibfnamefont {L.}~\bibnamefont {Rezzolla}},\
  and\ \bibinfo {author} {\bibfnamefont {L.}~\bibnamefont {Baiotti}},\
  }\bibfield  {title} {\bibinfo {title} {{Spectral properties of the
  post-merger gravitational-wave signal from binary neutron stars}},\ }\href
  {https://doi.org/10.1103/PhysRevD.91.064001} {\bibfield  {journal} {\bibinfo
  {journal} {Phys. Rev. D}\ }\textbf {\bibinfo {volume} {91}},\ \bibinfo
  {pages} {064001} (\bibinfo {year} {2015})},\ \Eprint
  {https://arxiv.org/abs/1412.3240} {arXiv:1412.3240 [gr-qc]} \BibitemShut
  {NoStop}%
\bibitem [{\citenamefont {Sotani}\ and\ \citenamefont
  {Kokkotas}(2004)}]{Sotani:2004rq}%
  \BibitemOpen
  \bibfield  {author} {\bibinfo {author} {\bibfnamefont {H.}~\bibnamefont
  {Sotani}}\ and\ \bibinfo {author} {\bibfnamefont {K.~D.}\ \bibnamefont
  {Kokkotas}},\ }\bibfield  {title} {\bibinfo {title} {{Probing strong-field
  scalar-tensor gravity with gravitational wave asteroseismology}},\ }\href
  {https://doi.org/10.1103/PhysRevD.70.084026} {\bibfield  {journal} {\bibinfo
  {journal} {Phys. Rev. D}\ }\textbf {\bibinfo {volume} {70}},\ \bibinfo
  {pages} {084026} (\bibinfo {year} {2004})},\ \Eprint
  {https://arxiv.org/abs/gr-qc/0409066} {arXiv:gr-qc/0409066} \BibitemShut
  {NoStop}%
\bibitem [{\citenamefont {Kr\"uger}\ and\ \citenamefont
  {Doneva}(2021)}]{Kruger:2021yay}%
  \BibitemOpen
  \bibfield  {author} {\bibinfo {author} {\bibfnamefont {C.~J.}\ \bibnamefont
  {Kr\"uger}}\ and\ \bibinfo {author} {\bibfnamefont {D.~D.}\ \bibnamefont
  {Doneva}},\ }\bibfield  {title} {\bibinfo {title} {{Oscillation dynamics of
  scalarized neutron stars}},\ }\href
  {https://doi.org/10.1103/PhysRevD.103.124034} {\bibfield  {journal} {\bibinfo
   {journal} {Phys. Rev. D}\ }\textbf {\bibinfo {volume} {103}},\ \bibinfo
  {pages} {124034} (\bibinfo {year} {2021})},\ \Eprint
  {https://arxiv.org/abs/2102.11698} {arXiv:2102.11698 [gr-qc]} \BibitemShut
  {NoStop}%
\bibitem [{\citenamefont {Bl\'azquez-Salcedo}\ \emph
  {et~al.}(2021)\citenamefont {Bl\'azquez-Salcedo}, \citenamefont {Khoo},
  \citenamefont {Kunz},\ and\ \citenamefont
  {Preut}}]{Blazquez-Salcedo:2021exm}%
  \BibitemOpen
  \bibfield  {author} {\bibinfo {author} {\bibfnamefont {J.~L.}\ \bibnamefont
  {Bl\'azquez-Salcedo}}, \bibinfo {author} {\bibfnamefont {F.~S.}\ \bibnamefont
  {Khoo}}, \bibinfo {author} {\bibfnamefont {J.}~\bibnamefont {Kunz}},\ and\
  \bibinfo {author} {\bibfnamefont {V.}~\bibnamefont {Preut}},\ }\bibfield
  {title} {\bibinfo {title} {{Polar Quasinormal Modes of Neutron Stars in
  Massive Scalar-Tensor Theories}},\ }\href
  {https://doi.org/10.3389/fphy.2021.741427} {\bibfield  {journal} {\bibinfo
  {journal} {Front. in Phys.}\ }\textbf {\bibinfo {volume} {9}},\ \bibinfo
  {pages} {741427} (\bibinfo {year} {2021})},\ \Eprint
  {https://arxiv.org/abs/2107.06726} {arXiv:2107.06726 [gr-qc]} \BibitemShut
  {NoStop}%
\bibitem [{\citenamefont {Clark}\ \emph {et~al.}(2014)\citenamefont {Clark},
  \citenamefont {Bauswein}, \citenamefont {Cadonati}, \citenamefont {Janka},
  \citenamefont {Pankow},\ and\ \citenamefont {Stergioulas}}]{Clark:2014wua}%
  \BibitemOpen
  \bibfield  {author} {\bibinfo {author} {\bibfnamefont {J.}~\bibnamefont
  {Clark}}, \bibinfo {author} {\bibfnamefont {A.}~\bibnamefont {Bauswein}},
  \bibinfo {author} {\bibfnamefont {L.}~\bibnamefont {Cadonati}}, \bibinfo
  {author} {\bibfnamefont {H.~T.}\ \bibnamefont {Janka}}, \bibinfo {author}
  {\bibfnamefont {C.}~\bibnamefont {Pankow}},\ and\ \bibinfo {author}
  {\bibfnamefont {N.}~\bibnamefont {Stergioulas}},\ }\bibfield  {title}
  {\bibinfo {title} {{Prospects For High Frequency Burst Searches Following
  Binary Neutron Star Coalescence With Advanced Gravitational Wave
  Detectors}},\ }\href {https://doi.org/10.1103/PhysRevD.90.062004} {\bibfield
  {journal} {\bibinfo  {journal} {Phys. Rev. D}\ }\textbf {\bibinfo {volume}
  {90}},\ \bibinfo {pages} {062004} (\bibinfo {year} {2014})},\ \Eprint
  {https://arxiv.org/abs/1406.5444} {arXiv:1406.5444 [astro-ph.HE]}
  \BibitemShut {NoStop}%
\bibitem [{\citenamefont {Chatziioannou}\ \emph {et~al.}(2017)\citenamefont
  {Chatziioannou}, \citenamefont {Clark}, \citenamefont {Bauswein},
  \citenamefont {Millhouse}, \citenamefont {Littenberg},\ and\ \citenamefont
  {Cornish}}]{Chatziioannou:2017ixj}%
  \BibitemOpen
  \bibfield  {author} {\bibinfo {author} {\bibfnamefont {K.}~\bibnamefont
  {Chatziioannou}}, \bibinfo {author} {\bibfnamefont {J.~A.}\ \bibnamefont
  {Clark}}, \bibinfo {author} {\bibfnamefont {A.}~\bibnamefont {Bauswein}},
  \bibinfo {author} {\bibfnamefont {M.}~\bibnamefont {Millhouse}}, \bibinfo
  {author} {\bibfnamefont {T.~B.}\ \bibnamefont {Littenberg}},\ and\ \bibinfo
  {author} {\bibfnamefont {N.}~\bibnamefont {Cornish}},\ }\bibfield  {title}
  {\bibinfo {title} {{Inferring the post-merger gravitational wave emission
  from binary neutron star coalescences}},\ }\href
  {https://doi.org/10.1103/PhysRevD.96.124035} {\bibfield  {journal} {\bibinfo
  {journal} {Phys. Rev. D}\ }\textbf {\bibinfo {volume} {96}},\ \bibinfo
  {pages} {124035} (\bibinfo {year} {2017})},\ \Eprint
  {https://arxiv.org/abs/1711.00040} {arXiv:1711.00040 [gr-qc]} \BibitemShut
  {NoStop}%
\bibitem [{\citenamefont {Pesios}\ \emph {et~al.}(2024)\citenamefont {Pesios},
  \citenamefont {Koutalios}, \citenamefont {Kugiumtzis},\ and\ \citenamefont
  {Stergioulas}}]{Pesios:2024bve}%
  \BibitemOpen
  \bibfield  {author} {\bibinfo {author} {\bibfnamefont {D.}~\bibnamefont
  {Pesios}}, \bibinfo {author} {\bibfnamefont {I.}~\bibnamefont {Koutalios}},
  \bibinfo {author} {\bibfnamefont {D.}~\bibnamefont {Kugiumtzis}},\ and\
  \bibinfo {author} {\bibfnamefont {N.}~\bibnamefont {Stergioulas}},\
  }\bibfield  {title} {\bibinfo {title} {{Predicting Binary Neutron Star
  Postmerger Spectra Using Artificial Neural Networks}},\ }\href@noop {}
  {\bibfield  {journal} {\bibinfo  {journal} {arXiv: 2405.09468}\ } (\bibinfo
  {year} {2024})},\ \Eprint {https://arxiv.org/abs/2405.09468}
  {arXiv:2405.09468 [gr-qc]} \BibitemShut {NoStop}%
\bibitem [{\citenamefont {Shao}(2019)}]{Shao:2019gjj}%
  \BibitemOpen
  \bibfield  {author} {\bibinfo {author} {\bibfnamefont {L.}~\bibnamefont
  {Shao}},\ }\bibfield  {title} {\bibinfo {title} {{Degeneracy in Studying the
  Supranuclear Equation of State and Modified Gravity with Neutron Stars}},\
  }\href {https://doi.org/10.1063/1.5117806} {\bibfield  {journal} {\bibinfo
  {journal} {AIP Conf. Proc.}\ }\textbf {\bibinfo {volume} {2127}},\ \bibinfo
  {pages} {020016} (\bibinfo {year} {2019})},\ \Eprint
  {https://arxiv.org/abs/1901.07546} {arXiv:1901.07546 [gr-qc]} \BibitemShut
  {NoStop}%
\bibitem [{\citenamefont {Yagi}\ and\ \citenamefont
  {Yunes}(2013{\natexlab{a}})}]{Yagi:2013awa}%
  \BibitemOpen
  \bibfield  {author} {\bibinfo {author} {\bibfnamefont {K.}~\bibnamefont
  {Yagi}}\ and\ \bibinfo {author} {\bibfnamefont {N.}~\bibnamefont {Yunes}},\
  }\bibfield  {title} {\bibinfo {title} {{I-Love-Q Relations in Neutron Stars
  and their Applications to Astrophysics, Gravitational Waves and Fundamental
  Physics}},\ }\href {https://doi.org/10.1103/PhysRevD.88.023009} {\bibfield
  {journal} {\bibinfo  {journal} {Phys. Rev. D}\ }\textbf {\bibinfo {volume}
  {88}},\ \bibinfo {pages} {023009} (\bibinfo {year} {2013}{\natexlab{a}})},\
  \Eprint {https://arxiv.org/abs/1303.1528} {arXiv:1303.1528 [gr-qc]}
  \BibitemShut {NoStop}%
\bibitem [{\citenamefont {Yagi}\ and\ \citenamefont
  {Yunes}(2013{\natexlab{b}})}]{Yagi:2013bca}%
  \BibitemOpen
  \bibfield  {author} {\bibinfo {author} {\bibfnamefont {K.}~\bibnamefont
  {Yagi}}\ and\ \bibinfo {author} {\bibfnamefont {N.}~\bibnamefont {Yunes}},\
  }\bibfield  {title} {\bibinfo {title} {{I-Love-Q}},\ }\href
  {https://doi.org/10.1126/science.1236462} {\bibfield  {journal} {\bibinfo
  {journal} {Science}\ }\textbf {\bibinfo {volume} {341}},\ \bibinfo {pages}
  {365} (\bibinfo {year} {2013}{\natexlab{b}})},\ \Eprint
  {https://arxiv.org/abs/1302.4499} {arXiv:1302.4499 [gr-qc]} \BibitemShut
  {NoStop}%
\bibitem [{\citenamefont {Maselli}\ \emph {et~al.}(2013)\citenamefont
  {Maselli}, \citenamefont {Cardoso}, \citenamefont {Ferrari}, \citenamefont
  {Gualtieri},\ and\ \citenamefont {Pani}}]{Maselli:2013mva}%
  \BibitemOpen
  \bibfield  {author} {\bibinfo {author} {\bibfnamefont {A.}~\bibnamefont
  {Maselli}}, \bibinfo {author} {\bibfnamefont {V.}~\bibnamefont {Cardoso}},
  \bibinfo {author} {\bibfnamefont {V.}~\bibnamefont {Ferrari}}, \bibinfo
  {author} {\bibfnamefont {L.}~\bibnamefont {Gualtieri}},\ and\ \bibinfo
  {author} {\bibfnamefont {P.}~\bibnamefont {Pani}},\ }\bibfield  {title}
  {\bibinfo {title} {{Equation-of-state-independent relations in neutron
  stars}},\ }\href {https://doi.org/10.1103/PhysRevD.88.023007} {\bibfield
  {journal} {\bibinfo  {journal} {Phys. Rev. D}\ }\textbf {\bibinfo {volume}
  {88}},\ \bibinfo {pages} {023007} (\bibinfo {year} {2013})},\ \Eprint
  {https://arxiv.org/abs/1304.2052} {arXiv:1304.2052 [gr-qc]} \BibitemShut
  {NoStop}%
\bibitem [{\citenamefont {Yagi}\ and\ \citenamefont
  {Yunes}(2016)}]{Yagi:2015pkc}%
  \BibitemOpen
  \bibfield  {author} {\bibinfo {author} {\bibfnamefont {K.}~\bibnamefont
  {Yagi}}\ and\ \bibinfo {author} {\bibfnamefont {N.}~\bibnamefont {Yunes}},\
  }\bibfield  {title} {\bibinfo {title} {{Binary Love Relations}},\ }\href
  {https://doi.org/10.1088/0264-9381/33/13/13LT01} {\bibfield  {journal}
  {\bibinfo  {journal} {Class. Quant. Grav.}\ }\textbf {\bibinfo {volume}
  {33}},\ \bibinfo {pages} {13LT01} (\bibinfo {year} {2016})},\ \Eprint
  {https://arxiv.org/abs/1512.02639} {arXiv:1512.02639 [gr-qc]} \BibitemShut
  {NoStop}%
\bibitem [{\citenamefont {Chatziioannou}\ \emph {et~al.}(2018)\citenamefont
  {Chatziioannou}, \citenamefont {Haster},\ and\ \citenamefont
  {Zimmerman}}]{Chatziioannou:2018vzf}%
  \BibitemOpen
  \bibfield  {author} {\bibinfo {author} {\bibfnamefont {K.}~\bibnamefont
  {Chatziioannou}}, \bibinfo {author} {\bibfnamefont {C.-J.}\ \bibnamefont
  {Haster}},\ and\ \bibinfo {author} {\bibfnamefont {A.}~\bibnamefont
  {Zimmerman}},\ }\bibfield  {title} {\bibinfo {title} {{Measuring the neutron
  star tidal deformability with equation-of-state-independent relations and
  gravitational waves}},\ }\href {https://doi.org/10.1103/PhysRevD.97.104036}
  {\bibfield  {journal} {\bibinfo  {journal} {Phys. Rev. D}\ }\textbf {\bibinfo
  {volume} {97}},\ \bibinfo {pages} {104036} (\bibinfo {year} {2018})},\
  \Eprint {https://arxiv.org/abs/1804.03221} {arXiv:1804.03221 [gr-qc]}
  \BibitemShut {NoStop}%
\bibitem [{\citenamefont {Torres-Rivas}\ \emph {et~al.}(2019)\citenamefont
  {Torres-Rivas}, \citenamefont {Chatziioannou}, \citenamefont {Bauswein},\
  and\ \citenamefont {Clark}}]{Torres-Rivas:2018svp}%
  \BibitemOpen
  \bibfield  {author} {\bibinfo {author} {\bibfnamefont {A.}~\bibnamefont
  {Torres-Rivas}}, \bibinfo {author} {\bibfnamefont {K.}~\bibnamefont
  {Chatziioannou}}, \bibinfo {author} {\bibfnamefont {A.}~\bibnamefont
  {Bauswein}},\ and\ \bibinfo {author} {\bibfnamefont {J.~A.}\ \bibnamefont
  {Clark}},\ }\bibfield  {title} {\bibinfo {title} {{Observing the post-merger
  signal of GW170817-like events with improved gravitational-wave detectors}},\
  }\href {https://doi.org/10.1103/PhysRevD.99.044014} {\bibfield  {journal}
  {\bibinfo  {journal} {Phys. Rev. D}\ }\textbf {\bibinfo {volume} {99}},\
  \bibinfo {pages} {044014} (\bibinfo {year} {2019})},\ \Eprint
  {https://arxiv.org/abs/1811.08931} {arXiv:1811.08931 [gr-qc]} \BibitemShut
  {NoStop}%
\bibitem [{\citenamefont {Breschi}\ \emph {et~al.}(2019)\citenamefont
  {Breschi}, \citenamefont {Bernuzzi}, \citenamefont {Zappa}, \citenamefont
  {Agathos}, \citenamefont {Perego}, \citenamefont {Radice},\ and\
  \citenamefont {Nagar}}]{Breschi:2019srl}%
  \BibitemOpen
  \bibfield  {author} {\bibinfo {author} {\bibfnamefont {M.}~\bibnamefont
  {Breschi}}, \bibinfo {author} {\bibfnamefont {S.}~\bibnamefont {Bernuzzi}},
  \bibinfo {author} {\bibfnamefont {F.}~\bibnamefont {Zappa}}, \bibinfo
  {author} {\bibfnamefont {M.}~\bibnamefont {Agathos}}, \bibinfo {author}
  {\bibfnamefont {A.}~\bibnamefont {Perego}}, \bibinfo {author} {\bibfnamefont
  {D.}~\bibnamefont {Radice}},\ and\ \bibinfo {author} {\bibfnamefont
  {A.}~\bibnamefont {Nagar}},\ }\bibfield  {title} {\bibinfo {title}
  {{kiloHertz gravitational waves from binary neutron star remnants:
  time-domain model and constraints on extreme matter}},\ }\href
  {https://doi.org/10.1103/PhysRevD.100.104029} {\bibfield  {journal} {\bibinfo
   {journal} {Phys. Rev. D}\ }\textbf {\bibinfo {volume} {100}},\ \bibinfo
  {pages} {104029} (\bibinfo {year} {2019})},\ \Eprint
  {https://arxiv.org/abs/1908.11418} {arXiv:1908.11418 [gr-qc]} \BibitemShut
  {NoStop}%
\bibitem [{\citenamefont {Breschi}\ \emph {et~al.}(2022)\citenamefont
  {Breschi}, \citenamefont {Bernuzzi}, \citenamefont {Godzieba}, \citenamefont
  {Perego},\ and\ \citenamefont {Radice}}]{Breschi:2021xrx}%
  \BibitemOpen
  \bibfield  {author} {\bibinfo {author} {\bibfnamefont {M.}~\bibnamefont
  {Breschi}}, \bibinfo {author} {\bibfnamefont {S.}~\bibnamefont {Bernuzzi}},
  \bibinfo {author} {\bibfnamefont {D.}~\bibnamefont {Godzieba}}, \bibinfo
  {author} {\bibfnamefont {A.}~\bibnamefont {Perego}},\ and\ \bibinfo {author}
  {\bibfnamefont {D.}~\bibnamefont {Radice}},\ }\bibfield  {title} {\bibinfo
  {title} {{Constraints on the Maximum Densities of Neutron Stars from
  Postmerger Gravitational Waves with Third-Generation Observations}},\ }\href
  {https://doi.org/10.1103/PhysRevLett.128.161102} {\bibfield  {journal}
  {\bibinfo  {journal} {Phys. Rev. Lett.}\ }\textbf {\bibinfo {volume} {128}},\
  \bibinfo {pages} {161102} (\bibinfo {year} {2022})},\ \Eprint
  {https://arxiv.org/abs/2110.06957} {arXiv:2110.06957 [gr-qc]} \BibitemShut
  {NoStop}%
\bibitem [{\citenamefont {Wijngaarden}\ \emph {et~al.}(2022)\citenamefont
  {Wijngaarden}, \citenamefont {Chatziioannou}, \citenamefont {Bauswein},
  \citenamefont {Clark},\ and\ \citenamefont {Cornish}}]{Wijngaarden:2022sah}%
  \BibitemOpen
  \bibfield  {author} {\bibinfo {author} {\bibfnamefont {M.}~\bibnamefont
  {Wijngaarden}}, \bibinfo {author} {\bibfnamefont {K.}~\bibnamefont
  {Chatziioannou}}, \bibinfo {author} {\bibfnamefont {A.}~\bibnamefont
  {Bauswein}}, \bibinfo {author} {\bibfnamefont {J.~A.}\ \bibnamefont
  {Clark}},\ and\ \bibinfo {author} {\bibfnamefont {N.~J.}\ \bibnamefont
  {Cornish}},\ }\bibfield  {title} {\bibinfo {title} {{Probing neutron stars
  with the full premerger and postmerger gravitational wave signal from binary
  coalescences}},\ }\href {https://doi.org/10.1103/PhysRevD.105.104019}
  {\bibfield  {journal} {\bibinfo  {journal} {Phys. Rev. D}\ }\textbf {\bibinfo
  {volume} {105}},\ \bibinfo {pages} {104019} (\bibinfo {year} {2022})},\
  \Eprint {https://arxiv.org/abs/2202.09382} {arXiv:2202.09382 [gr-qc]}
  \BibitemShut {NoStop}%
\bibitem [{\citenamefont {Puecher}\ \emph {et~al.}(2023)\citenamefont
  {Puecher}, \citenamefont {Dietrich}, \citenamefont {Tsang}, \citenamefont
  {Kalaghatgi}, \citenamefont {Roy}, \citenamefont {Setyawati},\ and\
  \citenamefont {Van Den~Broeck}}]{Puecher:2022oiz}%
  \BibitemOpen
  \bibfield  {author} {\bibinfo {author} {\bibfnamefont {A.}~\bibnamefont
  {Puecher}}, \bibinfo {author} {\bibfnamefont {T.}~\bibnamefont {Dietrich}},
  \bibinfo {author} {\bibfnamefont {K.~W.}\ \bibnamefont {Tsang}}, \bibinfo
  {author} {\bibfnamefont {C.}~\bibnamefont {Kalaghatgi}}, \bibinfo {author}
  {\bibfnamefont {S.}~\bibnamefont {Roy}}, \bibinfo {author} {\bibfnamefont
  {Y.}~\bibnamefont {Setyawati}},\ and\ \bibinfo {author} {\bibfnamefont
  {C.}~\bibnamefont {Van Den~Broeck}},\ }\bibfield  {title} {\bibinfo {title}
  {{Unraveling information about supranuclear-dense matter from the complete
  binary neutron star coalescence process using future gravitational-wave
  detector networks}},\ }\href {https://doi.org/10.1103/PhysRevD.107.124009}
  {\bibfield  {journal} {\bibinfo  {journal} {Phys. Rev. D}\ }\textbf {\bibinfo
  {volume} {107}},\ \bibinfo {pages} {124009} (\bibinfo {year} {2023})},\
  \Eprint {https://arxiv.org/abs/2210.09259} {arXiv:2210.09259 [gr-qc]}
  \BibitemShut {NoStop}%
\bibitem [{\citenamefont {Suleiman}\ and\ \citenamefont
  {Read}(2024)}]{Suleiman:2024ztn}%
  \BibitemOpen
  \bibfield  {author} {\bibinfo {author} {\bibfnamefont {L.}~\bibnamefont
  {Suleiman}}\ and\ \bibinfo {author} {\bibfnamefont {J.}~\bibnamefont
  {Read}},\ }\bibfield  {title} {\bibinfo {title} {{Quasiuniversal relations in
  the context of future neutron star detections}},\ }\href
  {https://doi.org/10.1103/PhysRevD.109.103029} {\bibfield  {journal} {\bibinfo
   {journal} {Phys. Rev. D}\ }\textbf {\bibinfo {volume} {109}},\ \bibinfo
  {pages} {103029} (\bibinfo {year} {2024})},\ \Eprint
  {https://arxiv.org/abs/2402.01948} {arXiv:2402.01948 [astro-ph.HE]}
  \BibitemShut {NoStop}%
\bibitem [{\citenamefont {Bernuzzi}\ \emph {et~al.}(2014)\citenamefont
  {Bernuzzi}, \citenamefont {Nagar}, \citenamefont {Balmelli}, \citenamefont
  {Dietrich},\ and\ \citenamefont {Ujevic}}]{Bernuzzi:2014kca}%
  \BibitemOpen
  \bibfield  {author} {\bibinfo {author} {\bibfnamefont {S.}~\bibnamefont
  {Bernuzzi}}, \bibinfo {author} {\bibfnamefont {A.}~\bibnamefont {Nagar}},
  \bibinfo {author} {\bibfnamefont {S.}~\bibnamefont {Balmelli}}, \bibinfo
  {author} {\bibfnamefont {T.}~\bibnamefont {Dietrich}},\ and\ \bibinfo
  {author} {\bibfnamefont {M.}~\bibnamefont {Ujevic}},\ }\bibfield  {title}
  {\bibinfo {title} {{Quasiuniversal properties of neutron star mergers}},\
  }\href {https://doi.org/10.1103/PhysRevLett.112.201101} {\bibfield  {journal}
  {\bibinfo  {journal} {Phys. Rev. Lett.}\ }\textbf {\bibinfo {volume} {112}},\
  \bibinfo {pages} {201101} (\bibinfo {year} {2014})},\ \Eprint
  {https://arxiv.org/abs/1402.6244} {arXiv:1402.6244 [gr-qc]} \BibitemShut
  {NoStop}%
\bibitem [{\citenamefont {Yagi}\ and\ \citenamefont
  {Yunes}(2017)}]{Yagi:2016bkt}%
  \BibitemOpen
  \bibfield  {author} {\bibinfo {author} {\bibfnamefont {K.}~\bibnamefont
  {Yagi}}\ and\ \bibinfo {author} {\bibfnamefont {N.}~\bibnamefont {Yunes}},\
  }\bibfield  {title} {\bibinfo {title} {{Approximate Universal Relations for
  Neutron Stars and Quark Stars}},\ }\href
  {https://doi.org/10.1016/j.physrep.2017.03.002} {\bibfield  {journal}
  {\bibinfo  {journal} {Phys. Rept.}\ }\textbf {\bibinfo {volume} {681}},\
  \bibinfo {pages} {1} (\bibinfo {year} {2017})},\ \Eprint
  {https://arxiv.org/abs/1608.02582} {arXiv:1608.02582 [gr-qc]} \BibitemShut
  {NoStop}%
\bibitem [{\citenamefont {Bauswein}\ and\ \citenamefont
  {Stergioulas}(2017)}]{Bauswein:2017aur}%
  \BibitemOpen
  \bibfield  {author} {\bibinfo {author} {\bibfnamefont {A.}~\bibnamefont
  {Bauswein}}\ and\ \bibinfo {author} {\bibfnamefont {N.}~\bibnamefont
  {Stergioulas}},\ }\bibfield  {title} {\bibinfo {title} {{Semi-analytic
  derivation of the threshold mass for prompt collapse in binary neutron star
  mergers}},\ }\href {https://doi.org/10.1093/mnras/stx1983} {\bibfield
  {journal} {\bibinfo  {journal} {Mon. Not. Roy. Astron. Soc.}\ }\textbf
  {\bibinfo {volume} {471}},\ \bibinfo {pages} {4956} (\bibinfo {year}
  {2017})},\ \Eprint {https://arxiv.org/abs/1702.02567} {arXiv:1702.02567
  [astro-ph.HE]} \BibitemShut {NoStop}%
\bibitem [{\citenamefont {Damour}\ and\ \citenamefont
  {Esposito-Farese}(1992)}]{Damour:1992we}%
  \BibitemOpen
  \bibfield  {author} {\bibinfo {author} {\bibfnamefont {T.}~\bibnamefont
  {Damour}}\ and\ \bibinfo {author} {\bibfnamefont {G.}~\bibnamefont
  {Esposito-Farese}},\ }\bibfield  {title} {\bibinfo {title} {{Tensor
  multiscalar theories of gravitation}},\ }\href
  {https://doi.org/10.1088/0264-9381/9/9/015} {\bibfield  {journal} {\bibinfo
  {journal} {Class. Quant. Grav.}\ }\textbf {\bibinfo {volume} {9}},\ \bibinfo
  {pages} {2093} (\bibinfo {year} {1992})}\BibitemShut {NoStop}%
\bibitem [{\citenamefont {Damour}\ and\ \citenamefont
  {Esposito-Farese}(1993)}]{Damour:1993hw}%
  \BibitemOpen
  \bibfield  {author} {\bibinfo {author} {\bibfnamefont {T.}~\bibnamefont
  {Damour}}\ and\ \bibinfo {author} {\bibfnamefont {G.}~\bibnamefont
  {Esposito-Farese}},\ }\bibfield  {title} {\bibinfo {title} {{Nonperturbative
  strong field effects in tensor - scalar theories of gravitation}},\ }\href
  {https://doi.org/10.1103/PhysRevLett.70.2220} {\bibfield  {journal} {\bibinfo
   {journal} {Phys. Rev. Lett.}\ }\textbf {\bibinfo {volume} {70}},\ \bibinfo
  {pages} {2220} (\bibinfo {year} {1993})}\BibitemShut {NoStop}%
\bibitem [{\citenamefont {Damour}\ and\ \citenamefont
  {Esposito-Farese}(1996)}]{Damour:1995kt}%
  \BibitemOpen
  \bibfield  {author} {\bibinfo {author} {\bibfnamefont {T.}~\bibnamefont
  {Damour}}\ and\ \bibinfo {author} {\bibfnamefont {G.}~\bibnamefont
  {Esposito-Farese}},\ }\bibfield  {title} {\bibinfo {title} {{Testing gravity
  to second postNewtonian order: A Field theory approach}},\ }\href
  {https://doi.org/10.1103/PhysRevD.53.5541} {\bibfield  {journal} {\bibinfo
  {journal} {Phys. Rev. D}\ }\textbf {\bibinfo {volume} {53}},\ \bibinfo
  {pages} {5541} (\bibinfo {year} {1996})},\ \Eprint
  {https://arxiv.org/abs/gr-qc/9506063} {arXiv:gr-qc/9506063} \BibitemShut
  {NoStop}%
\bibitem [{\citenamefont {Damour}\ and\ \citenamefont
  {Esposito-Farese}(1998)}]{Damour:1998jk}%
  \BibitemOpen
  \bibfield  {author} {\bibinfo {author} {\bibfnamefont {T.}~\bibnamefont
  {Damour}}\ and\ \bibinfo {author} {\bibfnamefont {G.}~\bibnamefont
  {Esposito-Farese}},\ }\bibfield  {title} {\bibinfo {title} {{Gravitational
  wave versus binary - pulsar tests of strong field gravity}},\ }\href
  {https://doi.org/10.1103/PhysRevD.58.042001} {\bibfield  {journal} {\bibinfo
  {journal} {Phys. Rev. D}\ }\textbf {\bibinfo {volume} {58}},\ \bibinfo
  {pages} {042001} (\bibinfo {year} {1998})},\ \Eprint
  {https://arxiv.org/abs/gr-qc/9803031} {arXiv:gr-qc/9803031} \BibitemShut
  {NoStop}%
\bibitem [{\citenamefont {Kuroda}\ and\ \citenamefont
  {Shibata}(2023)}]{Kuroda:2023zbz}%
  \BibitemOpen
  \bibfield  {author} {\bibinfo {author} {\bibfnamefont {T.}~\bibnamefont
  {Kuroda}}\ and\ \bibinfo {author} {\bibfnamefont {M.}~\bibnamefont
  {Shibata}},\ }\bibfield  {title} {\bibinfo {title} {{Spontaneous
  scalarization as a new core-collapse supernova mechanism and its
  multimessenger signals}},\ }\href
  {https://doi.org/10.1103/PhysRevD.107.103025} {\bibfield  {journal} {\bibinfo
   {journal} {Phys. Rev. D}\ }\textbf {\bibinfo {volume} {107}},\ \bibinfo
  {pages} {103025} (\bibinfo {year} {2023})},\ \Eprint
  {https://arxiv.org/abs/2302.09853} {arXiv:2302.09853 [astro-ph.HE]}
  \BibitemShut {NoStop}%
\bibitem [{\citenamefont {Kuan}\ \emph {et~al.}(2023)\citenamefont {Kuan},
  \citenamefont {Van~Aelst}, \citenamefont {Lam},\ and\ \citenamefont
  {Shibata}}]{Kuan:2023hrh}%
  \BibitemOpen
  \bibfield  {author} {\bibinfo {author} {\bibfnamefont {H.-J.}\ \bibnamefont
  {Kuan}}, \bibinfo {author} {\bibfnamefont {K.}~\bibnamefont {Van~Aelst}},
  \bibinfo {author} {\bibfnamefont {A.~T.-L.}\ \bibnamefont {Lam}},\ and\
  \bibinfo {author} {\bibfnamefont {M.}~\bibnamefont {Shibata}},\ }\bibfield
  {title} {\bibinfo {title} {{Binary neutron star mergers in massive
  scalar-tensor theory: Quasiequilibrium states and dynamical enhancement of
  the scalarization}},\ }\href {https://doi.org/10.1103/PhysRevD.108.064057}
  {\bibfield  {journal} {\bibinfo  {journal} {Phys. Rev. D}\ }\textbf {\bibinfo
  {volume} {108}},\ \bibinfo {pages} {064057} (\bibinfo {year} {2023})},\
  \Eprint {https://arxiv.org/abs/2309.01709} {arXiv:2309.01709 [gr-qc]}
  \BibitemShut {NoStop}%
\bibitem [{\citenamefont {Lam}\ \emph {et~al.}(2024)\citenamefont {Lam},
  \citenamefont {Kuan}, \citenamefont {Shibata}, \citenamefont {Van~Aelst},\
  and\ \citenamefont {Kiuchi}}]{Lam:2024wpq}%
  \BibitemOpen
  \bibfield  {author} {\bibinfo {author} {\bibfnamefont {A.~T.-L.}\
  \bibnamefont {Lam}}, \bibinfo {author} {\bibfnamefont {H.-J.}\ \bibnamefont
  {Kuan}}, \bibinfo {author} {\bibfnamefont {M.}~\bibnamefont {Shibata}},
  \bibinfo {author} {\bibfnamefont {K.}~\bibnamefont {Van~Aelst}},\ and\
  \bibinfo {author} {\bibfnamefont {K.}~\bibnamefont {Kiuchi}},\ }\bibfield
  {title} {\bibinfo {title} {{Binary neutron star mergers in massive
  scalar-tensor theory: Properties of post-merger remnants}},\ }\href@noop {}
  {\bibfield  {journal} {\bibinfo  {journal} {arXiv:2406.05211}\ } (\bibinfo
  {year} {2024})},\ \Eprint {https://arxiv.org/abs/2406.05211}
  {arXiv:2406.05211 [gr-qc]} \BibitemShut {NoStop}%
\bibitem [{\citenamefont {Shao}\ \emph {et~al.}(2017)\citenamefont {Shao},
  \citenamefont {Sennett}, \citenamefont {Buonanno}, \citenamefont {Kramer},\
  and\ \citenamefont {Wex}}]{Shao:2017gwu}%
  \BibitemOpen
  \bibfield  {author} {\bibinfo {author} {\bibfnamefont {L.}~\bibnamefont
  {Shao}}, \bibinfo {author} {\bibfnamefont {N.}~\bibnamefont {Sennett}},
  \bibinfo {author} {\bibfnamefont {A.}~\bibnamefont {Buonanno}}, \bibinfo
  {author} {\bibfnamefont {M.}~\bibnamefont {Kramer}},\ and\ \bibinfo {author}
  {\bibfnamefont {N.}~\bibnamefont {Wex}},\ }\bibfield  {title} {\bibinfo
  {title} {{Constraining nonperturbative strong-field effects in scalar-tensor
  gravity by combining pulsar timing and laser-interferometer
  gravitational-wave detectors}},\ }\href
  {https://doi.org/10.1103/PhysRevX.7.041025} {\bibfield  {journal} {\bibinfo
  {journal} {Phys. Rev. X}\ }\textbf {\bibinfo {volume} {7}},\ \bibinfo {pages}
  {041025} (\bibinfo {year} {2017})},\ \Eprint
  {https://arxiv.org/abs/1704.07561} {arXiv:1704.07561 [gr-qc]} \BibitemShut
  {NoStop}%
\bibitem [{\citenamefont {Anderson}\ \emph {et~al.}(2019)\citenamefont
  {Anderson}, \citenamefont {Freire},\ and\ \citenamefont
  {Yunes}}]{Anderson:2019eay}%
  \BibitemOpen
  \bibfield  {author} {\bibinfo {author} {\bibfnamefont {D.}~\bibnamefont
  {Anderson}}, \bibinfo {author} {\bibfnamefont {P.}~\bibnamefont {Freire}},\
  and\ \bibinfo {author} {\bibfnamefont {N.}~\bibnamefont {Yunes}},\ }\bibfield
   {title} {\bibinfo {title} {{Binary pulsar constraints on massless
  scalar\textendash{}tensor theories using Bayesian statistics}},\ }\href
  {https://doi.org/10.1088/1361-6382/ab3a1c} {\bibfield  {journal} {\bibinfo
  {journal} {Class. Quant. Grav.}\ }\textbf {\bibinfo {volume} {36}},\ \bibinfo
  {pages} {225009} (\bibinfo {year} {2019})},\ \Eprint
  {https://arxiv.org/abs/1901.00938} {arXiv:1901.00938 [gr-qc]} \BibitemShut
  {NoStop}%
\bibitem [{\citenamefont {Zhao}\ \emph {et~al.}(2022)\citenamefont {Zhao},
  \citenamefont {Freire}, \citenamefont {Kramer}, \citenamefont {Shao},\ and\
  \citenamefont {Wex}}]{Zhao:2022vig}%
  \BibitemOpen
  \bibfield  {author} {\bibinfo {author} {\bibfnamefont {J.}~\bibnamefont
  {Zhao}}, \bibinfo {author} {\bibfnamefont {P.~C.~C.}\ \bibnamefont {Freire}},
  \bibinfo {author} {\bibfnamefont {M.}~\bibnamefont {Kramer}}, \bibinfo
  {author} {\bibfnamefont {L.}~\bibnamefont {Shao}},\ and\ \bibinfo {author}
  {\bibfnamefont {N.}~\bibnamefont {Wex}},\ }\bibfield  {title} {\bibinfo
  {title} {{Closing a spontaneous-scalarization window with binary pulsars}},\
  }\href {https://doi.org/10.1088/1361-6382/ac69a3} {\bibfield  {journal}
  {\bibinfo  {journal} {Class. Quant. Grav.}\ }\textbf {\bibinfo {volume}
  {39}},\ \bibinfo {pages} {11LT01} (\bibinfo {year} {2022})},\ \Eprint
  {https://arxiv.org/abs/2201.03771} {arXiv:2201.03771 [astro-ph.HE]}
  \BibitemShut {NoStop}%
\bibitem [{\citenamefont {Ramazano\u{g}lu}\ and\ \citenamefont
  {Pretorius}(2016)}]{Ramazanoglu:2016kul}%
  \BibitemOpen
  \bibfield  {author} {\bibinfo {author} {\bibfnamefont {F.~M.}\ \bibnamefont
  {Ramazano\u{g}lu}}\ and\ \bibinfo {author} {\bibfnamefont {F.}~\bibnamefont
  {Pretorius}},\ }\bibfield  {title} {\bibinfo {title} {{Spontaneous
  Scalarization with Massive Fields}},\ }\href
  {https://doi.org/10.1103/PhysRevD.93.064005} {\bibfield  {journal} {\bibinfo
  {journal} {Phys. Rev. D}\ }\textbf {\bibinfo {volume} {93}},\ \bibinfo
  {pages} {064005} (\bibinfo {year} {2016})},\ \Eprint
  {https://arxiv.org/abs/1601.07475} {arXiv:1601.07475 [gr-qc]} \BibitemShut
  {NoStop}%
\bibitem [{\citenamefont {Yazadjiev}\ \emph {et~al.}(2016)\citenamefont
  {Yazadjiev}, \citenamefont {Doneva},\ and\ \citenamefont
  {Popchev}}]{Yazadjiev:2016pcb}%
  \BibitemOpen
  \bibfield  {author} {\bibinfo {author} {\bibfnamefont {S.~S.}\ \bibnamefont
  {Yazadjiev}}, \bibinfo {author} {\bibfnamefont {D.~D.}\ \bibnamefont
  {Doneva}},\ and\ \bibinfo {author} {\bibfnamefont {D.}~\bibnamefont
  {Popchev}},\ }\bibfield  {title} {\bibinfo {title} {{Slowly rotating neutron
  stars in scalar-tensor theories with a massive scalar field}},\ }\href
  {https://doi.org/10.1103/PhysRevD.93.084038} {\bibfield  {journal} {\bibinfo
  {journal} {Phys. Rev. D}\ }\textbf {\bibinfo {volume} {93}},\ \bibinfo
  {pages} {084038} (\bibinfo {year} {2016})},\ \Eprint
  {https://arxiv.org/abs/1602.04766} {arXiv:1602.04766 [gr-qc]} \BibitemShut
  {NoStop}%
\bibitem [{\citenamefont {Zhang}\ \emph {et~al.}(2021)\citenamefont {Zhang},
  \citenamefont {Lyu}, \citenamefont {Huang}, \citenamefont {Johnson},
  \citenamefont {Sagunski}, \citenamefont {Sakellariadou},\ and\ \citenamefont
  {Yang}}]{Zhang:2021mks}%
  \BibitemOpen
  \bibfield  {author} {\bibinfo {author} {\bibfnamefont {J.}~\bibnamefont
  {Zhang}}, \bibinfo {author} {\bibfnamefont {Z.}~\bibnamefont {Lyu}}, \bibinfo
  {author} {\bibfnamefont {J.}~\bibnamefont {Huang}}, \bibinfo {author}
  {\bibfnamefont {M.~C.}\ \bibnamefont {Johnson}}, \bibinfo {author}
  {\bibfnamefont {L.}~\bibnamefont {Sagunski}}, \bibinfo {author}
  {\bibfnamefont {M.}~\bibnamefont {Sakellariadou}},\ and\ \bibinfo {author}
  {\bibfnamefont {H.}~\bibnamefont {Yang}},\ }\bibfield  {title} {\bibinfo
  {title} {{First Constraints on Nuclear Coupling of Axionlike Particles from
  the Binary Neutron Star Gravitational Wave Event GW170817}},\ }\href
  {https://doi.org/10.1103/PhysRevLett.127.161101} {\bibfield  {journal}
  {\bibinfo  {journal} {Phys. Rev. Lett.}\ }\textbf {\bibinfo {volume} {127}},\
  \bibinfo {pages} {161101} (\bibinfo {year} {2021})},\ \Eprint
  {https://arxiv.org/abs/2105.13963} {arXiv:2105.13963 [hep-ph]} \BibitemShut
  {NoStop}%
\bibitem [{\citenamefont {Shibata}\ \emph {et~al.}(2014)\citenamefont
  {Shibata}, \citenamefont {Taniguchi}, \citenamefont {Okawa},\ and\
  \citenamefont {Buonanno}}]{Shibata:2013pra}%
  \BibitemOpen
  \bibfield  {author} {\bibinfo {author} {\bibfnamefont {M.}~\bibnamefont
  {Shibata}}, \bibinfo {author} {\bibfnamefont {K.}~\bibnamefont {Taniguchi}},
  \bibinfo {author} {\bibfnamefont {H.}~\bibnamefont {Okawa}},\ and\ \bibinfo
  {author} {\bibfnamefont {A.}~\bibnamefont {Buonanno}},\ }\bibfield  {title}
  {\bibinfo {title} {{Coalescence of binary neutron stars in a scalar-tensor
  theory of gravity}},\ }\href {https://doi.org/10.1103/PhysRevD.89.084005}
  {\bibfield  {journal} {\bibinfo  {journal} {Phys. Rev. D}\ }\textbf {\bibinfo
  {volume} {89}},\ \bibinfo {pages} {084005} (\bibinfo {year} {2014})},\
  \Eprint {https://arxiv.org/abs/1310.0627} {arXiv:1310.0627 [gr-qc]}
  \BibitemShut {NoStop}%
\bibitem [{\citenamefont {Taniguchi}\ \emph {et~al.}(2015)\citenamefont
  {Taniguchi}, \citenamefont {Shibata},\ and\ \citenamefont
  {Buonanno}}]{Taniguchi:2014fqa}%
  \BibitemOpen
  \bibfield  {author} {\bibinfo {author} {\bibfnamefont {K.}~\bibnamefont
  {Taniguchi}}, \bibinfo {author} {\bibfnamefont {M.}~\bibnamefont {Shibata}},\
  and\ \bibinfo {author} {\bibfnamefont {A.}~\bibnamefont {Buonanno}},\
  }\bibfield  {title} {\bibinfo {title} {{Quasiequilibrium sequences of binary
  neutron stars undergoing dynamical scalarization}},\ }\href
  {https://doi.org/10.1103/PhysRevD.91.024033} {\bibfield  {journal} {\bibinfo
  {journal} {Phys. Rev. D}\ }\textbf {\bibinfo {volume} {91}},\ \bibinfo
  {pages} {024033} (\bibinfo {year} {2015})},\ \Eprint
  {https://arxiv.org/abs/1410.0738} {arXiv:1410.0738 [gr-qc]} \BibitemShut
  {NoStop}%
\bibitem [{\citenamefont {Read}\ \emph {et~al.}(2009)\citenamefont {Read},
  \citenamefont {Lackey}, \citenamefont {Owen},\ and\ \citenamefont
  {Friedman}}]{Read:2008iy}%
  \BibitemOpen
  \bibfield  {author} {\bibinfo {author} {\bibfnamefont {J.~S.}\ \bibnamefont
  {Read}}, \bibinfo {author} {\bibfnamefont {B.~D.}\ \bibnamefont {Lackey}},
  \bibinfo {author} {\bibfnamefont {B.~J.}\ \bibnamefont {Owen}},\ and\
  \bibinfo {author} {\bibfnamefont {J.~L.}\ \bibnamefont {Friedman}},\
  }\bibfield  {title} {\bibinfo {title} {{Constraints on a phenomenologically
  parameterized neutron-star equation of state}},\ }\href
  {https://doi.org/10.1103/PhysRevD.79.124032} {\bibfield  {journal} {\bibinfo
  {journal} {Phys. Rev. D}\ }\textbf {\bibinfo {volume} {79}},\ \bibinfo
  {pages} {124032} (\bibinfo {year} {2009})},\ \Eprint
  {https://arxiv.org/abs/0812.2163} {arXiv:0812.2163 [astro-ph]} \BibitemShut
  {NoStop}%
\bibitem [{\citenamefont {Kiuchi}\ \emph {et~al.}(2020)\citenamefont {Kiuchi},
  \citenamefont {Kawaguchi}, \citenamefont {Kyutoku}, \citenamefont
  {Sekiguchi},\ and\ \citenamefont {Shibata}}]{Kiuchi:2019kzt}%
  \BibitemOpen
  \bibfield  {author} {\bibinfo {author} {\bibfnamefont {K.}~\bibnamefont
  {Kiuchi}}, \bibinfo {author} {\bibfnamefont {K.}~\bibnamefont {Kawaguchi}},
  \bibinfo {author} {\bibfnamefont {K.}~\bibnamefont {Kyutoku}}, \bibinfo
  {author} {\bibfnamefont {Y.}~\bibnamefont {Sekiguchi}},\ and\ \bibinfo
  {author} {\bibfnamefont {M.}~\bibnamefont {Shibata}},\ }\bibfield  {title}
  {\bibinfo {title} {{Sub-radian-accuracy gravitational waves from coalescing
  binary neutron stars in numerical relativity. II. Systematic study on the
  equation of state, binary mass, and mass ratio}},\ }\href
  {https://doi.org/10.1103/PhysRevD.101.084006} {\bibfield  {journal} {\bibinfo
   {journal} {Phys. Rev. D}\ }\textbf {\bibinfo {volume} {101}},\ \bibinfo
  {pages} {084006} (\bibinfo {year} {2020})},\ \Eprint
  {https://arxiv.org/abs/1907.03790} {arXiv:1907.03790 [astro-ph.HE]}
  \BibitemShut {NoStop}%
\bibitem [{\citenamefont {Flanagan}\ and\ \citenamefont
  {Hinderer}(2008)}]{Flanagan:2007ix}%
  \BibitemOpen
  \bibfield  {author} {\bibinfo {author} {\bibfnamefont {E.~E.}\ \bibnamefont
  {Flanagan}}\ and\ \bibinfo {author} {\bibfnamefont {T.}~\bibnamefont
  {Hinderer}},\ }\bibfield  {title} {\bibinfo {title} {{Constraining neutron
  star tidal Love numbers with gravitational wave detectors}},\ }\href
  {https://doi.org/10.1103/PhysRevD.77.021502} {\bibfield  {journal} {\bibinfo
  {journal} {Phys. Rev. D}\ }\textbf {\bibinfo {volume} {77}},\ \bibinfo
  {pages} {021502} (\bibinfo {year} {2008})},\ \Eprint
  {https://arxiv.org/abs/0709.1915} {arXiv:0709.1915 [astro-ph]} \BibitemShut
  {NoStop}%
\bibitem [{\citenamefont {Hinderer}(2008)}]{Hinderer:2007mb}%
  \BibitemOpen
  \bibfield  {author} {\bibinfo {author} {\bibfnamefont {T.}~\bibnamefont
  {Hinderer}},\ }\bibfield  {title} {\bibinfo {title} {{Tidal Love numbers of
  neutron stars}},\ }\href {https://doi.org/10.1086/533487} {\bibfield
  {journal} {\bibinfo  {journal} {Astrophys. J.}\ }\textbf {\bibinfo {volume}
  {677}},\ \bibinfo {pages} {1216} (\bibinfo {year} {2008})},\ \bibinfo {note}
  {[Erratum: Astrophys.J. 697, 964 (2009)]},\ \Eprint
  {https://arxiv.org/abs/0711.2420} {arXiv:0711.2420 [astro-ph]} \BibitemShut
  {NoStop}%
\bibitem [{\citenamefont {Favata}(2014)}]{Favata:2013rwa}%
  \BibitemOpen
  \bibfield  {author} {\bibinfo {author} {\bibfnamefont {M.}~\bibnamefont
  {Favata}},\ }\bibfield  {title} {\bibinfo {title} {{Systematic parameter
  errors in inspiraling neutron star binaries}},\ }\href
  {https://doi.org/10.1103/PhysRevLett.112.101101} {\bibfield  {journal}
  {\bibinfo  {journal} {Phys. Rev. Lett.}\ }\textbf {\bibinfo {volume} {112}},\
  \bibinfo {pages} {101101} (\bibinfo {year} {2014})},\ \Eprint
  {https://arxiv.org/abs/1310.8288} {arXiv:1310.8288 [gr-qc]} \BibitemShut
  {NoStop}%
\bibitem [{\citenamefont {Wade}\ \emph {et~al.}(2014)\citenamefont {Wade},
  \citenamefont {Creighton}, \citenamefont {Ochsner}, \citenamefont {Lackey},
  \citenamefont {Farr}, \citenamefont {Littenberg},\ and\ \citenamefont
  {Raymond}}]{Wade:2014vqa}%
  \BibitemOpen
  \bibfield  {author} {\bibinfo {author} {\bibfnamefont {L.}~\bibnamefont
  {Wade}}, \bibinfo {author} {\bibfnamefont {J.~D.~E.}\ \bibnamefont
  {Creighton}}, \bibinfo {author} {\bibfnamefont {E.}~\bibnamefont {Ochsner}},
  \bibinfo {author} {\bibfnamefont {B.~D.}\ \bibnamefont {Lackey}}, \bibinfo
  {author} {\bibfnamefont {B.~F.}\ \bibnamefont {Farr}}, \bibinfo {author}
  {\bibfnamefont {T.~B.}\ \bibnamefont {Littenberg}},\ and\ \bibinfo {author}
  {\bibfnamefont {V.}~\bibnamefont {Raymond}},\ }\bibfield  {title} {\bibinfo
  {title} {{Systematic and statistical errors in a bayesian approach to the
  estimation of the neutron-star equation of state using advanced gravitational
  wave detectors}},\ }\href {https://doi.org/10.1103/PhysRevD.89.103012}
  {\bibfield  {journal} {\bibinfo  {journal} {Phys. Rev. D}\ }\textbf {\bibinfo
  {volume} {89}},\ \bibinfo {pages} {103012} (\bibinfo {year} {2014})},\
  \Eprint {https://arxiv.org/abs/1402.5156} {arXiv:1402.5156 [gr-qc]}
  \BibitemShut {NoStop}%
\bibitem [{\citenamefont {Hu}\ \emph {et~al.}(2021)\citenamefont {Hu},
  \citenamefont {Gao}, \citenamefont {Xu},\ and\ \citenamefont
  {Shao}}]{Hu:2021tyw}%
  \BibitemOpen
  \bibfield  {author} {\bibinfo {author} {\bibfnamefont {Z.}~\bibnamefont
  {Hu}}, \bibinfo {author} {\bibfnamefont {Y.}~\bibnamefont {Gao}}, \bibinfo
  {author} {\bibfnamefont {R.}~\bibnamefont {Xu}},\ and\ \bibinfo {author}
  {\bibfnamefont {L.}~\bibnamefont {Shao}},\ }\bibfield  {title} {\bibinfo
  {title} {{Scalarized neutron stars in massive scalar-tensor gravity: X-ray
  pulsars and tidal deformability}},\ }\href
  {https://doi.org/10.1103/PhysRevD.104.104014} {\bibfield  {journal} {\bibinfo
   {journal} {Phys. Rev. D}\ }\textbf {\bibinfo {volume} {104}},\ \bibinfo
  {pages} {104014} (\bibinfo {year} {2021})},\ \Eprint
  {https://arxiv.org/abs/2109.13453} {arXiv:2109.13453 [gr-qc]} \BibitemShut
  {NoStop}%
\bibitem [{\citenamefont {Read}\ \emph {et~al.}(2013)\citenamefont {Read},
  \citenamefont {Baiotti}, \citenamefont {Creighton}, \citenamefont {Friedman},
  \citenamefont {Giacomazzo}, \citenamefont {Kyutoku}, \citenamefont
  {Markakis}, \citenamefont {Rezzolla}, \citenamefont {Shibata},\ and\
  \citenamefont {Taniguchi}}]{Read:2013zra}%
  \BibitemOpen
  \bibfield  {author} {\bibinfo {author} {\bibfnamefont {J.~S.}\ \bibnamefont
  {Read}}, \bibinfo {author} {\bibfnamefont {L.}~\bibnamefont {Baiotti}},
  \bibinfo {author} {\bibfnamefont {J.~D.~E.}\ \bibnamefont {Creighton}},
  \bibinfo {author} {\bibfnamefont {J.~L.}\ \bibnamefont {Friedman}}, \bibinfo
  {author} {\bibfnamefont {B.}~\bibnamefont {Giacomazzo}}, \bibinfo {author}
  {\bibfnamefont {K.}~\bibnamefont {Kyutoku}}, \bibinfo {author} {\bibfnamefont
  {C.}~\bibnamefont {Markakis}}, \bibinfo {author} {\bibfnamefont
  {L.}~\bibnamefont {Rezzolla}}, \bibinfo {author} {\bibfnamefont
  {M.}~\bibnamefont {Shibata}},\ and\ \bibinfo {author} {\bibfnamefont
  {K.}~\bibnamefont {Taniguchi}},\ }\bibfield  {title} {\bibinfo {title}
  {{Matter effects on binary neutron star waveforms}},\ }\href
  {https://doi.org/10.1103/PhysRevD.88.044042} {\bibfield  {journal} {\bibinfo
  {journal} {Phys. Rev. D}\ }\textbf {\bibinfo {volume} {88}},\ \bibinfo
  {pages} {044042} (\bibinfo {year} {2013})},\ \Eprint
  {https://arxiv.org/abs/1306.4065} {arXiv:1306.4065 [gr-qc]} \BibitemShut
  {NoStop}%
\bibitem [{\citenamefont {Rezzolla}\ and\ \citenamefont
  {Takami}(2016)}]{Rezzolla:2016nxn}%
  \BibitemOpen
  \bibfield  {author} {\bibinfo {author} {\bibfnamefont {L.}~\bibnamefont
  {Rezzolla}}\ and\ \bibinfo {author} {\bibfnamefont {K.}~\bibnamefont
  {Takami}},\ }\bibfield  {title} {\bibinfo {title} {{Gravitational-wave signal
  from binary neutron stars: a systematic analysis of the spectral
  properties}},\ }\href {https://doi.org/10.1103/PhysRevD.93.124051} {\bibfield
   {journal} {\bibinfo  {journal} {Phys. Rev. D}\ }\textbf {\bibinfo {volume}
  {93}},\ \bibinfo {pages} {124051} (\bibinfo {year} {2016})},\ \Eprint
  {https://arxiv.org/abs/1604.00246} {arXiv:1604.00246 [gr-qc]} \BibitemShut
  {NoStop}%
\bibitem [{\citenamefont {Bernuzzi}\ \emph
  {et~al.}(2015{\natexlab{a}})\citenamefont {Bernuzzi}, \citenamefont {Nagar},
  \citenamefont {Dietrich},\ and\ \citenamefont {Damour}}]{Bernuzzi:2014owa}%
  \BibitemOpen
  \bibfield  {author} {\bibinfo {author} {\bibfnamefont {S.}~\bibnamefont
  {Bernuzzi}}, \bibinfo {author} {\bibfnamefont {A.}~\bibnamefont {Nagar}},
  \bibinfo {author} {\bibfnamefont {T.}~\bibnamefont {Dietrich}},\ and\
  \bibinfo {author} {\bibfnamefont {T.}~\bibnamefont {Damour}},\ }\bibfield
  {title} {\bibinfo {title} {{Modeling the Dynamics of Tidally Interacting
  Binary Neutron Stars up to the Merger}},\ }\href
  {https://doi.org/10.1103/PhysRevLett.114.161103} {\bibfield  {journal}
  {\bibinfo  {journal} {Phys. Rev. Lett.}\ }\textbf {\bibinfo {volume} {114}},\
  \bibinfo {pages} {161103} (\bibinfo {year} {2015}{\natexlab{a}})},\ \Eprint
  {https://arxiv.org/abs/1412.4553} {arXiv:1412.4553 [gr-qc]} \BibitemShut
  {NoStop}%
\bibitem [{\citenamefont {Kiuchi}\ \emph {et~al.}(2017)\citenamefont {Kiuchi},
  \citenamefont {Kawaguchi}, \citenamefont {Kyutoku}, \citenamefont
  {Sekiguchi}, \citenamefont {Shibata},\ and\ \citenamefont
  {Taniguchi}}]{Kiuchi:2017pte}%
  \BibitemOpen
  \bibfield  {author} {\bibinfo {author} {\bibfnamefont {K.}~\bibnamefont
  {Kiuchi}}, \bibinfo {author} {\bibfnamefont {K.}~\bibnamefont {Kawaguchi}},
  \bibinfo {author} {\bibfnamefont {K.}~\bibnamefont {Kyutoku}}, \bibinfo
  {author} {\bibfnamefont {Y.}~\bibnamefont {Sekiguchi}}, \bibinfo {author}
  {\bibfnamefont {M.}~\bibnamefont {Shibata}},\ and\ \bibinfo {author}
  {\bibfnamefont {K.}~\bibnamefont {Taniguchi}},\ }\bibfield  {title} {\bibinfo
  {title} {{Sub-radian-accuracy gravitational waveforms of coalescing binary
  neutron stars in numerical relativity}},\ }\href
  {https://doi.org/10.1103/PhysRevD.96.084060} {\bibfield  {journal} {\bibinfo
  {journal} {Phys. Rev. D}\ }\textbf {\bibinfo {volume} {96}},\ \bibinfo
  {pages} {084060} (\bibinfo {year} {2017})},\ \Eprint
  {https://arxiv.org/abs/1708.08926} {arXiv:1708.08926 [astro-ph.HE]}
  \BibitemShut {NoStop}%
\bibitem [{\citenamefont {Bernuzzi}\ \emph
  {et~al.}(2015{\natexlab{b}})\citenamefont {Bernuzzi}, \citenamefont
  {Dietrich},\ and\ \citenamefont {Nagar}}]{Bernuzzi:2015rla}%
  \BibitemOpen
  \bibfield  {author} {\bibinfo {author} {\bibfnamefont {S.}~\bibnamefont
  {Bernuzzi}}, \bibinfo {author} {\bibfnamefont {T.}~\bibnamefont {Dietrich}},\
  and\ \bibinfo {author} {\bibfnamefont {A.}~\bibnamefont {Nagar}},\ }\bibfield
   {title} {\bibinfo {title} {{Modeling the complete gravitational wave
  spectrum of neutron star mergers}},\ }\href
  {https://doi.org/10.1103/PhysRevLett.115.091101} {\bibfield  {journal}
  {\bibinfo  {journal} {Phys. Rev. Lett.}\ }\textbf {\bibinfo {volume} {115}},\
  \bibinfo {pages} {091101} (\bibinfo {year} {2015}{\natexlab{b}})},\ \Eprint
  {https://arxiv.org/abs/1504.01764} {arXiv:1504.01764 [gr-qc]} \BibitemShut
  {NoStop}%
\bibitem [{\citenamefont {Breschi}\ \emph {et~al.}(2024)\citenamefont
  {Breschi}, \citenamefont {Bernuzzi}, \citenamefont {Chakravarti},
  \citenamefont {Camilletti}, \citenamefont {Prakash},\ and\ \citenamefont
  {Perego}}]{Breschi:2022xnc}%
  \BibitemOpen
  \bibfield  {author} {\bibinfo {author} {\bibfnamefont {M.}~\bibnamefont
  {Breschi}}, \bibinfo {author} {\bibfnamefont {S.}~\bibnamefont {Bernuzzi}},
  \bibinfo {author} {\bibfnamefont {K.}~\bibnamefont {Chakravarti}}, \bibinfo
  {author} {\bibfnamefont {A.}~\bibnamefont {Camilletti}}, \bibinfo {author}
  {\bibfnamefont {A.}~\bibnamefont {Prakash}},\ and\ \bibinfo {author}
  {\bibfnamefont {A.}~\bibnamefont {Perego}},\ }\bibfield  {title} {\bibinfo
  {title} {{Kilohertz gravitational waves from binary neutron star mergers:
  Numerical-relativity informed postmerger model}},\ }\href
  {https://doi.org/10.1103/PhysRevD.109.064009} {\bibfield  {journal} {\bibinfo
   {journal} {Phys. Rev. D}\ }\textbf {\bibinfo {volume} {109}},\ \bibinfo
  {pages} {064009} (\bibinfo {year} {2024})},\ \Eprint
  {https://arxiv.org/abs/2205.09112} {arXiv:2205.09112 [gr-qc]} \BibitemShut
  {NoStop}%
\bibitem [{\citenamefont {Bauswein}\ \emph {et~al.}(2012)\citenamefont
  {Bauswein}, \citenamefont {Janka}, \citenamefont {Hebeler},\ and\
  \citenamefont {Schwenk}}]{Bauswein:2012ya}%
  \BibitemOpen
  \bibfield  {author} {\bibinfo {author} {\bibfnamefont {A.}~\bibnamefont
  {Bauswein}}, \bibinfo {author} {\bibfnamefont {H.~T.}\ \bibnamefont {Janka}},
  \bibinfo {author} {\bibfnamefont {K.}~\bibnamefont {Hebeler}},\ and\ \bibinfo
  {author} {\bibfnamefont {A.}~\bibnamefont {Schwenk}},\ }\bibfield  {title}
  {\bibinfo {title} {{Equation-of-state dependence of the gravitational-wave
  signal from the ring-down phase of neutron-star mergers}},\ }\href
  {https://doi.org/10.1103/PhysRevD.86.063001} {\bibfield  {journal} {\bibinfo
  {journal} {Phys. Rev. D}\ }\textbf {\bibinfo {volume} {86}},\ \bibinfo
  {pages} {063001} (\bibinfo {year} {2012})},\ \Eprint
  {https://arxiv.org/abs/1204.1888} {arXiv:1204.1888 [astro-ph.SR]}
  \BibitemShut {NoStop}%
\bibitem [{\citenamefont {Bauswein}\ \emph {et~al.}(2014)\citenamefont
  {Bauswein}, \citenamefont {Stergioulas},\ and\ \citenamefont
  {Janka}}]{Bauswein:2014qla}%
  \BibitemOpen
  \bibfield  {author} {\bibinfo {author} {\bibfnamefont {A.}~\bibnamefont
  {Bauswein}}, \bibinfo {author} {\bibfnamefont {N.}~\bibnamefont
  {Stergioulas}},\ and\ \bibinfo {author} {\bibfnamefont {H.~T.}\ \bibnamefont
  {Janka}},\ }\bibfield  {title} {\bibinfo {title} {{Revealing the high-density
  equation of state through binary neutron star mergers}},\ }\href
  {https://doi.org/10.1103/PhysRevD.90.023002} {\bibfield  {journal} {\bibinfo
  {journal} {Phys. Rev. D}\ }\textbf {\bibinfo {volume} {90}},\ \bibinfo
  {pages} {023002} (\bibinfo {year} {2014})},\ \Eprint
  {https://arxiv.org/abs/1403.5301} {arXiv:1403.5301 [astro-ph.SR]}
  \BibitemShut {NoStop}%
\bibitem [{\citenamefont {Lioutas}\ \emph {et~al.}(2021)\citenamefont
  {Lioutas}, \citenamefont {Bauswein},\ and\ \citenamefont
  {Stergioulas}}]{Lioutas:2021jbl}%
  \BibitemOpen
  \bibfield  {author} {\bibinfo {author} {\bibfnamefont {G.}~\bibnamefont
  {Lioutas}}, \bibinfo {author} {\bibfnamefont {A.}~\bibnamefont {Bauswein}},\
  and\ \bibinfo {author} {\bibfnamefont {N.}~\bibnamefont {Stergioulas}},\
  }\bibfield  {title} {\bibinfo {title} {{Frequency deviations in universal
  relations of isolated neutron stars and postmerger remnants}},\ }\href
  {https://doi.org/10.1103/PhysRevD.104.043011} {\bibfield  {journal} {\bibinfo
   {journal} {Phys. Rev. D}\ }\textbf {\bibinfo {volume} {104}},\ \bibinfo
  {pages} {043011} (\bibinfo {year} {2021})},\ \Eprint
  {https://arxiv.org/abs/2102.12455} {arXiv:2102.12455 [astro-ph.HE]}
  \BibitemShut {NoStop}%
\bibitem [{\citenamefont {Bauswein}\ \emph {et~al.}(2016)\citenamefont
  {Bauswein}, \citenamefont {Stergioulas},\ and\ \citenamefont
  {Janka}}]{Bauswein:2015vxa}%
  \BibitemOpen
  \bibfield  {author} {\bibinfo {author} {\bibfnamefont {A.}~\bibnamefont
  {Bauswein}}, \bibinfo {author} {\bibfnamefont {N.}~\bibnamefont
  {Stergioulas}},\ and\ \bibinfo {author} {\bibfnamefont {H.-T.}\ \bibnamefont
  {Janka}},\ }\bibfield  {title} {\bibinfo {title} {{Exploring properties of
  high-density matter through remnants of neutron-star mergers}},\ }\href
  {https://doi.org/10.1140/epja/i2016-16056-7} {\bibfield  {journal} {\bibinfo
  {journal} {Eur. Phys. J. A}\ }\textbf {\bibinfo {volume} {52}},\ \bibinfo
  {pages} {56} (\bibinfo {year} {2016})},\ \Eprint
  {https://arxiv.org/abs/1508.05493} {arXiv:1508.05493 [astro-ph.HE]}
  \BibitemShut {NoStop}%
\bibitem [{\citenamefont {Vretinaris}\ \emph {et~al.}(2020)\citenamefont
  {Vretinaris}, \citenamefont {Stergioulas},\ and\ \citenamefont
  {Bauswein}}]{Vretinaris:2019spn}%
  \BibitemOpen
  \bibfield  {author} {\bibinfo {author} {\bibfnamefont {S.}~\bibnamefont
  {Vretinaris}}, \bibinfo {author} {\bibfnamefont {N.}~\bibnamefont
  {Stergioulas}},\ and\ \bibinfo {author} {\bibfnamefont {A.}~\bibnamefont
  {Bauswein}},\ }\bibfield  {title} {\bibinfo {title} {{Empirical relations for
  gravitational-wave asteroseismology of binary neutron star mergers}},\ }\href
  {https://doi.org/10.1103/PhysRevD.101.084039} {\bibfield  {journal} {\bibinfo
   {journal} {Phys. Rev. D}\ }\textbf {\bibinfo {volume} {101}},\ \bibinfo
  {pages} {084039} (\bibinfo {year} {2020})},\ \Eprint
  {https://arxiv.org/abs/1910.10856} {arXiv:1910.10856 [gr-qc]} \BibitemShut
  {NoStop}%
\bibitem [{\citenamefont {Dietrich}\ \emph {et~al.}(2018)\citenamefont
  {Dietrich}, \citenamefont {Radice}, \citenamefont {Bernuzzi}, \citenamefont
  {Zappa}, \citenamefont {Perego}, \citenamefont {Br\"ugmann}, \citenamefont
  {Chaurasia}, \citenamefont {Dudi}, \citenamefont {Tichy},\ and\ \citenamefont
  {Ujevic}}]{Dietrich:2018phi}%
  \BibitemOpen
  \bibfield  {author} {\bibinfo {author} {\bibfnamefont {T.}~\bibnamefont
  {Dietrich}}, \bibinfo {author} {\bibfnamefont {D.}~\bibnamefont {Radice}},
  \bibinfo {author} {\bibfnamefont {S.}~\bibnamefont {Bernuzzi}}, \bibinfo
  {author} {\bibfnamefont {F.}~\bibnamefont {Zappa}}, \bibinfo {author}
  {\bibfnamefont {A.}~\bibnamefont {Perego}}, \bibinfo {author} {\bibfnamefont
  {B.}~\bibnamefont {Br\"ugmann}}, \bibinfo {author} {\bibfnamefont {S.~V.}\
  \bibnamefont {Chaurasia}}, \bibinfo {author} {\bibfnamefont {R.}~\bibnamefont
  {Dudi}}, \bibinfo {author} {\bibfnamefont {W.}~\bibnamefont {Tichy}},\ and\
  \bibinfo {author} {\bibfnamefont {M.}~\bibnamefont {Ujevic}},\ }\bibfield
  {title} {\bibinfo {title} {{CoRe database of binary neutron star merger
  waveforms}},\ }\href {https://doi.org/10.1088/1361-6382/aaebc0} {\bibfield
  {journal} {\bibinfo  {journal} {Class. Quant. Grav.}\ }\textbf {\bibinfo
  {volume} {35}},\ \bibinfo {pages} {24LT01} (\bibinfo {year} {2018})},\
  \Eprint {https://arxiv.org/abs/1806.01625} {arXiv:1806.01625 [gr-qc]}
  \BibitemShut {NoStop}%
\bibitem [{\citenamefont {Raithel}\ and\ \citenamefont
  {Most}(2022)}]{Raithel:2022orm}%
  \BibitemOpen
  \bibfield  {author} {\bibinfo {author} {\bibfnamefont {C.~A.}\ \bibnamefont
  {Raithel}}\ and\ \bibinfo {author} {\bibfnamefont {E.~R.}\ \bibnamefont
  {Most}},\ }\bibfield  {title} {\bibinfo {title} {{Characterizing the
  Breakdown of Quasi-universality in Postmerger Gravitational Waves from Binary
  Neutron Star Mergers}},\ }\href {https://doi.org/10.3847/2041-8213/ac7c75}
  {\bibfield  {journal} {\bibinfo  {journal} {Astrophys. J. Lett.}\ }\textbf
  {\bibinfo {volume} {933}},\ \bibinfo {pages} {L39} (\bibinfo {year}
  {2022})},\ \Eprint {https://arxiv.org/abs/2201.03594} {arXiv:2201.03594
  [astro-ph.HE]} \BibitemShut {NoStop}%
\bibitem [{\citenamefont {Most}\ \emph {et~al.}(2019)\citenamefont {Most},
  \citenamefont {Papenfort}, \citenamefont {Dexheimer}, \citenamefont
  {Hanauske}, \citenamefont {Schramm}, \citenamefont {St\"ocker},\ and\
  \citenamefont {Rezzolla}}]{Most:2018eaw}%
  \BibitemOpen
  \bibfield  {author} {\bibinfo {author} {\bibfnamefont {E.~R.}\ \bibnamefont
  {Most}}, \bibinfo {author} {\bibfnamefont {L.~J.}\ \bibnamefont {Papenfort}},
  \bibinfo {author} {\bibfnamefont {V.}~\bibnamefont {Dexheimer}}, \bibinfo
  {author} {\bibfnamefont {M.}~\bibnamefont {Hanauske}}, \bibinfo {author}
  {\bibfnamefont {S.}~\bibnamefont {Schramm}}, \bibinfo {author} {\bibfnamefont
  {H.}~\bibnamefont {St\"ocker}},\ and\ \bibinfo {author} {\bibfnamefont
  {L.}~\bibnamefont {Rezzolla}},\ }\bibfield  {title} {\bibinfo {title}
  {{Signatures of quark-hadron phase transitions in general-relativistic
  neutron-star mergers}},\ }\href
  {https://doi.org/10.1103/PhysRevLett.122.061101} {\bibfield  {journal}
  {\bibinfo  {journal} {Phys. Rev. Lett.}\ }\textbf {\bibinfo {volume} {122}},\
  \bibinfo {pages} {061101} (\bibinfo {year} {2019})},\ \Eprint
  {https://arxiv.org/abs/1807.03684} {arXiv:1807.03684 [astro-ph.HE]}
  \BibitemShut {NoStop}%
\bibitem [{\citenamefont {Bauswein}\ \emph {et~al.}(2019)\citenamefont
  {Bauswein}, \citenamefont {Bastian}, \citenamefont {Blaschke}, \citenamefont
  {Chatziioannou}, \citenamefont {Clark}, \citenamefont {Fischer},\ and\
  \citenamefont {Oertel}}]{Bauswein:2018bma}%
  \BibitemOpen
  \bibfield  {author} {\bibinfo {author} {\bibfnamefont {A.}~\bibnamefont
  {Bauswein}}, \bibinfo {author} {\bibfnamefont {N.-U.~F.}\ \bibnamefont
  {Bastian}}, \bibinfo {author} {\bibfnamefont {D.~B.}\ \bibnamefont
  {Blaschke}}, \bibinfo {author} {\bibfnamefont {K.}~\bibnamefont
  {Chatziioannou}}, \bibinfo {author} {\bibfnamefont {J.~A.}\ \bibnamefont
  {Clark}}, \bibinfo {author} {\bibfnamefont {T.}~\bibnamefont {Fischer}},\
  and\ \bibinfo {author} {\bibfnamefont {M.}~\bibnamefont {Oertel}},\
  }\bibfield  {title} {\bibinfo {title} {{Identifying a first-order phase
  transition in neutron star mergers through gravitational waves}},\ }\href
  {https://doi.org/10.1103/PhysRevLett.122.061102} {\bibfield  {journal}
  {\bibinfo  {journal} {Phys. Rev. Lett.}\ }\textbf {\bibinfo {volume} {122}},\
  \bibinfo {pages} {061102} (\bibinfo {year} {2019})},\ \Eprint
  {https://arxiv.org/abs/1809.01116} {arXiv:1809.01116 [astro-ph.HE]}
  \BibitemShut {NoStop}%
\bibitem [{\citenamefont {Weih}\ \emph {et~al.}(2020)\citenamefont {Weih},
  \citenamefont {Hanauske},\ and\ \citenamefont {Rezzolla}}]{Weih:2019xvw}%
  \BibitemOpen
  \bibfield  {author} {\bibinfo {author} {\bibfnamefont {L.~R.}\ \bibnamefont
  {Weih}}, \bibinfo {author} {\bibfnamefont {M.}~\bibnamefont {Hanauske}},\
  and\ \bibinfo {author} {\bibfnamefont {L.}~\bibnamefont {Rezzolla}},\
  }\bibfield  {title} {\bibinfo {title} {{Postmerger Gravitational-Wave
  Signatures of Phase Transitions in Binary Mergers}},\ }\href
  {https://doi.org/10.1103/PhysRevLett.124.171103} {\bibfield  {journal}
  {\bibinfo  {journal} {Phys. Rev. Lett.}\ }\textbf {\bibinfo {volume} {124}},\
  \bibinfo {pages} {171103} (\bibinfo {year} {2020})},\ \Eprint
  {https://arxiv.org/abs/1912.09340} {arXiv:1912.09340 [gr-qc]} \BibitemShut
  {NoStop}%
\bibitem [{\citenamefont {Blacker}\ \emph {et~al.}(2020)\citenamefont
  {Blacker}, \citenamefont {Bastian}, \citenamefont {Bauswein}, \citenamefont
  {Blaschke}, \citenamefont {Fischer}, \citenamefont {Oertel}, \citenamefont
  {Soultanis},\ and\ \citenamefont {Typel}}]{Blacker:2020nlq}%
  \BibitemOpen
  \bibfield  {author} {\bibinfo {author} {\bibfnamefont {S.}~\bibnamefont
  {Blacker}}, \bibinfo {author} {\bibfnamefont {N.-U.~F.}\ \bibnamefont
  {Bastian}}, \bibinfo {author} {\bibfnamefont {A.}~\bibnamefont {Bauswein}},
  \bibinfo {author} {\bibfnamefont {D.~B.}\ \bibnamefont {Blaschke}}, \bibinfo
  {author} {\bibfnamefont {T.}~\bibnamefont {Fischer}}, \bibinfo {author}
  {\bibfnamefont {M.}~\bibnamefont {Oertel}}, \bibinfo {author} {\bibfnamefont
  {T.}~\bibnamefont {Soultanis}},\ and\ \bibinfo {author} {\bibfnamefont
  {S.}~\bibnamefont {Typel}},\ }\bibfield  {title} {\bibinfo {title}
  {{Constraining the onset density of the hadron-quark phase transition with
  gravitational-wave observations}},\ }\href
  {https://doi.org/10.1103/PhysRevD.102.123023} {\bibfield  {journal} {\bibinfo
   {journal} {Phys. Rev. D}\ }\textbf {\bibinfo {volume} {102}},\ \bibinfo
  {pages} {123023} (\bibinfo {year} {2020})},\ \Eprint
  {https://arxiv.org/abs/2006.03789} {arXiv:2006.03789 [astro-ph.HE]}
  \BibitemShut {NoStop}%
\bibitem [{\citenamefont {Liebling}\ \emph {et~al.}(2021)\citenamefont
  {Liebling}, \citenamefont {Palenzuela},\ and\ \citenamefont
  {Lehner}}]{Liebling:2020dhf}%
  \BibitemOpen
  \bibfield  {author} {\bibinfo {author} {\bibfnamefont {S.~L.}\ \bibnamefont
  {Liebling}}, \bibinfo {author} {\bibfnamefont {C.}~\bibnamefont
  {Palenzuela}},\ and\ \bibinfo {author} {\bibfnamefont {L.}~\bibnamefont
  {Lehner}},\ }\bibfield  {title} {\bibinfo {title} {{Effects of High Density
  Phase Transitions on Neutron Star Dynamics}},\ }\href
  {https://doi.org/10.1088/1361-6382/abf898} {\bibfield  {journal} {\bibinfo
  {journal} {Class. Quant. Grav.}\ }\textbf {\bibinfo {volume} {38}},\ \bibinfo
  {pages} {115007} (\bibinfo {year} {2021})},\ \Eprint
  {https://arxiv.org/abs/2010.12567} {arXiv:2010.12567 [gr-qc]} \BibitemShut
  {NoStop}%
\bibitem [{\citenamefont {Prakash}\ \emph {et~al.}(2021)\citenamefont
  {Prakash}, \citenamefont {Radice}, \citenamefont {Logoteta}, \citenamefont
  {Perego}, \citenamefont {Nedora}, \citenamefont {Bombaci}, \citenamefont
  {Kashyap}, \citenamefont {Bernuzzi},\ and\ \citenamefont
  {Endrizzi}}]{Prakash:2021wpz}%
  \BibitemOpen
  \bibfield  {author} {\bibinfo {author} {\bibfnamefont {A.}~\bibnamefont
  {Prakash}}, \bibinfo {author} {\bibfnamefont {D.}~\bibnamefont {Radice}},
  \bibinfo {author} {\bibfnamefont {D.}~\bibnamefont {Logoteta}}, \bibinfo
  {author} {\bibfnamefont {A.}~\bibnamefont {Perego}}, \bibinfo {author}
  {\bibfnamefont {V.}~\bibnamefont {Nedora}}, \bibinfo {author} {\bibfnamefont
  {I.}~\bibnamefont {Bombaci}}, \bibinfo {author} {\bibfnamefont
  {R.}~\bibnamefont {Kashyap}}, \bibinfo {author} {\bibfnamefont
  {S.}~\bibnamefont {Bernuzzi}},\ and\ \bibinfo {author} {\bibfnamefont
  {A.}~\bibnamefont {Endrizzi}},\ }\bibfield  {title} {\bibinfo {title}
  {{Signatures of deconfined quark phases in binary neutron star mergers}},\
  }\href {https://doi.org/10.1103/PhysRevD.104.083029} {\bibfield  {journal}
  {\bibinfo  {journal} {Phys. Rev. D}\ }\textbf {\bibinfo {volume} {104}},\
  \bibinfo {pages} {083029} (\bibinfo {year} {2021})},\ \Eprint
  {https://arxiv.org/abs/2106.07885} {arXiv:2106.07885 [astro-ph.HE]}
  \BibitemShut {NoStop}%
\bibitem [{\citenamefont {Raithel}\ and\ \citenamefont
  {Most}(2023)}]{Raithel:2022efm}%
  \BibitemOpen
  \bibfield  {author} {\bibinfo {author} {\bibfnamefont {C.~A.}\ \bibnamefont
  {Raithel}}\ and\ \bibinfo {author} {\bibfnamefont {E.~R.}\ \bibnamefont
  {Most}},\ }\bibfield  {title} {\bibinfo {title} {{Degeneracy in the Inference
  of Phase Transitions in the Neutron Star Equation of State from Gravitational
  Wave Data}},\ }\href {https://doi.org/10.1103/PhysRevLett.130.201403}
  {\bibfield  {journal} {\bibinfo  {journal} {Phys. Rev. Lett.}\ }\textbf
  {\bibinfo {volume} {130}},\ \bibinfo {pages} {201403} (\bibinfo {year}
  {2023})},\ \Eprint {https://arxiv.org/abs/2208.04294} {arXiv:2208.04294
  [astro-ph.HE]} \BibitemShut {NoStop}%
\bibitem [{\citenamefont {Fujimoto}\ \emph {et~al.}(2023)\citenamefont
  {Fujimoto}, \citenamefont {Fukushima}, \citenamefont {Hotokezaka},\ and\
  \citenamefont {Kyutoku}}]{Fujimoto:2022xhv}%
  \BibitemOpen
  \bibfield  {author} {\bibinfo {author} {\bibfnamefont {Y.}~\bibnamefont
  {Fujimoto}}, \bibinfo {author} {\bibfnamefont {K.}~\bibnamefont {Fukushima}},
  \bibinfo {author} {\bibfnamefont {K.}~\bibnamefont {Hotokezaka}},\ and\
  \bibinfo {author} {\bibfnamefont {K.}~\bibnamefont {Kyutoku}},\ }\bibfield
  {title} {\bibinfo {title} {{Gravitational Wave Signal for Quark Matter with
  Realistic Phase Transition}},\ }\href
  {https://doi.org/10.1103/PhysRevLett.130.091404} {\bibfield  {journal}
  {\bibinfo  {journal} {Phys. Rev. Lett.}\ }\textbf {\bibinfo {volume} {130}},\
  \bibinfo {pages} {091404} (\bibinfo {year} {2023})},\ \Eprint
  {https://arxiv.org/abs/2205.03882} {arXiv:2205.03882 [astro-ph.HE]}
  \BibitemShut {NoStop}%
\bibitem [{\citenamefont {Haque}\ \emph {et~al.}(2023)\citenamefont {Haque},
  \citenamefont {Mallick},\ and\ \citenamefont {Thakur}}]{Haque:2022dsc}%
  \BibitemOpen
  \bibfield  {author} {\bibinfo {author} {\bibfnamefont {S.}~\bibnamefont
  {Haque}}, \bibinfo {author} {\bibfnamefont {R.}~\bibnamefont {Mallick}},\
  and\ \bibinfo {author} {\bibfnamefont {S.~K.}\ \bibnamefont {Thakur}},\
  }\bibfield  {title} {\bibinfo {title} {{Effects of onset of phase transition
  on binary neutron star mergers}},\ }\href
  {https://doi.org/10.1093/mnras/stad3839} {\bibfield  {journal} {\bibinfo
  {journal} {Mon. Not. Roy. Astron. Soc.}\ }\textbf {\bibinfo {volume} {527}},\
  \bibinfo {pages} {11575} (\bibinfo {year} {2023})},\ \Eprint
  {https://arxiv.org/abs/2207.14485} {arXiv:2207.14485 [astro-ph.HE]}
  \BibitemShut {NoStop}%
\bibitem [{\citenamefont {Prakash}\ \emph {et~al.}(2024)\citenamefont
  {Prakash}, \citenamefont {Gupta}, \citenamefont {Breschi}, \citenamefont
  {Kashyap}, \citenamefont {Radice}, \citenamefont {Bernuzzi}, \citenamefont
  {Logoteta},\ and\ \citenamefont {Sathyaprakash}}]{Prakash:2023afe}%
  \BibitemOpen
  \bibfield  {author} {\bibinfo {author} {\bibfnamefont {A.}~\bibnamefont
  {Prakash}}, \bibinfo {author} {\bibfnamefont {I.}~\bibnamefont {Gupta}},
  \bibinfo {author} {\bibfnamefont {M.}~\bibnamefont {Breschi}}, \bibinfo
  {author} {\bibfnamefont {R.}~\bibnamefont {Kashyap}}, \bibinfo {author}
  {\bibfnamefont {D.}~\bibnamefont {Radice}}, \bibinfo {author} {\bibfnamefont
  {S.}~\bibnamefont {Bernuzzi}}, \bibinfo {author} {\bibfnamefont
  {D.}~\bibnamefont {Logoteta}},\ and\ \bibinfo {author} {\bibfnamefont
  {B.~S.}\ \bibnamefont {Sathyaprakash}},\ }\bibfield  {title} {\bibinfo
  {title} {{Detectability of QCD phase transitions in binary neutron star
  mergers: Bayesian inference with the next generation gravitational wave
  detectors}},\ }\href {https://doi.org/10.1103/PhysRevD.109.103008} {\bibfield
   {journal} {\bibinfo  {journal} {Phys. Rev. D}\ }\textbf {\bibinfo {volume}
  {109}},\ \bibinfo {pages} {103008} (\bibinfo {year} {2024})},\ \Eprint
  {https://arxiv.org/abs/2310.06025} {arXiv:2310.06025 [gr-qc]} \BibitemShut
  {NoStop}%
\bibitem [{\citenamefont {Huang}\ \emph {et~al.}(2022)\citenamefont {Huang},
  \citenamefont {Baiotti}, \citenamefont {Kojo}, \citenamefont {Takami},
  \citenamefont {Sotani}, \citenamefont {Togashi}, \citenamefont {Hatsuda},
  \citenamefont {Nagataki},\ and\ \citenamefont {Fan}}]{Huang:2022mqp}%
  \BibitemOpen
  \bibfield  {author} {\bibinfo {author} {\bibfnamefont {Y.-J.}\ \bibnamefont
  {Huang}}, \bibinfo {author} {\bibfnamefont {L.}~\bibnamefont {Baiotti}},
  \bibinfo {author} {\bibfnamefont {T.}~\bibnamefont {Kojo}}, \bibinfo {author}
  {\bibfnamefont {K.}~\bibnamefont {Takami}}, \bibinfo {author} {\bibfnamefont
  {H.}~\bibnamefont {Sotani}}, \bibinfo {author} {\bibfnamefont
  {H.}~\bibnamefont {Togashi}}, \bibinfo {author} {\bibfnamefont
  {T.}~\bibnamefont {Hatsuda}}, \bibinfo {author} {\bibfnamefont
  {S.}~\bibnamefont {Nagataki}},\ and\ \bibinfo {author} {\bibfnamefont
  {Y.-Z.}\ \bibnamefont {Fan}},\ }\bibfield  {title} {\bibinfo {title} {{Merger
  and Postmerger of Binary Neutron Stars with a Quark-Hadron Crossover Equation
  of State}},\ }\href {https://doi.org/10.1103/PhysRevLett.129.181101}
  {\bibfield  {journal} {\bibinfo  {journal} {Phys. Rev. Lett.}\ }\textbf
  {\bibinfo {volume} {129}},\ \bibinfo {pages} {181101} (\bibinfo {year}
  {2022})},\ \Eprint {https://arxiv.org/abs/2203.04528} {arXiv:2203.04528
  [astro-ph.HE]} \BibitemShut {NoStop}%
\bibitem [{\citenamefont {Baym}\ \emph {et~al.}(2018)\citenamefont {Baym},
  \citenamefont {Hatsuda}, \citenamefont {Kojo}, \citenamefont {Powell},
  \citenamefont {Song},\ and\ \citenamefont {Takatsuka}}]{Baym:2017whm}%
  \BibitemOpen
  \bibfield  {author} {\bibinfo {author} {\bibfnamefont {G.}~\bibnamefont
  {Baym}}, \bibinfo {author} {\bibfnamefont {T.}~\bibnamefont {Hatsuda}},
  \bibinfo {author} {\bibfnamefont {T.}~\bibnamefont {Kojo}}, \bibinfo {author}
  {\bibfnamefont {P.~D.}\ \bibnamefont {Powell}}, \bibinfo {author}
  {\bibfnamefont {Y.}~\bibnamefont {Song}},\ and\ \bibinfo {author}
  {\bibfnamefont {T.}~\bibnamefont {Takatsuka}},\ }\bibfield  {title} {\bibinfo
  {title} {{From hadrons to quarks in neutron stars: a review}},\ }\href
  {https://doi.org/10.1088/1361-6633/aaae14} {\bibfield  {journal} {\bibinfo
  {journal} {Rept. Prog. Phys.}\ }\textbf {\bibinfo {volume} {81}},\ \bibinfo
  {pages} {056902} (\bibinfo {year} {2018})},\ \Eprint
  {https://arxiv.org/abs/1707.04966} {arXiv:1707.04966 [astro-ph.HE]}
  \BibitemShut {NoStop}%
\bibitem [{\citenamefont {Baym}\ \emph {et~al.}(2019)\citenamefont {Baym},
  \citenamefont {Furusawa}, \citenamefont {Hatsuda}, \citenamefont {Kojo},\
  and\ \citenamefont {Togashi}}]{Baym:2019iky}%
  \BibitemOpen
  \bibfield  {author} {\bibinfo {author} {\bibfnamefont {G.}~\bibnamefont
  {Baym}}, \bibinfo {author} {\bibfnamefont {S.}~\bibnamefont {Furusawa}},
  \bibinfo {author} {\bibfnamefont {T.}~\bibnamefont {Hatsuda}}, \bibinfo
  {author} {\bibfnamefont {T.}~\bibnamefont {Kojo}},\ and\ \bibinfo {author}
  {\bibfnamefont {H.}~\bibnamefont {Togashi}},\ }\bibfield  {title} {\bibinfo
  {title} {{New Neutron Star Equation of State with Quark-Hadron Crossover}},\
  }\href {https://doi.org/10.3847/1538-4357/ab441e} {\bibfield  {journal}
  {\bibinfo  {journal} {Astrophys. J.}\ }\textbf {\bibinfo {volume} {885}},\
  \bibinfo {pages} {42} (\bibinfo {year} {2019})},\ \Eprint
  {https://arxiv.org/abs/1903.08963} {arXiv:1903.08963 [astro-ph.HE]}
  \BibitemShut {NoStop}%
\bibitem [{\citenamefont {Blacker}\ and\ \citenamefont
  {Bauswein}(2024)}]{Blacker:2024tet}%
  \BibitemOpen
  \bibfield  {author} {\bibinfo {author} {\bibfnamefont {S.}~\bibnamefont
  {Blacker}}\ and\ \bibinfo {author} {\bibfnamefont {A.}~\bibnamefont
  {Bauswein}},\ }\bibfield  {title} {\bibinfo {title} {{Comprehensive survey of
  hybrid equations of state in neutron star mergers and constraints on the
  hadron-quark phase transition}},\ }\href@noop {} {\bibfield  {journal}
  {\bibinfo  {journal} {arXiv: 2406.14669}\ } (\bibinfo {year} {2024})},\
  \Eprint {https://arxiv.org/abs/2406.14669} {arXiv:2406.14669 [astro-ph.HE]}
  \BibitemShut {NoStop}%
\bibitem [{\citenamefont {Tonetto}\ and\ \citenamefont
  {Lugones}(2020)}]{Tonetto:2020bie}%
  \BibitemOpen
  \bibfield  {author} {\bibinfo {author} {\bibfnamefont {L.}~\bibnamefont
  {Tonetto}}\ and\ \bibinfo {author} {\bibfnamefont {G.}~\bibnamefont
  {Lugones}},\ }\bibfield  {title} {\bibinfo {title} {{Discontinuity gravity
  modes in hybrid stars: assessing the role of rapid and slow phase
  conversions}},\ }\href {https://doi.org/10.1103/PhysRevD.101.123029}
  {\bibfield  {journal} {\bibinfo  {journal} {Phys. Rev. D}\ }\textbf {\bibinfo
  {volume} {101}},\ \bibinfo {pages} {123029} (\bibinfo {year} {2020})},\
  \Eprint {https://arxiv.org/abs/2003.01259} {arXiv:2003.01259 [astro-ph.HE]}
  \BibitemShut {NoStop}%
\bibitem [{\citenamefont {Lau}\ and\ \citenamefont {Yagi}(2021)}]{Lau:2020bfq}%
  \BibitemOpen
  \bibfield  {author} {\bibinfo {author} {\bibfnamefont {S.~Y.}\ \bibnamefont
  {Lau}}\ and\ \bibinfo {author} {\bibfnamefont {K.}~\bibnamefont {Yagi}},\
  }\bibfield  {title} {\bibinfo {title} {{Probing hybrid stars with
  gravitational waves via interfacial modes}},\ }\href
  {https://doi.org/10.1103/PhysRevD.103.063015} {\bibfield  {journal} {\bibinfo
   {journal} {Phys. Rev. D}\ }\textbf {\bibinfo {volume} {103}},\ \bibinfo
  {pages} {063015} (\bibinfo {year} {2021})},\ \Eprint
  {https://arxiv.org/abs/2012.13000} {arXiv:2012.13000 [astro-ph.HE]}
  \BibitemShut {NoStop}%
\bibitem [{\citenamefont {Zhu}\ \emph {et~al.}(2023)\citenamefont {Zhu},
  \citenamefont {Wang}, \citenamefont {Xia}, \citenamefont {Zhou},\ and\
  \citenamefont {Ma}}]{Zhu:2022pja}%
  \BibitemOpen
  \bibfield  {author} {\bibinfo {author} {\bibfnamefont {J.}~\bibnamefont
  {Zhu}}, \bibinfo {author} {\bibfnamefont {C.}~\bibnamefont {Wang}}, \bibinfo
  {author} {\bibfnamefont {C.}~\bibnamefont {Xia}}, \bibinfo {author}
  {\bibfnamefont {E.}~\bibnamefont {Zhou}},\ and\ \bibinfo {author}
  {\bibfnamefont {Y.}~\bibnamefont {Ma}},\ }\bibfield  {title} {\bibinfo
  {title} {{Probing phase transitions in neutron stars via the crust-core
  interfacial mode}},\ }\href {https://doi.org/10.1103/PhysRevD.107.083023}
  {\bibfield  {journal} {\bibinfo  {journal} {Phys. Rev. D}\ }\textbf {\bibinfo
  {volume} {107}},\ \bibinfo {pages} {083023} (\bibinfo {year} {2023})},\
  \Eprint {https://arxiv.org/abs/2211.11529} {arXiv:2211.11529 [astro-ph.HE]}
  \BibitemShut {NoStop}%
\bibitem [{\citenamefont {Mendes}\ and\ \citenamefont
  {Ortiz}(2018)}]{Mendes:2018qwo}%
  \BibitemOpen
  \bibfield  {author} {\bibinfo {author} {\bibfnamefont {R.~F.~P.}\
  \bibnamefont {Mendes}}\ and\ \bibinfo {author} {\bibfnamefont
  {N.}~\bibnamefont {Ortiz}},\ }\bibfield  {title} {\bibinfo {title} {{New
  class of quasinormal modes of neutron stars in scalar-tensor gravity}},\
  }\href {https://doi.org/10.1103/PhysRevLett.120.201104} {\bibfield  {journal}
  {\bibinfo  {journal} {Phys. Rev. Lett.}\ }\textbf {\bibinfo {volume} {120}},\
  \bibinfo {pages} {201104} (\bibinfo {year} {2018})},\ \Eprint
  {https://arxiv.org/abs/1802.07847} {arXiv:1802.07847 [gr-qc]} \BibitemShut
  {NoStop}%
\bibitem [{\citenamefont {Figura}\ \emph {et~al.}(2020)\citenamefont {Figura},
  \citenamefont {Lu}, \citenamefont {Burgio}, \citenamefont {Li},\ and\
  \citenamefont {Schulze}}]{Figura:2020fkj}%
  \BibitemOpen
  \bibfield  {author} {\bibinfo {author} {\bibfnamefont {A.}~\bibnamefont
  {Figura}}, \bibinfo {author} {\bibfnamefont {J.~J.}\ \bibnamefont {Lu}},
  \bibinfo {author} {\bibfnamefont {G.~F.}\ \bibnamefont {Burgio}}, \bibinfo
  {author} {\bibfnamefont {Z.~H.}\ \bibnamefont {Li}},\ and\ \bibinfo {author}
  {\bibfnamefont {H.~J.}\ \bibnamefont {Schulze}},\ }\bibfield  {title}
  {\bibinfo {title} {{Hybrid equation of state approach in binary neutron-star
  merger simulations}},\ }\href {https://doi.org/10.1103/PhysRevD.102.043006}
  {\bibfield  {journal} {\bibinfo  {journal} {Phys. Rev. D}\ }\textbf {\bibinfo
  {volume} {102}},\ \bibinfo {pages} {043006} (\bibinfo {year} {2020})},\
  \Eprint {https://arxiv.org/abs/2005.08691} {arXiv:2005.08691 [gr-qc]}
  \BibitemShut {NoStop}%
\bibitem [{\citenamefont {Raithel}\ and\ \citenamefont
  {Paschalidis}(2024)}]{Raithel:2023gct}%
  \BibitemOpen
  \bibfield  {author} {\bibinfo {author} {\bibfnamefont {C.~A.}\ \bibnamefont
  {Raithel}}\ and\ \bibinfo {author} {\bibfnamefont {V.}~\bibnamefont
  {Paschalidis}},\ }\bibfield  {title} {\bibinfo {title} {{Detectability of
  finite-temperature effects from neutron star mergers with next-generation
  gravitational wave detectors}},\ }\href
  {https://doi.org/10.1103/PhysRevD.110.043002} {\bibfield  {journal} {\bibinfo
   {journal} {Phys. Rev. D}\ }\textbf {\bibinfo {volume} {110}},\ \bibinfo
  {pages} {043002} (\bibinfo {year} {2024})},\ \Eprint
  {https://arxiv.org/abs/2312.14046} {arXiv:2312.14046 [astro-ph.HE]}
  \BibitemShut {NoStop}%
\bibitem [{\citenamefont {Fields}\ \emph {et~al.}(2023)\citenamefont {Fields},
  \citenamefont {Prakash}, \citenamefont {Breschi}, \citenamefont {Radice},
  \citenamefont {Bernuzzi},\ and\ \citenamefont
  {da~Silva~Schneider}}]{Fields:2023bhs}%
  \BibitemOpen
  \bibfield  {author} {\bibinfo {author} {\bibfnamefont {J.}~\bibnamefont
  {Fields}}, \bibinfo {author} {\bibfnamefont {A.}~\bibnamefont {Prakash}},
  \bibinfo {author} {\bibfnamefont {M.}~\bibnamefont {Breschi}}, \bibinfo
  {author} {\bibfnamefont {D.}~\bibnamefont {Radice}}, \bibinfo {author}
  {\bibfnamefont {S.}~\bibnamefont {Bernuzzi}},\ and\ \bibinfo {author}
  {\bibfnamefont {A.}~\bibnamefont {da~Silva~Schneider}},\ }\bibfield  {title}
  {\bibinfo {title} {{Thermal Effects in Binary Neutron Star Mergers}},\ }\href
  {https://doi.org/10.3847/2041-8213/ace5b2} {\bibfield  {journal} {\bibinfo
  {journal} {Astrophys. J. Lett.}\ }\textbf {\bibinfo {volume} {952}},\
  \bibinfo {pages} {L36} (\bibinfo {year} {2023})},\ \Eprint
  {https://arxiv.org/abs/2302.11359} {arXiv:2302.11359 [astro-ph.HE]}
  \BibitemShut {NoStop}%
\bibitem [{\citenamefont {Miravet-Ten\'es}\ \emph {et~al.}(2024)\citenamefont
  {Miravet-Ten\'es}, \citenamefont {Guerra}, \citenamefont {Ruiz},
  \citenamefont {Cerd\'a-Dur\'an},\ and\ \citenamefont
  {Font}}]{Miravet-Tenes:2024vba}%
  \BibitemOpen
  \bibfield  {author} {\bibinfo {author} {\bibfnamefont {M.}~\bibnamefont
  {Miravet-Ten\'es}}, \bibinfo {author} {\bibfnamefont {D.}~\bibnamefont
  {Guerra}}, \bibinfo {author} {\bibfnamefont {M.}~\bibnamefont {Ruiz}},
  \bibinfo {author} {\bibfnamefont {P.}~\bibnamefont {Cerd\'a-Dur\'an}},\ and\
  \bibinfo {author} {\bibfnamefont {J.~A.}\ \bibnamefont {Font}},\ }\bibfield
  {title} {\bibinfo {title} {{Identifying thermal effects in neutron star
  merger remnants with model-agnostic waveform reconstructions and
  third-generation detectors}},\ }\href@noop {} {\bibfield  {journal} {\bibinfo
   {journal} {arXiv:2401.02493}\ } (\bibinfo {year} {2024})},\ \Eprint
  {https://arxiv.org/abs/2401.02493} {arXiv:2401.02493 [gr-qc]} \BibitemShut
  {NoStop}%
\bibitem [{\citenamefont {Creci}\ \emph {et~al.}(2023)\citenamefont {Creci},
  \citenamefont {Hinderer},\ and\ \citenamefont {Steinhoff}}]{Creci:2023cfx}%
  \BibitemOpen
  \bibfield  {author} {\bibinfo {author} {\bibfnamefont {G.}~\bibnamefont
  {Creci}}, \bibinfo {author} {\bibfnamefont {T.}~\bibnamefont {Hinderer}},\
  and\ \bibinfo {author} {\bibfnamefont {J.}~\bibnamefont {Steinhoff}},\
  }\bibfield  {title} {\bibinfo {title} {{Tidal properties of neutron stars in
  scalar-tensor theories of gravity}},\ }\href
  {https://doi.org/10.1103/PhysRevD.108.124073} {\bibfield  {journal} {\bibinfo
   {journal} {Phys. Rev. D}\ }\textbf {\bibinfo {volume} {108}},\ \bibinfo
  {pages} {124073} (\bibinfo {year} {2023})},\ \Eprint
  {https://arxiv.org/abs/2308.11323} {arXiv:2308.11323 [gr-qc]} \BibitemShut
  {NoStop}%
\bibitem [{\citenamefont {Pani}\ and\ \citenamefont
  {Berti}(2014)}]{Pani:2014jra}%
  \BibitemOpen
  \bibfield  {author} {\bibinfo {author} {\bibfnamefont {P.}~\bibnamefont
  {Pani}}\ and\ \bibinfo {author} {\bibfnamefont {E.}~\bibnamefont {Berti}},\
  }\bibfield  {title} {\bibinfo {title} {{Slowly rotating neutron stars in
  scalar-tensor theories}},\ }\href
  {https://doi.org/10.1103/PhysRevD.90.024025} {\bibfield  {journal} {\bibinfo
  {journal} {Phys. Rev. D}\ }\textbf {\bibinfo {volume} {90}},\ \bibinfo
  {pages} {024025} (\bibinfo {year} {2014})},\ \Eprint
  {https://arxiv.org/abs/1405.4547} {arXiv:1405.4547 [gr-qc]} \BibitemShut
  {NoStop}%
\bibitem [{\citenamefont {Brown}(2023)}]{Brown:2022kbw}%
  \BibitemOpen
  \bibfield  {author} {\bibinfo {author} {\bibfnamefont {S.~M.}\ \bibnamefont
  {Brown}},\ }\bibfield  {title} {\bibinfo {title} {{Tidal Deformability of
  Neutron Stars in Scalar-tensor Theories of Gravity}},\ }\href
  {https://doi.org/10.3847/1538-4357/acfbe5} {\bibfield  {journal} {\bibinfo
  {journal} {Astrophys. J.}\ }\textbf {\bibinfo {volume} {958}},\ \bibinfo
  {pages} {125} (\bibinfo {year} {2023})},\ \Eprint
  {https://arxiv.org/abs/2210.14025} {arXiv:2210.14025 [gr-qc]} \BibitemShut
  {NoStop}%
\end{thebibliography}%

\end{document}